\documentclass[letterpaper,11pt]{article}

\usepackage{jheppub}
\usepackage{subfig}
\usepackage{slashed}
\usepackage[T1]{fontenc}

\def\Fig#1{Fig.~{\ref{#1}}}
\def\eq#1{Eq.~(\ref{#1})}

\DeclareRobustCommand{\Sec}[1]{Sec.~\ref{#1}}
\DeclareRobustCommand{\Eq}[1]{Eq.~(\ref{#1})}

\newcommand{\Sl}[1]{\slashed{#1}}

\newcommand{\bea}{\begin{eqnarray}}
\newcommand{\eea}{\end{eqnarray}}

\def\cB{\mathcal{B}}

\def\cL{\mathcal{L}}
\def\cM{\mathcal{M}}

\def\cP{\mathcal{P}}

\def \as {\relax\ifmmode\alpha_s\else{$\alpha_s${ }}\fi}

\newcommand{\df}{\mathrm{d}}
\newcommand{\nn}{\nonumber}

\allowdisplaybreaks[4]

\definecolor{pur}{rgb}{0.6,0.,1.}
\definecolor{orange}{rgb}{1,0.5,0}

\begin{document}

\title{Energy Correlators in $V+X$ as a Benchmark Observable for Precision QCD}

\author[a]{Terry Generet,}
\affiliation[a]{Cavendish Laboratory, University of Cambridge, Cambridge CB3 0HE, UK}
\emailAdd{generet@hep.phy.cam.ac.uk}

\author[b]{Kyle Lee,}
\affiliation[b]{High Energy Physics Division, Argonne National Laboratory, Lemont, IL 60439, USA}
\emailAdd{kyle@anl.gov}

\author[c]{Ian Moult,}
\affiliation[c]{Department of Physics, Yale University, New Haven, CT 06511, USA}
\emailAdd{ian.moult@yale.edu}

\author[d]{Rene Poncelet,}
\affiliation[d]{Institute of Nuclear Physics, ul. Radzikowskiego 152, 31--342 Krakow, Poland}
\emailAdd{rene.poncelet@ifj.edu.pl}

\author[e]{and Xiaoyuan Zhang}
\affiliation[e]{Center for Theoretical Physics - a Leinweber Institute, Massachusetts Institute of Technology, Cambridge, MA 02139, USA}
\emailAdd{xyz2@mit.edu}
\preprint{\vbox{%
    \hbox{Cavendish-HEP-26/04}
    \hbox{MIT-CTP 6030}
    \hbox{IFJPAN-IV-2026-14}
}}

\abstract{
We propose projected energy correlators measured on the recoiling QCD radiation of a $Z/\gamma$ as a benchmark observable for precision QCD at the LHC. Using the $Z/\gamma$ as a hard scale prevents the need for a jet algorithm, simplifying both the perturbative and non-perturbative corrections.
We develop a framework to combine state-of-the-art fixed-order amplitudes, high order resummation, and universal non-perturbative matrix elements.
Our approach is based on numerically computed inclusive hard functions, allowing flexibility in the process and the inclusion of realistic experimental cuts.
We perform detailed numerical studies of the projected energy correlators at next-to-leading order + next-to-next-to-leading logarithm (NLO+NNLL) to verify the stability of our setup. 
We present numerical results at NLO+NNLL, which are the first complete matched predictions for energy correlators at the LHC at this order.
We discuss the prospects for extensions to higher orders, outlining a path to NNLO calculations of energy correlators at the LHC.

}

\maketitle

\section{Introduction}\label{sec:intro}

The Large Hadron Collider (LHC), together with its high-luminosity upgrade (HL-LHC), will deliver an enormous dataset with the potential to sharpen our understanding of the dynamics of Quantum Chromodynamics (QCD) at the TeV scale, probing increasingly extreme regions of phase space with ever-greater precision. Turning this data into precise tests of the Standard Model is theoretically demanding. Precision calculations of QCD observables at hadron colliders rest on three ingredients: high-multiplicity, high-loop scattering amplitudes; efficient subtraction schemes that turn these amplitudes into predictions for physical, infrared-safe observables; and factorization theorems that organize the dynamics across the disparate scales involved. Over the past decade each of these directions has advanced dramatically, and they are now beginning to converge. In this paper we argue that they come together in a particularly clean way for a single observable: the projected energy correlator measured on the QCD radiation that recoils against an electroweak boson. We develop the theoretical framework needed to predict it, and present the first complete matched results for energy correlators at the LHC with unparalleled accuracy.

This convergence is most visible in the recent conversion of two-loop amplitudes for $2\to 3$ scattering into NNLO phenomenology. Fully differential predictions are now available for a wide range of final states, including $pp\to\gamma\gamma\gamma$ at leading~\cite{Chawdhry:2019bji} and full~\cite{Czakon:2025wgs} color, $pp\to\gamma jj$~\cite{Badger:2023mgf}, and $pp\to jjj$~\cite{Czakon:2021mjy,Alvarez:2023fhi}, and they have been used to compute hadron-collider event shapes in both multijet~\cite{Alvarez:2023fhi} and diphoton~\cite{Buccioni:2025bkl} final states. The reach has extended to processes with a massive leg, such as the associated production of an electroweak boson with a bottom-quark pair, $pp\to Wb\bar b$~\cite{Hartanto:2022qhh,Buonocore:2022pqq}, with related processes matched to parton showers~\cite{Mazzitelli:2024ura,Biello:2024pgo}, and even to two massive legs in $pp\to t\bar t j$~\cite{Badger:2025ilt}. This mirrors the history of $e^+e^-$ colliders, where the two-loop three-parton amplitudes had long been known and it was the construction of suitable infrared subtraction schemes that finally enabled the calculation of event shapes at NNLO~\cite{Gehrmann-DeRidder:2007foh,Gehrmann-DeRidder:2007vsv,Gehrmann-DeRidder:2007nzq,Weinzierl:2008iv}, ushering in an era of precision QCD. It is natural to ask whether these same ingredients can now be brought to bear on collider processes at the LHC.

In parallel, jet substructure at the LHC has grown into one of the most active areas of collider physics~\cite{Larkoski:2017jix,Asquith:2018igt}. Building on the extensive study of event shapes in the controlled environment of $e^+e^-$ colliders, it has greatly broadened the set of available observables and adapted them to the hadron collider environment, probing QCD in regimes that were previously inaccessible. Combining this progress with modern fixed-order calculations has proven difficult: many jet substructure observables are either too intricate to compute at high orders, or are contaminated by large contributions from the underlying event and other soft hadronic activity. The soft-drop groomed jet mass~\cite{Frye:2016aiz,Frye:2016okc} was among the first jet substructure observables to be matched to fixed order at next-to-leading order. However, the grooming procedure complicates the perturbative description and the treatment of non-perturbative corrections~\cite{Hoang:2019ceu,Ferdinand:2023vaf}, and prospect of going to even higher order and possibility of precision comparison with data remain uncertain.

Energy correlators sidestep several of these difficulties. Building on the energy-energy correlator of Refs.~\cite{Basham:1978zq,Basham:1979gh,Basham:1977iq,Basham:1978bw} applied to $e^+e^-$ colliders, it was proposed in Refs.~\cite{Chen:2020vvp,Lee:2022ige} to reformulate jet substructure in terms of correlation functions of energy flow operators; see Ref.~\cite{Moult:2025nhu} for a recent review. They admit a clean field-theoretic definition and can be computed on charged-particle (track) information with correspondingly improved angular resolution~\cite{Chen:2020vvp,Komiske:2022enw,Lee:2026hub}. Of particular interest are the projected $N$-point correlators~\cite{Chen:2020vvp}, which retain only the largest pairwise angle among the $N$ particles; varying $N$ yields a family of observables whose ratios furnish additional, theoretically robust probes~\cite{Lee:2022ige}, in analogy with the classic $3/2$ jet ratios. Because they weight final-state particles by their energy, energy correlators are dominated by the energetic core of the radiation and are correspondingly insensitive to soft contamination from pileup and the underlying event. Furthermore, for precision QCD, in the collinear limit these projected correlators exhibit a simple, approximately power-law scaling in the opening angle, with an exponent set by the timelike anomalous dimensions that is proportional to $\alpha_s$ at leading order~\cite{Dixon:2019uzg,Chen:2020vvp}. This makes the slope of the distribution a theoretically well-controlled handle on the strong coupling constant $\alpha_s$ at hadron colliders.

The promise of this approach has already been borne out in data. Projected energy correlators and their ratios were measured inside high-energy jets at the LHC by CMS~\cite{CMS:2024mlf}, enabling one of the most precise extractions of $\alpha_s$ from jet substructure to date, and they have also been measured in $Z$-boson-tagged jets in heavy-ion collisions~\cite{CMS:2025jam}. Complementary information has come from the transverse energy-energy correlator, which is defined on the full hadronic final state and has been used to determine $\alpha_s$ at the LHC~\cite{ATLAS:2023tgo}, although accessing the collinear scaling regime that makes energy correlators such precise probes of $\alpha_s$ is difficult for an observable defined on the entire event. Together, these measurements make the experimental case for a precision jet substructure program at the LHC. Realizing it requires a theoretical description with control over both the perturbative and the non-perturbative components in the hadron collider environment.

\begin{figure}
\centering
\includegraphics[width=0.45\textwidth]{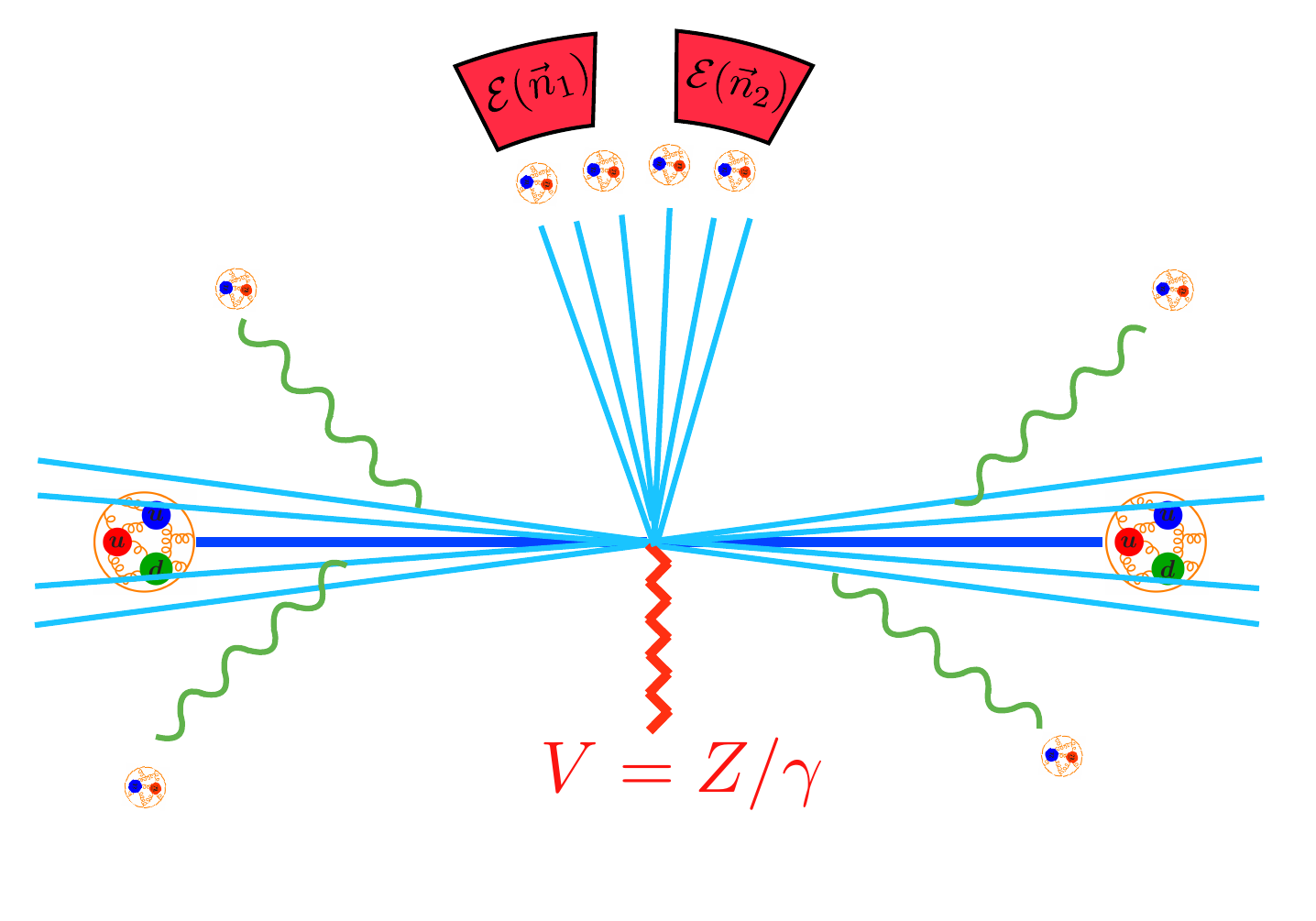}
\includegraphics[width=0.45\textwidth]{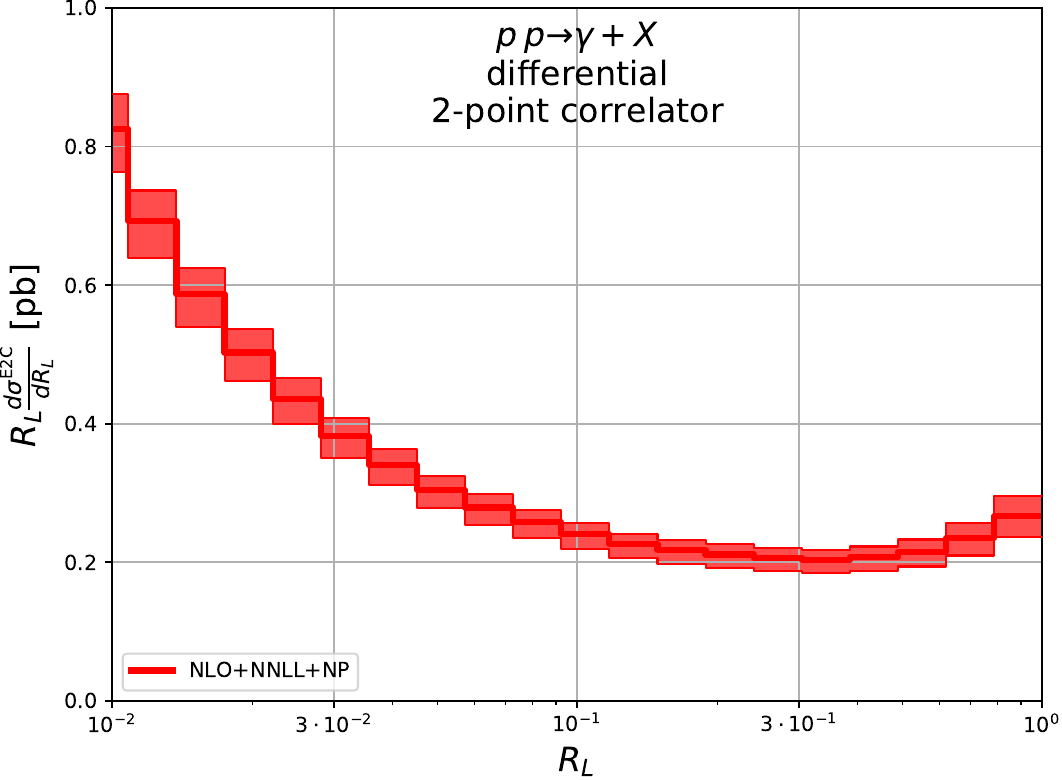}
\caption{Left: Illustration of the projected energy correlator in $V+X$. An electroweak boson $V=\gamma,Z$ recoils against the QCD radiation produced in the collision, and the correlation of energy flow in the recoiling radiation is measured by the energy flow operators $\mathcal{E}(\vec n_i)$. Right: State of the art calculation of the two-point energy correlator, at NLO+NNLL, and including leading non-perturbative corrections.}
\label{fig:intro_fig}
\end{figure}

The observable we study in this paper is especially well suited to closing this gap. We propose measuring projected energy correlators on the QCD radiation recoiling against an electroweak boson, $pp\to V+X$ with $V=\gamma,Z$. The key point is that the $V$ supplies the hard scale of the problem on its own. As a result, no jet algorithm is needed to reconstruct the hard scattering, and the correlator can be measured directly on the hadronic recoil without reference to a jet axis. This greatly simplifies the observable: there is no jet algorithm to complicate the perturbative description, and none of the power corrections in the jet radius that accompany substructure measured on a reconstructed jet. The result is an especially clean setting in which to combine the three ingredients of precision QCD identified above. The observable is illustrated in Fig.~\ref{fig:intro_fig}: the boson recoils against the QCD radiation produced in the collision, and the correlation of energy flow is measured on that recoiling radiation. The process also places jet substructure in direct contact with the frontier of multi-leg scattering amplitudes. Reaching NNLO accuracy, for instance, requires the two-loop amplitude for the boson together with four partons, a five-point amplitude with a single off-shell color-singlet leg, of the kind that has recently become available~\cite{DeLaurentis:2025dxw}.

At the heart of this program is a factorization theorem for the energy correlators in the collinear limit, derived for the two-point case in Ref.~\cite{Dixon:2019uzg} and generalized to the projected $N$-point correlators in Ref.~\cite{Chen:2020vvp}, with the extension to hadron colliders following Ref.~\cite{Lee:2022uwt,Lee:2024icn}. It expresses the observable as a convolution of a jet function, which captures the collinear dynamics and is universal across processes, with an inclusive hard function, the same partonic coefficient function that governs single-inclusive identified-hadron production. The entire process dependence resides in the hard function. In $e^+e^-$ collisions both ingredients are known analytically to high order, which has enabled high-precision calculations of the collinear energy correlator~\cite{Electron-PositronAlliance:2025fhk,Lee:2026zyl}. At a hadron collider, and in the presence of realistic fiducial cuts, the hard function can no longer be obtained analytically; what is needed instead is a numerical framework able to compute it for a generic initial state and arbitrary selection cuts, and to perform the matching to fixed order. Such a framework is already at hand. Within the \textsc{Stripper} subtraction scheme~\cite{Czakon:2010td,Czakon:2014oma,Czakon:2019tmo}, extended to fragmentation processes in Refs.~\cite{Czakon:2021ohs,Czakon:2022pyz,Czakon:2024tjr,Czakon:2025yti}, one can compute both the inclusive hard functions and the full fixed-order cross sections required for the matching. In Ref.~\cite{Generet:2025vth} we initiated a framework that interfaces these numerically computed ingredients with the analytic factorization theorems for jet observables, illustrating it with single-inclusive small-radius jet production at NNLO+NNLL accuracy. 

In the present work we build on this framework and apply it to projected energy correlators in $V+X$. We assemble a complete calculational framework that combines state-of-the-art fixed-order amplitudes and subtractions, collinear resummation, and non-perturbative power corrections. The perturbative and non-perturbative components can be fed in together and evolved consistently: the leading non-perturbative correction is controlled by the same universal matrix element that governs $e^+e^-$ event shapes~\cite{Lee:2024esz}, and we include its renormalization-group running alongside higher-order hard functions. We then present detailed predictions for the projected energy correlators at NLO+NNLL. These are the first complete matched predictions for energy correlators at the LHC built on exact fixed-order hard functions, and they reach a level of perturbative control comparable to the best-understood jet substructure observables. Achieving genuine NNLL accuracy requires the two-loop constants of the energy-correlator jet functions, which we incorporate~\cite{Chen:2023zlx,Lee:2026zyl}. Our numerical studies are carried out for the photon channel, $pp\to\gamma+X$, where the hard scale is set by a directly reconstructed, color-neutral boson. Furthermore, we study in detail the structure of the perturbative power corrections to the energy correlator, that is, the terms suppressed by powers of the opening angle relative to the leading-power singular behavior. The resummation does not capture these terms, which are instead supplied by the matching to fixed order; their size controls how deep into the singular region the fixed-order computation must be carried, and they can modify the slope of the distribution sensitive to $\alpha_s$. We find that they are under good control and take a simple form, with coefficients that depend linearly on the number of points $N$. This is encouraging for an extension to NNLO, since the fixed-order computation then need not be pushed deep into the singular region where subtractions are most delicate; the remaining ingredient, the two-loop five-point one-mass ($V+$jets) amplitudes, is now available~\cite{DeLaurentis:2025dxw}, which would also provide a striking illustration of their phenomenological relevance.

Together, these results chart a path towards genuine precision jet substructure at the LHC, uniting high-precision data with modern scattering amplitudes, subtraction schemes, and factorization. With perturbative and non-perturbative control already on par with the most thoroughly studied jet substructure observables, and a clear route to NNLO, the energy correlator in $V+X$ is poised to become a benchmark observable for precision QCD at the LHC.

The remainder of this paper is organized as follows. In \Sec{sec:def} we define the projected energy correlators and the fiducial selection cuts that we impose for the $V+X$ process. In \Sec{sec:theory} we develop the theoretical framework, discussing the fixed-order calculation, the collinear factorization theorem for energy correlators in $V+X$, the numerical evaluation of the inclusive hard functions, the matching of the resummed and fixed-order results, and the inclusion of non-perturbative power corrections. In \Sec{sec:pheno} we present our results for the photon channel, studying in detail the structure of perturbative power corrections and presenting NLO+NNLL predictions for the energy correlator in $\gamma+X$. We conclude and discuss future directions in \Sec{sec:conc}.

\section{Projected Energy Correlators in $V+X$ Processes}\label{sec:def}
In this section, we define the projected energy correlators for the $pp\to V+X$ process, together with the fiducial selection cuts imposed in our analysis. We first briefly review the projected correlators in $e^+e^-$ collisions, where they were introduced, and their generalization to hadron colliders using jets. 

Energy correlators characterize the distribution of energy in the final state of a collision through correlation functions of energy flow operators~\cite{Sveshnikov:1995vi,Tkachov:1995kk,Hofman:2008ar},
\begin{align}
\label{eq:eflow_def}
\mathcal{E}(\vec n) = \int_0^\infty dt\, \lim_{r\to\infty} r^2\, n^i\, T_{0i}(t, r\vec n)\,,
\end{align}
each of which acts as an idealized calorimeter, recording the energy deposited over time in the lightlike direction $\vec n$. The simplest such observable is the two-point energy-energy correlator (EEC), introduced long ago in the context of $e^+e^-$ annihilation~\cite{Basham:1978zq,Basham:1979gh,Basham:1977iq,Basham:1978bw}. This was generalized to the projected $N$-point correlators (ENC) also in $e^+e^-$~\cite{Chen:2020vvp}, obtained by correlating $N$ energy flow operators and integrating over all pairwise angles while keeping the largest one, $x_L$, fixed:
\begin{align}
\label{eq:enc_operator_def}
\frac{d\sigma^{[N]}_{e^+e^-}}{dx_L}
=
\int \prod_{i=1}^N d\Omega_{\vec n_i}\,
\delta\!\left(x_L - \max_{1\le i<j\le N} x_{ij}\right)
\int d^4 x\,
\frac{e^{iq\cdot x}}{Q^N}\,
\langle 0 | \mathcal{O}^\dagger(x)\, \mathcal{E}(\vec n_1)\cdots \mathcal{E}(\vec n_N)\, \mathcal{O}(0) | 0\rangle\,.
\end{align}
Here $x_{ij} = (1-\vec n_i\cdot\vec n_j)/2$ is the angular distance between detectors $i$ and $j$, $d \Omega_{\vec n} = \frac{1}{4 \pi} \sin\theta\, d \theta\, d \phi$ is the area element on the celestial sphere, and $\mathcal{O}$ is the source operator for the process under consideration. In $e^+e^-$ annihilation the source is the electroweak current, whose virtuality $q^2=Q^2$ fixes the hard scale of the measurement. In perturbation theory, Eq.~\eqref{eq:enc_operator_def} can be rewritten as a sum of energy-weighted cross sections over final states,
\begin{align}
\label{eq:enc_particle_def}
\frac{d\sigma^{[N]}_{e^+e^-}}{dx_L}
=\frac{1}{2Q^2}&
\sum_m \int
d\Phi_m \sum_{X_m}\, \left|\mathcal{M}_{X_m}\right|^2\notag\\&\times
\sum_{1\le i_1,\dots,i_N\le m}
\frac{\prod_{a=1}^N E_{i_a}}{Q^N}\,
\delta\!\left(x_L - \max\{x_{i_1 i_2},\, x_{i_1 i_3},\, \dots,\, x_{i_{N-1}i_N}\}\right)\,,
\end{align}
where $d\Phi_m$ is the massless Lorentz-invariant phase space of the $m$-particle final state $X_m$ and $\left|\mathcal{M}_{X_m}\right|^2$ is the corresponding squared amplitude. The indices $i_a$ are allowed to coincide, producing contact terms. In $e^+e^-$, the projected correlator is defined over the entire angular region, from the collinear limit $x_L \to 0$ to the back-to-back limit $x_L \to 1$. For later convenience, we also define the cumulant of ENC by integrating the distribution,
\begin{align}
\label{eq:enc_cumulant_def}
\Sigma^{[N]}_{e^+e^-}\left(x_L\right)\equiv \int_0^{x_L} d x_L^{\prime}\, \frac{d\sigma^{[N]}_{e^+e^-}}{dx_L^\prime}\,.
\end{align}
Two ingredients implicit in Eq.~\eqref{eq:enc_particle_def} are especially worth preserving at a hadron collider: the energy weights are normalized to the hard scale $Q$, which the colliding beams fix once and for all, and the observable is a function of the final state alone, requiring no auxiliary algorithmic construction.

At a hadron collider there is no direct analogue of the global scale $Q$, since the scale of the underlying hard partonic scattering varies event by event. Jets provide the standard means of tagging the hard scattering: a high transverse momentum jet, reconstructed with a jet algorithm such as anti-$k_T$~\cite{Cacciari:2008gp} with radius parameter $R$, serves as a proxy for a hard parton, and its substructure has grown into a rich field in its own right~\cite{Larkoski:2017jix,Asquith:2018igt}. Energy correlators were brought to the LHC within this setting, as one such substructure observable, by measuring the correlations of energy flow among the particles inside the jet~\cite{Lee:2022ige},
\begin{align}
\label{eq:enc_jet_def}
\frac{d\sigma^{[N]}_{\text{jet}}}{dR_L}
\;=\;&
\int d\sigma_{pp\to \text{jet}+X}
\sum_{i_1,\dots,i_N \in \text{jet}}
\frac{\prod_{a=1}^N p_{i_a,T}}{\big(p_{\text{jet},T}\big)^N}\,
\delta\!\left(R_L - \max\{R_{i_1 i_2},\, R_{i_1 i_3},\, \dots,\, R_{i_{N-1}i_N}\}\right)\,,
\end{align}
where the angles are now measured with the boost-invariant rapidity-azimuth distance $R_{ij}=\sqrt{\Delta \eta_{ij}^2+\Delta \phi_{ij}^2}$, with $\Delta \eta_{ij}=\eta_i-\eta_j$ and $\Delta \phi_{ij}=\phi_i-\phi_j$ the rapidity and azimuthal differences between particles $i$ and $j$, and $R_L$ denotes the largest such distance. Here $d\sigma_{pp\to \text{jet}+X}$ is shorthand for the fully differential single-inclusive jet production cross section, the analogue of the explicit phase-space integrals and squared amplitudes in Eq.~\eqref{eq:enc_particle_def}, now including the parton distribution functions, flux factors and a summation over all jets passing the kinematic selection of the measurement. The jet transverse momentum $p_T^{\text{jet}}$ plays the role of the hard scale, both selecting the hard scattering and normalizing the energy weights. This strategy has proven remarkably powerful and underlies the success of the jet substructure program. For energy correlators in particular, the enormous rates of high-$p_T$ jets give access to the collinear scaling regime at TeV energies, across a wide range of angular scales. It has driven a broad and successful measurement program by several collaborations~\cite{CMS:2024mlf,ALICE:2024dfl,STAR:2025jut} in $pp$ and even to growing measurements in heavy-ion collisions as well~\cite{CMS:2025jam,CMS:2025ydi,ALICE:2026giw}.

At the same time, measuring correlators inside jets introduces structure that is not intrinsic to the scattering itself. The jet algorithm and the radius $R$ are parameters of the analysis rather than of the process: nothing in the underlying scattering singles out a particular value of $R$, which enters instead as an external choice that defines the object whose constituents are correlated. For a precision program this has several consequences. First, the jet introduces the additional scale $p_T R$, and the theoretical description acquires algorithm-dependent ingredients: in the factorized description of in-jet correlators, the hard production is connected to the collinear dynamics by matching coefficients that depend explicitly on the jet algorithm, the radius $R$, and the scale $p_T R$~\cite{Kang:2016mcy,Lee:2024icn}. Second, the normalization of the energy weights in Eq.~\eqref{eq:enc_jet_def} is the jet $p_T$ itself, an algorithm-dependent quantity with its own perturbative and non-perturbative corrections, rather than a scale set by the hard scattering. Third, the jet boundary limits the angular range: as $R_L$ approaches $R$, the collinear power expansion breaks down and the correlator is distorted by clustering effects at the jet edge, introducing power corrections in $R_L/R$, potentially together with clustering and non-global logarithms, so that only the region $R_L \ll R$ is cleanly accessible, in contrast to the full angular coverage available in $e^+e^-$. 

The $V+X$ process removes this auxiliary structure. In $pp\to V+X$ with $V=\gamma, Z$, the electroweak boson is reconstructed independently of the hadronic activity and its transverse momentum $p_{V,T}$ is a genuine scale of the hard scattering, determined by the interaction itself rather than imposed as a parameter of the analysis. In this sense, $V+X$ is the closest hadron-collider analogue of $e^+e^-$ annihilation, which likewise ``tags'' the hard interaction with an electroweak boson, and can be viewed as a crossed version of it. In $e^+e^-$, the hadronic final state is sourced by a timelike virtual $\gamma^*/Z$ of virtuality $q^2=Q^2$, which sets the hard scale. In $V+X$, the same electroweak current appears in the final state as an on-shell boson, a real photon ($q^2=0$) or a $Z$ near $M_Z$, that recoils against the QCD radiation whose correlations we measure, with the hard scale now set by the transverse momentum $p_{V,T}$, as illustrated in Fig.~\ref{fig:intro_fig}. We therefore define the projected energy correlators in $V+X$ on the recoiling radiation, with the energy weights normalized to the boson transverse momentum:
\begin{align}
\label{eq:enc_particle_def_lhc}
\frac{d\sigma^{[N]}}{dR_L}
\;=\;
\sum_m\;&\sum_{X_m}\;\int d\sigma_{pp\to V+X_m}\notag\\&\times
\sum_{1\le i_1,\dots,i_N\le m}
\frac{\prod_{a=1}^N p_{i_a,T}}{p_{V,T}^N}\,
\delta\!\left(R_L - \max\{R_{i_1 i_2},\, R_{i_1 i_3},\, \dots,\, R_{i_{N-1}i_N}\}\right)\,.
\end{align}
Here $d\sigma_{pp\to V+X_m}$ is the fully differential cross section for producing the boson together with an $m$-particle hadronic final state $X_m$, including the parton distribution functions and flux factors, and the correlator sums run over all final-state particles except the boson (or its decay products). We give the explicit perturbative expansion of Eq.~\eqref{eq:enc_particle_def_lhc} in \Sec{sec:theory}.

No jet algorithm enters this definition: there is no clustering, no radius, and no scale $p_T R$. As in $e^+e^-$, the correlator is a property of the final state alone, tagged by a hard scale of the scattering, and it is not partitioned by any jet boundary. The collinear regime $R_L \to 0$ connects continuously out to wide angles. Finally, $p_{V,T}$ is fixed purely by the boson kinematics and is therefore independent of the reconstruction of hadronic energy. Relative to the in-jet case, this removes the jet-radius power corrections and the jet-energy-scale non-perturbative corrections and simplifies the perturbative description, while the leading non-perturbative correction to the correlator shape itself remains the universal collinear matrix element shared with $e^+e^-$ event shapes.

In this work, we focus our numerical studies on the photon channel, $pp\to\gamma+X$, which can be resolved with precise resolution even at the highest transverse momenta. The identical framework applies to $Z+X$ with $Z \to \ell^+\ell^-$, which avoids the photon isolation requirements discussed below and provides an exceptionally clean tag; we foresee it as a central target for future measurements. For the correlator measurement, an important consideration is that the energy weights in Eq.~\eqref{eq:enc_particle_def_lhc} enhance configurations in which the hadronic system is much harder than the photon. In particular, dijet-like configurations (though can be computed within our framework and are included), in which the photon is relatively soft and the remaining final-state particles recoil against each other in the partonic center-of-mass frame, receive weights $\prod_a^N p_{i_a,T}/p_{\gamma,T}^N > 1$, and their contribution grows with $N$. To prevent these dijet configurations completely dominating, we impose a stringent lower cut on the photon transverse momentum, $p_{\gamma,T}>500$~GeV, together with $|\eta_\gamma|<2.4$. All other final-state particles are also required to satisfy $|\eta|<2.4$. Furthermore, the perturbative definition of a prompt photon requires an isolation prescription, which suppresses contributions in which the photon is produced collinear to QCD radiation and which would otherwise require non-perturbative photon fragmentation functions. We require both hard cone and smooth cone isolation, in the so-called hybrid cone approach~\cite{Siegert:2016bre,Chen:2019zmr}. Only particles with $|\eta|<2.4$ contribute to the isolation requirements. In the hard cone, the scalar sum of the transverse momenta of all particles with $\Delta R_{\gamma,\text{particle}} < 0.2$ must be less than $0.09\, p_{\gamma,T}$. In the smooth cone, the scalar sum of the transverse momenta of all particles with $\Delta R_{\gamma,\text{particle}} < r$ must be smaller than $0.15\,(1-\cos r)/(1-\cos 0.1)\, p_{\gamma,T}$ for all $r<0.1$. All of these isolation criteria can be implemented within the fixed-order computation using Monte Carlo integration, which is described in detail in \Sec{sec:FO}.

It is also interesting to study the ratios of the projected $N$-point correlators to the $2$-point correlator,
\begin{equation}
    \frac{d\sigma^{[N]}/dR_L}{d\sigma^{[2]}/dR_L}\,.
\end{equation}
There are several advantages to using ratios of correlators as observables. First, uncertainties that affect only the overall normalization of a measurement, such as the luminosity uncertainty, cancel between numerator and denominator. Second, the leading non-perturbative corrections, which are controlled by the same universal matrix element in the numerator and the denominator, partially cancel in the ratio. Finally, the ratios increase the relative sensitivity to the value of the strong coupling~\cite{Lee:2026zyl}. These features make the ratio correlators prime candidates for extracting the strong coupling from LHC data.

\section{Theoretical Framework}\label{sec:theory}
In this section, we will introduce the theoretical framework for the calculation of energy correlators in hadronic processes. While we will focus on the process, $pp\to \gamma+X$ at the LHC, our framework is designed to enable the calculation of the energy correlator in a wide variety of processes at the LHC and other colliders. Our aim is a prediction that is accurate across the range of the observable, from the small-angle region $R_L\to 0$ to moderate angles $R_L\sim\mathcal{O}(1)$, which we obtain by combining a fixed-order calculation of the correlator with the resummation of the logarithms that dominate at small angles.

At fixed order in $\alpha_s$, the projected correlator is computed directly from the QCD matrix elements for a photon and partons, with the energy weighting of \Eq{eq:enc_particle_def_lhc} imposed on the final state. This provides the complete result at moderate angles, $R_L\sim\mathcal{O}(1)$, where the distribution is not dominated by the collinear singularity and the power corrections in $R_L$ are substantial rather than negligible. In \Sec{sec:FO}, we describe how the \textsc{Stripper} subtraction scheme~\cite{Czakon:2010td, Czakon:2014oma, Czakon:2019tmo} was used to carry out these computations.

In the small-angle collinear region, $R_L\to 0$, the observable obeys a factorization theorem into a perturbatively calculable jet function, which captures the collinear dynamics and is universal across processes, and a hard function that can also be computed in  \textsc{Stripper}. We present the factorization and its resummation in \Sec{sec:fact} and the numerical evaluation of the process-dependent hard functions in \Sec{sec:hard}. Non-perturbative power corrections are incorporated in \Sec{sec:NP_corrections}.

\subsection{Fixed Order Calculations Using the \textsc{Stripper} Framework}\label{sec:FO}
\begin{figure}
    \centering
    \includegraphics[width=0.7\linewidth]{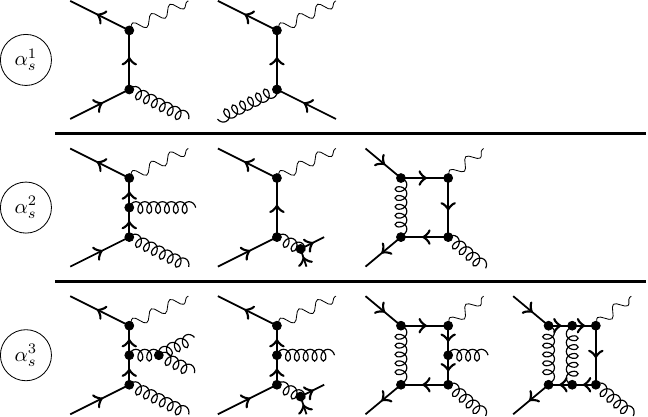}
    \caption{Example Feynman diagrams contributing to the fixed-order predictions at $\mathcal{O}(\alpha_s)$ (top row), $\mathcal{O}(\alpha_s^2)$ (middle row), and $\mathcal{O}(\alpha_s^3)$ (bottom row). With a single parton in the final state, the diagrams in the top row contribute only to the $\delta(R_L)$ endpoint, so the middle and bottom rows define the LO and NLO of the differential projected correlators at $R_L\neq 0$.}
    \label{fig:fo-diag}
\end{figure}

The fixed-order expansion in $\alpha_s$ of the projected energy-energy correlator defined in Eq.~\eqref{eq:enc_particle_def_lhc} for the process $pp \to \gamma + X$ reads:
\begin{equation}
    \frac{d \sigma^{[N]}}{d R_L} = \frac{d \sigma^{[N] (0)}}{d R_L} + \frac{d \sigma^{[N] (1)}}{d R_L} + \frac{d \sigma^{[N] (2)}}{d R_L} + \mathcal{O}\left(\alpha_s^4\right)\;,
\end{equation}
where the superscript $(i)$ labels successive orders in the strong coupling, beginning with the lowest, $\mathcal{O}(\alpha_s)$, term $d\sigma^{[N](0)}/dR_L$. An important feature of the projected correlator is that, away from the endpoint at $R_L=0$, its differential distribution starts one order in $\alpha_s$ higher than the cross section. At the lowest order, a single parton recoils against the photon and carries the entire hadronic transverse momentum, so all of the energy weight is deposited at zero separation and contributes only to a contact piece $\propto\delta(R_L)$. The first genuine contribution at $R_L\neq0$, which we take to define the leading order (LO) of the correlator, therefore appears at $\mathcal{O}(\alpha_s^2)$, and its next-to-leading order (NLO) at $\mathcal{O}(\alpha_s^3)$. Equivalently, the LO and NLO of the differential correlator $d\sigma^{[N]}/dR_L$ correspond to the NLO and NNLO of the $pp\to\gamma+X$ cross section, or of its cumulant $\Sigma^{[N]}$.

In this subsection, we follow closely the notation in Ref.~\cite{Czakon:2014oma}. The perturbative coefficients of the hadronic correlator $\frac{d \sigma^{[N] (i)}}{d R_L}$ expressed in terms of partonic coefficients $\frac{d \hat{\sigma}^{[N] (i)}}{d R_L}$ are given by
\begin{equation}
    \frac{d \sigma^{[N] (i)}}{d R_L} = \sum_{ab} \iint_0^1 \mathrm{d}x_1
    \mathrm{d}x_2 \, f_{a}(x_1, \mu_F) \,
    f_{b}(x_2, \mu_F) \, \frac{d\hat{\sigma}^{[N] (i)}_{ab}}{d R_L} (x_1,\,x_2,\,\mu_F, \,\mu_R,\,\alpha_s(\mu_R)) \; .
\end{equation}
Here we made explicit the dependence on the renormalization and factorization scales ($\mu_R$ and $\mu_F$, respectively) and on the initial-state momentum fractions $x_1$ and $x_2$. For brevity, these are omitted in the following expressions. At lowest (B)orn order, i.e. $\mathcal{O}(\alpha_s)$, we have
\begin{equation}
    \frac{d \hat{\sigma}^{[N](0)}_{ab}}{d R_L} = \frac{d \hat{\sigma}^{[N]\,{\rm B}}_{ab}}{d R_L} =
    \frac{1}{2\hat{s}} \frac{1}{N_{ab}} \int \mathrm{d} \Phi_2 \, \langle
    \mathcal{M}_{ab \to \gamma k}^{(0)} | \mathcal{M}_{ab \to \gamma k}^{(0)} \rangle \mathcal{F}^{[N]}_1 (R_L)\, 
\end{equation}
with implied sum over all partons $k$ allowed by the initial state partons $ab$\footnote{At LO, there is only one allowed $k$ for each $ab$.}. We used the abbreviation
\begin{equation}
    \mathcal{F}^{[N]}_m (R_L) \equiv \sum_{1\le i_1,\dots,i_N\le m}\frac{\prod_{a=1}^N p_{i_a,T}}{p_{\gamma,T}^N}\,
\delta\!\left(R_L - \max\{R_{i_1 i_2},\, R_{i_1 i_3},\, \dots,\, R_{i_{N-1}i_N}\}\right)\;,
\end{equation}
for the definition of $N$-point projected energy correlators. We also use $\langle\mathcal{M}^{(j)} | \mathcal{M}^{(i)} \rangle$ to denote the interference between the $i$- and $j$-loop matrix elements. Finally, $N_{ab}$ denotes the number of incoming partonic spin and color degrees of freedom, whereas $\hat{s} = (p_a+p_b)^2$. Examples for contributing Feynman diagrams at this order are shown in the first row in Fig.~\ref{fig:fo-diag}. With only one parton in the final state, this only contributes to a $\delta(R_L)$ in projected correlators.

The leading contribution of the projected correlator for $R_L\neq 0$ starts at $\mathcal{O}(\alpha_s^2)$ and receives contributions from tree-level $ij\to \gamma+kl$ matrix elements. One-loop corrections of $ij\to \gamma+k$ again contribute to the coefficient of $\delta(R_L)$ only. The partonic coefficient reads
\begin{align}
\frac{d \hat{\sigma}^{[N](1)}_{ab}}{d R_L} &= \frac{d \hat{\sigma}^{[N]\,\rm R}_{ab}}{d R_L} + \frac{d \hat{\sigma}^{[N]\,\rm V}_{ab}}{d R_L} + \frac{d \hat{\sigma}^{[N]\,\rm C}_{ab}}{d R_L}\;,
\end{align}
where $\hat{\sigma}^{\rm R}_{ab}$ and $\hat{\sigma}^{\rm V}_{ab}$ refer to real emission and virtual corrections, respectively, and $\hat{\sigma}^{\rm C}_{ab}$ to collinear beam factorization terms to cancel initial state collinear singularities. Explicitly, these contributions are given by
\begin{align}
    \frac{d \hat{\sigma}^{[N]\,\rm R}_{ab}}{d R_L}  &=
    \frac{1}{2\hat{s}} \frac{1}{N_{ab}} \int \mathrm{d} \Phi_{3} \, \langle
    \mathcal{M}_{ab \to \gamma kl}^{(0)} | \mathcal{M}_{ab \to \gamma kl}^{(0)} \rangle \, \mathcal{F}^{[N]}_2 (R_L) \; , \\[0.2cm] 
    \frac{d \hat{\sigma}^{[N]\,\rm V}_{ab}}{d R_L}  &= \frac{1}{2\hat{s}} \frac{1}{N_{ab}} \int
    \mathrm{d} \Phi_2 \, 2 \Re \, \langle \mathcal{M}_{ab \to \gamma k}^{(0)} | \mathcal{M}_{ab \to \gamma k}^{(1)} \rangle \,
    \mathcal{F}^{[N]}_1 (R_L) \; ,\nonumber \\[0.2cm] 
    \frac{d \hat{\sigma}^{[N]\,\rm C}_{ab}}{d R_L}  &= \frac{\alpha_s}{2\pi} \frac{1}{\epsilon}
    \left( \frac{\mu_R^2}{\mu_F^2} \right)^{\epsilon} \sum_c \int_0^1 \mathrm{d}z
    \left[P^{(0)}_{ca}(z) \, \hat{\sigma}^{[N]\,\mathrm{B}}_{cb}(zx_1,x_2) +
      P^{(0)}_{cb}(z) \, \hat{\sigma}^{[N]\,\mathrm{B}}_{ac}(x_1,zx_2) \right]\nonumber
    \; .
\end{align}
Again, we imply the sum over all partonic states $k$ and $kl$ allowed by the initial state $ab$. The virtual and collinear factorization counterterms contribute only a $\delta(R_L)$ because they have only one parton in the final state. Example Feynman diagrams for the real and virtual corrections are shown in the middle row of Fig.~\ref{fig:fo-diag}.

At NLO for the projected correlators (which corresponds to NNLO for the $pp \to \gamma + X$ process), corrections from double real emissions, virtual corrections to single real emissions, and two-loop contributions have to be taken into account. The NNLO coefficient can decompose in
\begin{align}
\frac{d \hat{\sigma}^{[N](2)}_{ab}}{d R_L} &= \frac{d \hat{\sigma}^{[N]\,\rm RR}_{ab}}{d R_L} + \frac{d \hat{\sigma}^{[N]\,\rm RV}_{ab}}{d R_L} + \frac{d \hat{\sigma}^{[N]\,\rm VV}_{ab}}{d R_L} + \frac{d \hat{\sigma}^{[N]\,\rm C1}_{ab}}{d R_L}+ \frac{d \hat{\sigma}^{[N]\,\rm C2}_{ab}}{d R_L}\;.
\end{align}
The expressions for the double real (RR), real virtual (RV), and double virtual (VV) contributions are
\begin{align}
\frac{d \hat{\sigma}^{[N]\,\rm RR}_{ab}}{d R_L} &=
\frac{1}{2\hat{s}} \frac{1}{N_{ab}} \int \mathrm{d} \Phi_{4}
\, \langle \mathcal{M}_{ab\to \gamma klm}^{(0)} | \mathcal{M}_{ab\to \gamma klm}^{(0)} \rangle \,\mathcal{F}^{[N]}_3 (R_L) \; ,
\\ \frac{d \hat{\sigma}^{[N]\,\rm RV}_{ab}}{d R_L} &=
\frac{1}{2\hat{s}} \frac{1}{N_{ab}} \int \mathrm{d} \Phi_{3} \, 2 \Re
\, \langle \mathcal{M}_{ab\to \gamma kl}^{(0)} | \mathcal{M}_{ab\to \gamma kl}^{(1)} \rangle \, \mathcal{F}^{[N]}_2 (R_L) \; ,\nonumber
\\ \frac{d \hat{\sigma}^{[N]\,\rm VV}_{ab}}{d R_L} &=
\frac{1}{2\hat{s}} \frac{1}{N_{ab}} \int \mathrm{d} {\Phi}_2 \, \Big( 2
\Re \, \langle \mathcal{M}_{ab\to \gamma k}^{(0)} | \mathcal{M}_{ab\to \gamma k}^{(2)} \rangle + \langle \mathcal{M}_{ab\to \gamma k}^{(1)} |
\mathcal{M}_{ab\to \gamma k}^{(1)} \rangle \Big) \, \mathcal{F}^{[N]}_1 (R_L)\;.\nonumber
\end{align}
In the interest of brevity, we do not spell out the expression for the collinear factorization contributions (C1 and C2), which can easily be extracted from Ref.~\cite{Czakon:2014oma}. Here, the double virtual corrections and the C2 contribution contribute only to the endpoint. Example diagrams for the real emission and virtual correction at this order can be found in the lowest row in Fig.~\ref{fig:fo-diag}. Following this, we can define all N${}^k$LO orders for the projected correlators at fixed order. While the differential ENC does not receive any contributions from terms proportional to $\delta(R_L)$, the cumulant, defined in Eq.~\eqref{eq:enc_cumulant_def}, does. In practice, we evaluate the N${}^{k+1}$LO cumulant distribution $\Sigma^{[N]}$ by keeping all contributions that are proportional to $\delta(R_L)$, i.e., subtraction terms and virtual contributions with tree-level kinematics. Using a NNLO subtraction scheme thus allows us to evaluate the NLO differential and NNLO cumulative distributions at fixed order in perturbation theory. Ingredients to push the differential ENC one order higher are available~\cite{Badger:2023mgf} but their incorporation is computationally very challenging and beyond the scope of the current work. The incorporation of the progress made on three-loop amplitudes for $pp \to \gamma +\text{jet}$~\cite{Gehrmann:2023jyv}, needed for the cumulative distribution, is lacking a method of computing N$^3$LO cross sections at hadron colliders, i.e.~an N$^3$LO subtraction scheme. While such a scheme is available for jet production in lepton collisions~\cite{Chen:2025kez}, so far, no such scheme has been constructed for hadronic collisions.

We employ the sector-improved residue subtraction scheme, as implemented in the \textsc{Stripper} framework \cite{Czakon:2010td, Czakon:2014oma, Czakon:2019tmo}, to handle infrared divergences up to NNLO QCD for the $pp \to \gamma+X$ process, or equivalently, next-to-leading order for the projected energy-energy correlator. All tree-level matrix elements are provided by the \textsc{avhlib} library~\cite{avhlib, Bury:2015dla}, while all required one-loop matrix elements are publicly available in the \textsc{OpenLoops2} library~\cite{Cascioli:2011va, Buccioni:2019sur}. The two-loop matrix elements have been taken from Ref.~\cite{Glover:2003cm} and newly implemented in a numerical code. 
For the numerical results presented in the following sections, we use the PDF4LHC21 PDF set~\cite{PDF4LHCWorkingGroup:2022cjn} to provide $\alpha_s$ and the PDF values. We work in $n_f=5$ massless QCD, i.e., we don't include contributions from top-quark loops. Further, we are using the dynamical scale choice $\mu_R = \mu_F = p_{\gamma,T}$. The fiducial phase space of the photon was described in Sec.~\ref{sec:def}. To estimate uncertainties arising from missing higher-order terms in the fixed-order predictions, we exploit a 7-point scale variation around the central scale choice by a factor of 2. The quoted uncertainties are derived from the envelope of these variations. For ratios, we use correlated variations in the numerator and denominator. 

While this fixed-order expansion is well behaved at moderate $R_L$, in the small-angle limit $R_L\to0$ each order develops large singular logarithms of $R_L$. These logarithms must be resummed to all orders to obtain a reliable prediction, which we achieve through the collinear factorization theorem described next.

\subsection{Collinear Factorization Theorem}\label{sec:fact}

In this subsection, we review the collinear factorization theorem for projected energy correlators. In fixed-order predictions, the ENC distribution receives large logarithmic enhancements at each order:
\begin{align}
    \frac{d\hat{\sigma}^{[N]}}{dR_L}=\sum_{L=1}^{\infty} \sum_{j=-1}^{L-1}\left(\frac{\alpha_s}{4\pi}\right)^L &e_{L,j} \cL^j (R_L)\,,\\
    \text{where }\quad \cL^{-1}(R_L)&=\delta(R_L),\,  \cL^{j}(R_L)= \left[\ln^j(R_L)/R_L \right]_+\nonumber
\end{align}
Here $e_{L,j}$ are the constants that are associated with the logarithmic terms.

For ENC cumulants,
the plus distributions $\cL^j(R_L)$ in the partonic coefficient $d\hat{\sigma}/dR_L$ will be mapped to $\ln^{j+1}R_L$ in the partonic coefficient cumulant $\hat{\Sigma}(R_L)$. The existence of these large logarithms can spoil the convergence of perturbation theory when $\alpha_s\ln R_L\sim 1$, and thus the resummation to all orders is necessary.

The collinear factorization for the 2-point correlator, EEC, is first derived in Ref.~\cite{Dixon:2019uzg} 
and generalized to the $N$-point projected correlator in Ref.~\cite{Chen:2020vvp} for $e^+e^-$ collisions and hadronic Higgs decays.  
Using the largest angle $x_L$ at $e^+e^-$ annihilation, the ENC cumulant can be written as (see~\eq{eq:enc_cumulant_def})

\begin{align}
\label{eq:enc_factorization_ee}
 \Sigma^{[N]}_{e^+e^-}(x_L)= \int_0^1 dx\, x^N \vec{J}^{[N]} \left(\ln\frac{x_L x^2 Q^2}{\mu^2},\mu\right)
   \cdot  \vec{H}_{e^+e^-} \left(x,\ln\frac{Q^2}{\mu^2},\mu\right) \,.
\end{align}
Here the hard function $\vec{H}_{e^+e^-}=\{H_{e^+e^-,\, q},H_{e^+e^-,\, g}\}$ is the single-inclusive fragmentation function fixing the energy fraction of one final-state parton. $\vec{J}^{[N]}$ is the ENC jet function, which can be defined in terms of operators in Soft-Collinear Effective Theory (SCET)~\cite{Bauer:2000yr,Bauer:2000ew,Bauer:2001yt,Bauer:2001ct,Beneke:2002ph}:
\begin{align}
\label{eq:jet_func_quark}
J_q^{[N]}(x_L,Q, \mu^2)&=\notag\\
&\hspace{-0.1cm}\int \frac{dl^+}{2\pi}\frac{1}{2N_C} \text{Tr} \int d^4x e^{i l\cdot x} \langle 0 | \frac{\Sl{\bar n}}{2} \chi_n(x) \widehat\cM_{\text{ENC}}^{[N]} ~ \delta (Q+\bar n \cdot \cP) \delta^2(\cP_\perp) \bar \chi_n(0) |0\rangle\,,\notag\\
J_g^{[N]}(x_L,Q,\mu^2)&=\notag\\
&\hspace{-1.5cm}\int \frac{dl^+}{2\pi}\frac{1}{2 (N^2_C-1)} \text{Tr} \int d^4x e^{i l\cdot x} \langle 0 |  \cB^{a,\mu}_{n,\perp}(x) \widehat\cM_{\text{ENC}}^{[N]} ~ \delta (Q+\bar n \cdot \cP) \delta^2(\cP_\perp) \cB^{a,\mu}_{n,\perp}(0) |0\rangle \,,
\end{align}
where $\chi_n\equiv W_n^\dagger \xi_n$ is  the collinear quark and $\cB^{\mu}_{n,\perp}\equiv \frac{1}{g}\left[\frac{1}{\bar n\cdot \cP}W_n^\dagger [i\bar n\cdot D_n, iD_{n\perp}^\mu] W_n\right]$ is the collinear gluon, and $\cP_{n\perp}^{\mu}$ form a complete set of collinear gauge invariant building blocks \cite{Marcantonini:2008qn} in SCET. Here, $\widehat\cM_{\text{ENC}}^{[N]}$ is the measurement operator for the projected
correlators, defined by its action on an arbitrary $m$-particle state $|X_m\rangle$,
\begin{equation}
    \widehat\cM_{\text{ENC}}^{[N]}\,|X_m\rangle
    =\sum_{1\le i_1,\dots,i_N\le m}\frac{\prod_{a=1}^N E_{i_a}}{Q^N}\,
    \Theta\!\left(x_L-\max\{x_{i_1 i_2},\, x_{i_1 i_3},\, \dots,\, x_{i_{N-1}i_N}\}\right)
    |X_m\rangle\,,
\end{equation}
where $x_{i_1 i_2}=(1-\cos\theta_{i_1 i_2})/2\approx \theta_{i_1 i_2}^2/4$ is the angular distance between particle $i_1$ and $i_2$. In perturbation theory, such jet functions can be either computed via SCET Feynman rules or extracted from fixed-order ENC distributions. 

At hadron colliders, the factorization theorem is similar to Eq.~\eqref{eq:enc_factorization_ee}, except that the angle $x_L$ is replaced by the rapidity-azimuthal distance $R_L$, i.e. $x_L \to R_L^2/4$, and one trades energies for transverse momenta as $E_i \to p_{i,T}$. We also need to convolve with the initial PDFs.
This allows us to write the factorization formula as
\begin{multline}
\label{eq:pp_fact_formula}
    \frac{d \sigma^{[N]}}{d R_L} =\frac{d}{dR_L}\bigg[ \sum_{ab} \iint_0^1 \mathrm{d}x_1
    \mathrm{d}x_2\,\, f_{a}(x_1, \mu_F) \,
    f_{b}(x_2, \mu_F)\times \sum_{c=q,g}\int \mathrm{d}z \,z^N \\
    \times d\hat{\sigma}_{ab\to c\gamma+X}(x_1,x_2,z,\mu_F,\mu_R,\mu_J,\alpha_s(\mu_R)) \times J_c^{[N]}\left(\ln\frac{R_L^2 z^2 p_{\gamma,T}^2}{\mu_J^2},\ln\frac{\mu_R}{\mu_J},\alpha_s(\mu_R)\right)\bigg]\,.
\end{multline}
Here $d\hat{\sigma}_{ab\to c\gamma+X}$ is the hard function for producing a parton $c$ in association with a photon and it is fully differential in the \textsc{Stripper} framework. Note that the fiducial phase space introduced in Sec.~\ref{sec:def} is implicitly included in the hard functions and
$z=p_{i,T}/p_{\gamma,T}$ is the energy fraction.
Since the factorization is performed in the cumulant level, we take the derivative with respect to $R_L$ to obtain the distribution.
In addition to the renormalization scale $\mu_R$ and PDF factorization scale $\mu_F$, we also introduce the $\mu_J$ scale for ENC factorization.

The ENC jet functions satisfy the modified DGLAP evolution,
\begin{align}
	\label{eq:jet_evo}
	\frac{d}{d \ln\mu^2}\vec{J}^{[N]}\left( \ln \frac{R_L^2 z^2 p_{\gamma,T}^2}{\mu^2},\alpha_s(\mu)\right) = \int_0^1  dy\, y^N \vec{J}^{[N]}\left( \ln \frac{R_L^2 y^2 z^2 p_{\gamma,T}^2}{\mu^2},\alpha_s(\mu)\right)
	\cdot 
	\widehat{P}(y,\alpha_s(\mu)) \,,
\end{align}
where $\widehat{P}(y,\alpha_s(\mu))$ is the singlet timelike splitting matrix, 
now known in full through three loops~\cite{Almasy:2011eq,Chen:2020uvt} 
and partially at four-loop~\cite{Gehrmann:2023cqm,Falcioni:2023luc,Falcioni:2023vqq,Falcioni:2024xyt,Falcioni:2024qpd,Falcioni:2025hfz}.
Since obtaining a closed-form solution for DGLAP evolution is challenging, we use the expanded expression. In practice, we use the following vector ansatz in the flavor space at some scale $\mu$
\begin{align}
\label{eq:jet_ansatz}
\vec{J}^{[N]}\left(\ln\frac{R_L^2 z^2 p_{\gamma,T}^2}{\mu^2},\alpha_s(\mu) \right)= \sum_{n=0}^\infty \sum_{m=0}^n \left(\frac{\alpha_s(\mu)}{4\pi}\right)^n \frac{\vec{j}_{n,m}^{[N]}}{m!} \ln^m \frac{R_L^2 z^2 p_{\gamma,T}^2}{\mu^2}\,,
\end{align}
plug into the RG equation in Eq.~\eqref{eq:jet_evo} and solve the coefficients $\vec{j}_{n,m}^{[N]}$ by collecting the expression at each $(n,m)$. Note that $\vec{j}_{n,0}^{[N]}$ are the jet boundary constants obtained from fixed-order calculations. The NLO result $\vec{j}_{1,0}^{[N]}$ for arbitrary $N$ is calculated in Ref.~\cite{Chen:2020vvp}. At NNLO, $\vec{j}_{2,0}^{[N]}$, the $N=3$ constant is derived in Ref.~\cite{Chen:2023zlx}, and the $4\leq N\leq 6$ is computed recently in Ref.~\cite{Lee:2026zyl}. 
With these ingredients at hand, we now obtain the complete NNLL jet functions. In App.~\ref{sec:pert_ingredients}, we provide the expressions for the ENC jet functions.

For resummation, we set $\mu=\mu_J$ in the solution in Eq.~\eqref{eq:jet_ansatz} and re-express $\alpha_s(\mu_J)$ in terms of $\alpha_s(\mu_R)$ via $\beta$ function. Plugging into Eq.~\eqref{eq:pp_fact_formula}, the remaining step for resummation is to perform the convolution with PDFs and hard functions, which we will discuss in the next section. For resummation order, we define N${}^k$LL through $(k+1)$-loop timelike splitting kernels, $k$-loop hard and jet boundary functions and $(k+1)$-loop $\beta$ function, and match it to N${}^{k-1}$LO in fixed-order. Note that for the log counting, we rewrite $\ln\left(R_L^2 z^2 p_{\gamma,T}^2/\mu^2\right)$ as $\ln R_L^2$ and $\ln\left(z^2 p_{\gamma,T}^2/\mu^2\right)$, and count $\ln R_L^2$ as a large log, while keeping $\ln\left(z^2 p_{\gamma,T}^2/\mu^2\right)$ as $\mathcal{O}(1)$ contribution. To the logarithmic order we claim, this will not change the predicted logarithms, but only change the higher-order terms, which is captured by the theory uncertainties.

\begin{figure}
    \centering
    \includegraphics[width=0.7\linewidth]{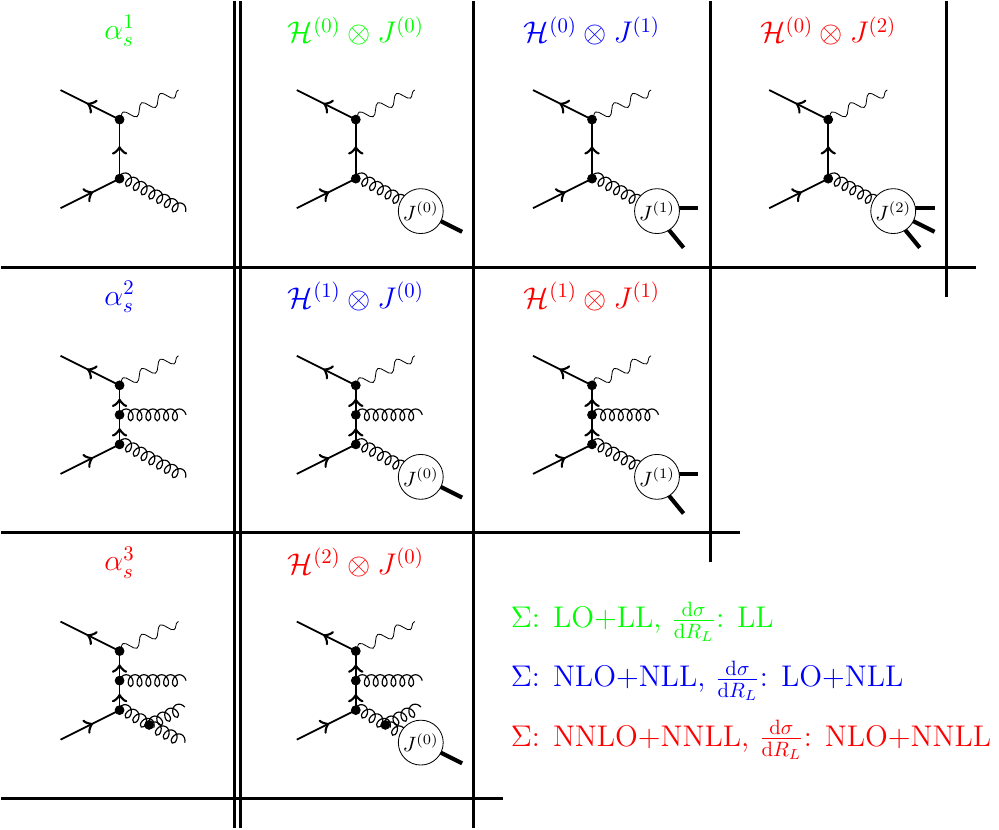}
    \caption{Illustration of the perturbative orders for the projected correlators. The first column shows example fixed-order contributions at $\mathcal{O}(\alpha_s^{1,2,3})$, and the remaining columns show their factorized counterparts, organized by the loop orders of the hard and jet functions, $\mathcal{H}^{(i)}\otimes J^{(j)}$. Contributions of the same color combine into the fixed-order and resummation accuracies listed in the legend, for both the cumulant $\Sigma$ and the differential distribution $d\sigma/dR_L$.}
    \label{fig:orders}
\end{figure}

A diagrammatic summary of the contributions needed to achieve a certain fixed-order or resummation precision within our perturbative counting is shown in Fig.~\ref{fig:orders}.

\subsection{Numerical Evaluation of Hard Functions}\label{sec:hard}

Now let us discuss the definition of the hard function and its numerical evaluation for $pp\to V+X$ process.
In $e^+e^-$ annihilation, the hard function for projected energy correlators is the single-inclusive hadron fragmentation function~\cite{Mitov:2006ic}, which is also a vector in the flavor space and depends on an outgoing parton energy fraction $x=\frac{2p\cdot q}{Q^2}$,
where $q$ is the total momentum and $p$ is the parton momentum.
At leading order, the $e^+e^-$ hard function is $\vec{H}_{ee}^{(0)} (x) = \{2\delta(1-x), 0\}$, obtained from the Born process $e^+e^-\to q\bar q$.
At one-loop, it starts to have non-trivial energy dependence
\begin{align}
\frac{1}{2}  H_{ee,q}^{(1)}(x) = &\  \frac{\alpha_s(Q)}{4 \pi} C_F 
\Bigg[
\left(\frac{4 \pi ^2}{3}-9\right) \delta(1-x) +4 \left[\frac{\ln(1-x)}{1-x} \right]_+
\nn\\
&\
+\left(4 \ln
   (x)-\frac{3}{2}\right)\left(2 \frac{1}{[1-x]}_+-x-1\right)-\frac{9 x}{2}-2 (x+1) \ln
   (1-x)+\frac{7}{2}
\Bigg] \,,
\nn\\
H_{ee,g}^{(1)}(x) = &\, \frac{\alpha_s(Q)}{4 \pi}C_F \Bigg[
\frac{4 \left(x^2-2 x+2\right) \ln (1-x)}{x}+\frac{8 \left(x^2-2 x+2\right) \ln   (x)}{x}
\Bigg] \,.
\end{align}
The factor $1/2$ in front of the quark channel indicates for identical contribution from anti-quark, since projected ENC jet function does not distinguish quark and anti-quark flavor. The two-loop correction is computed in Ref.~\cite{Mitov:2006ic}, and very recently, the three-loop result is also available~\cite{He:2025hin}.

To compute the hard functions for $pp\to V+X$ process, one can use the implementation of fragmentation in \textsc{Stripper} \cite{Czakon:2021ohs,Czakon:2022pyz,Czakon:2024tjr,Generet:2025vth}. This implementation can be used to calculate convolutions of hard functions with arbitrary distribution-valued fragmentation functions. By taking the fragmentation functions to be $\delta$-distributions, one can numerically obtain all the partonic hard functions separately. 
Implementing various fiducial cuts as we wish using the \textsc{Stripper} implementation, we can compute the hard function $d\sigma_{Vi+X}$, the fiducial single-inclusive cross section for producing fragmenting parton $i$ in association with a vector boson $V$, fully differential in the phase space of the final state. Here, we take $V=\gamma$ with the fiducial cuts described in~\Sec{sec:def}.

In practice, what we really need is this hard function convolved with the jet functions discussed in~\Sec{sec:fact}. After the generalization performed in Ref.~\cite{Generet:2025vth}, the convolution can involve arbitrary functions of the renormalization and final-state factorization scales and a characteristic scale which can be an arbitrary function of the kinematics. 
As discussed in~\Sec{sec:fact}, the jet RGEs are solved iteratively and truncated at a higher order. We have properly restored the dependence on the renormalization scale $\mu_R$ and the ENC factorization scale $\mu_J$.

So in the end, the leading-power computation reduces to computing weighted fiducial cross sections of the type
\begin{align}
\label{eq:weightedfiducial}
    \int d\sigma_{\gamma\:i+X}\bigg(\frac{p_{i,T}}{p_{\gamma,T}}\bigg)^N\alpha_s^l(\mu_R)\ln^m\bigg(\frac{p_{i,T}^2}{\mu_J^2}\bigg)\ln^n\bigg(\frac{\mu_R^2}{\mu_J^2}\bigg)\;,
\end{align}
where $d\sigma_{\gamma\:i+X}$ is the fiducial single-inclusive cross section for producing parton $i$ in association with a photon, fully differential in the phase space of the final state, and we used $z=p_{i,T}/p_{\gamma,T}$. We will be interested in $N$-point correlators up to $N=6$. For NNLL accuracy with our log counting, only $m,n=0,1,2$ are needed. In fact, since we will only be interested in values of $\mu_R$ and $\mu_J$ which are identical up to a constant factor, the weighting by arbitrary powers of $\ln(\mu_R^2/\mu_J^2)$ can be easily performed a posteriori, and only $n=0$ needs to be explicitly computed during the Monte Carlo integration. $l$ starts at 0 and we truncate it at $l=14$.
We checked that this leads to a truncation error well below 1 permille for the values of $R_L$ considered here.

Once all of these weighted cross sections are computed, they can be combined with the appropriate powers of $\ln R_L$ and the corresponding values of the perturbative coefficients of the jet functions to obtain the final results for the projected energy correlators. 
Note that although we compute the generalized moments numerically through the \textsc{Stripper} framework, the final resummation results are  nevertheless analytic functions of $R_L$, 

allowing us to take derivatives and change the values of $R_L$ considered in post. We further note that, due to the very inclusive nature of the moments, their Monte Carlo integration converges very quickly: only about 40k core-h were needed for the unmatched results presented in this work.

\subsection{Non-perturbative Power Corrections}\label{sec:NP_corrections}

To obtain final predictions for the projected energy correlators, we must also include non-pertur-bative power corrections. The structure of the leading non-perturbative corrections to event-shape observables was understood in the seminal work of Korchemsky and Sterman~\cite{Korchemsky:1999kt}, and this analysis has recently been generalized to the projected $N$-point energy correlators~\cite{Lee:2024esz,Chen:2024nyc}.
The leading non-perturbative power correction, suppressed by $\Lambda_\text{QCD}/p_T$, is characterized by the soft matrix element
\begin{equation}
\label{eq:Omega1_def}
\Omega_{1\kappa}\equiv
\frac{1}{N_\kappa}
\langle 0|
\mathrm{tr}\!\left[
\overline{Y}_{\bar n}^{\dagger \kappa}
Y_n^{\dagger \kappa}
\mathcal{E}_T(0)
Y_n^\kappa
\overline{Y}_{\bar n}^\kappa
\right]
|0\rangle\,,
\qquad
\kappa=q,g\,.
\end{equation}
Here $Y_n$ and $\overline{Y}_{\bar{n}}$ are soft Wilson lines along the two back-to-back lightlike directions $n^\mu$ and $\bar{n}^\mu$, and $\mathcal{E}_T(0)$ is the transverse energy flow operator evaluated at the origin. The color normalization factors are $N_q=N_c$ and $N_g=N_c^2-1$ for the quark and gluon channels, respectively.
Note that the $\Omega_{1\kappa}$ are the same non-perturbative parameters that enter $e^+e^-$ event shapes such as thrust, heavy jet mass, and the $C$-parameter~\cite{Hoang:2014wka,Benitez:2024nav,Benitez:2025vsp}. They are scheme dependent; throughout this work we adopt the $\overline{\text{MS}}$ scheme and denote them by $\overline{\Omega}_{1\kappa}$.

In the collinear factorization of \Sec{sec:fact},
the leading power correction can be incorporated into the jet functions as
\begin{align}
\label{eq:np_jet} J^{\kappa[N]}\!\left(\ln\frac{R_L^2 x^2 p_T^2}{\mu^2},\mu\right)
=
\hat{J}^{\kappa[N]}\!\left(\ln\frac{R_L^2 x^2 p_T^2}{\mu^2},\mu\right)
-
\frac{N}{x\, p_T R_L} \hat{J}_{\rm NP}^{\kappa[N-1]}\!\left(\ln\frac{R_L^2 x^2 p_T^2}{\mu^2},\mu\right)\,,
\end{align}
where $\hat{J}$ and $\hat{J}_{\rm NP}$ are the perturbative and non-perturbative jet functions, respectively. Due to the prefactor $1/x$ in the second term, $\hat{J}_{\rm NP}^{\kappa[N-1]}$ obeys the same evolution equation as the $(N-1)$-point perturbative jet function $\hat{J}^{\kappa[N-1]}$.
At leading order in $\alpha_s$, the non-perturbative jet function at its canonical scale $\mu_J^0 = p_T R_L$ are given by
\begin{align}
\hat{J}_{\rm NP}^{\kappa[N-1]}\!\left(\ln\frac{R_L^2 p_T^2}{(\mu_J^0)^2},\mu_J^0\right)=\overline{\Omega}_{1\kappa} + \mathcal{O}(\alpha_s)\,.
\end{align}
In this work, we use $\overline{\Omega}_{1q}=0.305$ GeV, obtained from the global thrust fit of Ref.~\cite{Benitez:2024nav} and converted to the $\overline{\text{MS}}$ scheme, and apply Casimir scaling for the gluon channel: $\overline{\Omega}_{1g}=C_A/C_F\, \overline{\Omega}_{1q}$. Since $\overline{\Omega}_{1g}$ has not been determined from data, we additionally vary it by a factor of two around this Casimir-scaling value and display the resulting uncertainty as a separate band in our final predictions below.
The constants of the non-perturbative jet functions beyond leading order are unknown; we therefore model them by the corresponding constants of the $(N-1)$-point perturbative jet function, rescaled by an overall factor of $\overline{\Omega}_{1\kappa}$, and assign each coefficient an independent 100\% uncertainty. Including these $\mathcal{O}(\alpha_s)$ corrections to the non-perturbative constants is particularly important and consistent for our observable. As discussed in \Sec{sec:def}, the energy weights enhance configurations in which the boson is relatively soft compared to the hadronic system, and such configurations first appear in the $\mathcal{O}(\alpha_s)$ corrections. In the fixed-order singular expansion of the non-perturbative power correction, they therefore enter through the NLO hard function convolved with the LO non-perturbative jet function, and consistency of the order counting then requires the NLO non-perturbative jet function to be included as well. Evolving the non-perturbative jet functions with Eq.~\eqref{eq:jet_evo} and convolving them with the hard functions as in Eq.~\eqref{eq:pp_fact_formula}, we obtain the final prediction for the non-perturbative power correction to the ENC in the collinear limit.

In practice, these non-perturbative power corrections can be obtained using the same strategy of computing generalized moments used to compute the singular perturbative contribution described around~\eq{eq:weightedfiducial}. There are only two differences. First, the non-perturbative corrections to the $N$-point correlator should be computed using the weight $z^{N-1}$, instead of $z^N$. Second, they should be computed using an additional factor $1/p_{\gamma,T}$. While the first difference is easily taken into account by just including one moment lower than needed for the purely perturbative parts (i.e.~by also computing $N=1$ in our case), the second difference can only be properly taken into account by computing a second set of moments, since $p_{\gamma,T}$ changes from event to event, and this additional factor can only be included on the fly. We therefore compute all moments for $N=1,...,6$, both with and without an additional factor $1/p_{\gamma,T}$. The inclusion of the non-perturbative corrections is then fully analogous to the computation of the resummed singular contribution.

So far we have only considered the non-perturbative corrections to the singular distribution. In principle, one should also include the corrections to the non-singular contribution in the fixed-order region. However, the $R_L$ range studied in this work lies mostly in the collinear regime, where the non-singular piece is very small (see Sec.~\ref{sec:singular}). We therefore simply extend the non-perturbative corrections obtained from the jet functions above across the fixed-order region. The final matched predictions for the projected energy correlators can then be written as
\begin{equation}
\label{eq:matching}
\frac{d\sigma^{[N]}}{dR_L}=\frac{d\hat{\sigma}^{[N]}_{\rm match}}{dR_L}+\frac{d\sigma^{[N]}_{\rm NP}}{dR_L},\qquad \text{with }\quad
\frac{d\hat{\sigma}^{[N]}_{\rm match}}{dR_L}
=
\frac{d\hat{\sigma}^{[N]}_{\rm resum}}{dR_L}
+
\frac{d\hat{\sigma}^{[N]}_{\rm ns}}{dR_L}
\,.
\end{equation}

\section{Numerical Results}\label{sec:pheno}

\begin{figure}
	\centering

	\includegraphics[width=0.49\textwidth]{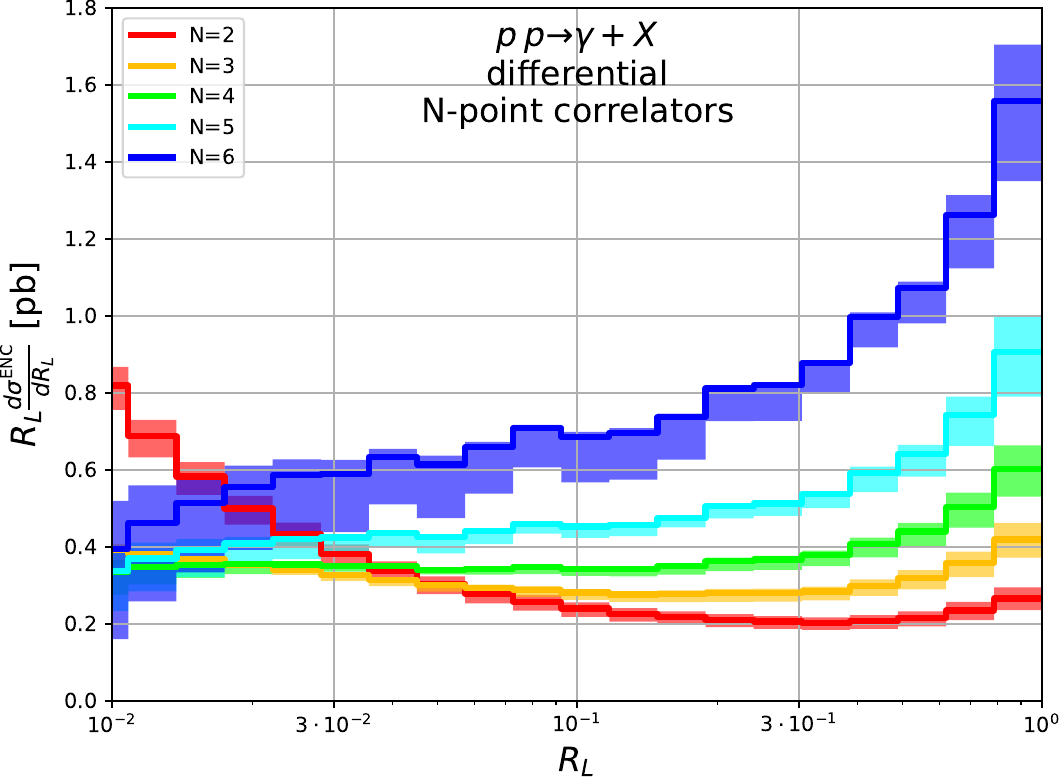}
	\includegraphics[width=0.49\textwidth]{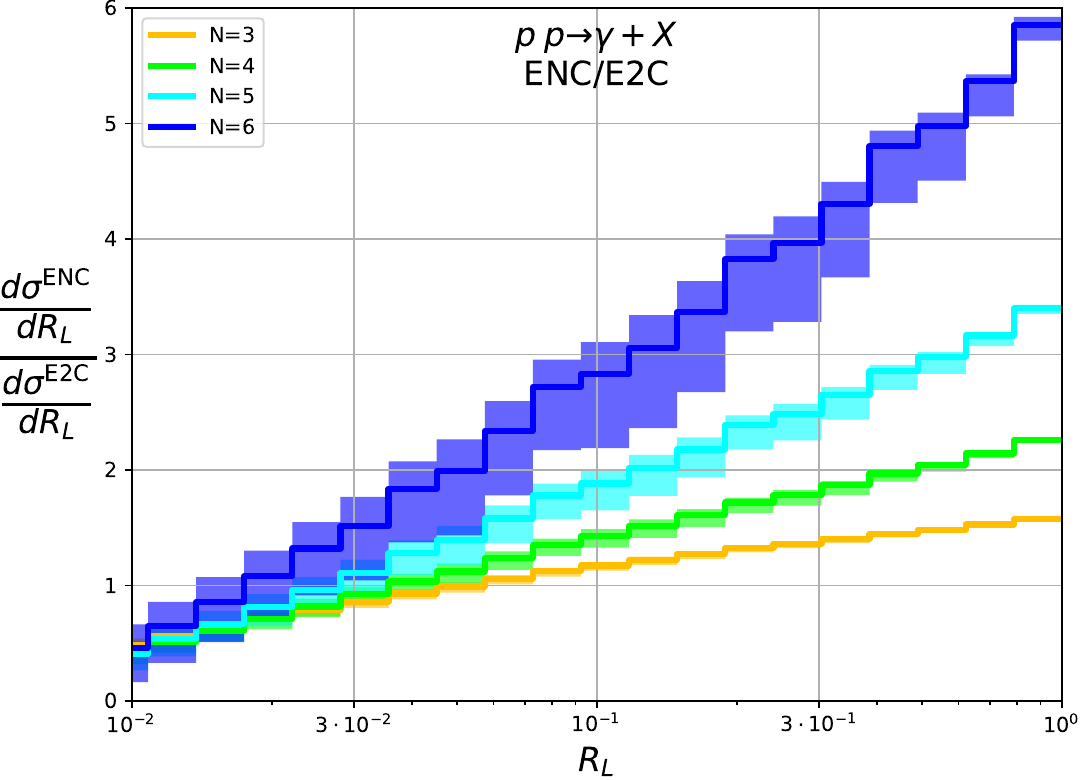}
	\caption{Left panel: NLO+NNLL predictions for projected energy correlators with all values of $N$, where the leading non-perturbative corrections are included. Right panel: The ratio of the $N$-point correlators with respect to the 2-point correlator.}
	\label{fig:BestPredictions}
\end{figure}

In this section, we apply our approach to derive numerical results for the projected energy correlators in $\gamma+X$ events at the LHC. Our main results are shown in \Fig{fig:BestPredictions}, presenting both the differential distributions for the energy correlators and their ratios. In the following sections, we will present the detailed contributions from resummation, fixed-order matching, and non-perturbative corrections, as well as study the perturbative convergence.

As a caveat throughout this section, all uncertainties are estimated through scale variations. While this is consistent with what is done in the literature, care must be taken in interpreting these estimates, particularly for the case of ratio observables. A better approach is provided by Theory Nuisance Parameters (TNPs)~\cite{Tackmann:2024kci,Cridge:2025wwo}, whose application we plan to investigate in the future.

\subsection{Singular and Non-Singular Distributions}\label{sec:singular}

\begin{figure}
	\centering
	\includegraphics[width=0.47\textwidth]{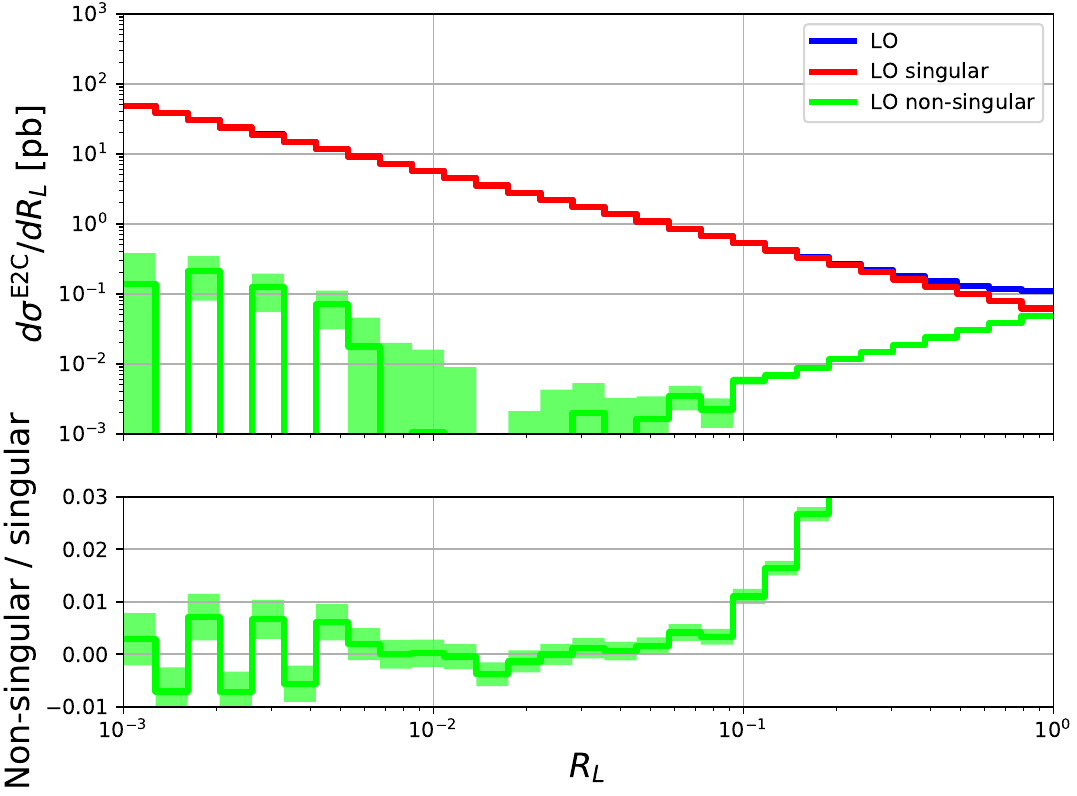}
	\includegraphics[width=0.47\textwidth]{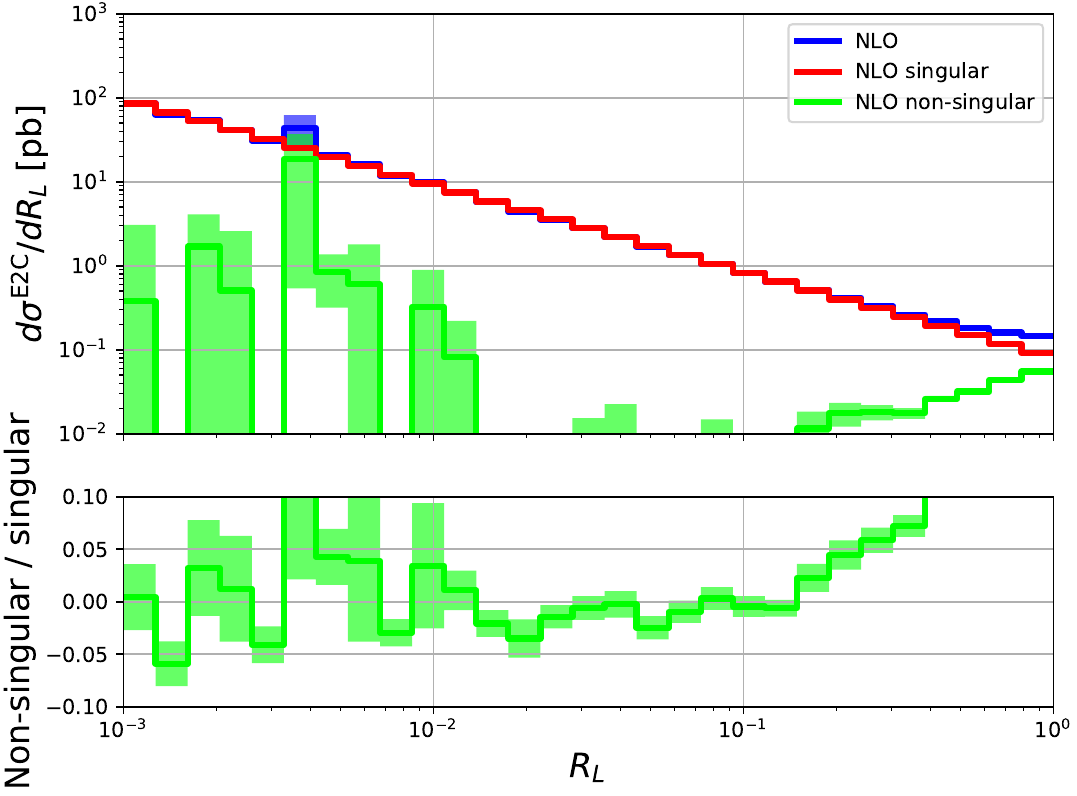}
	\caption{The full, singular and non-singular two-point energy correlators at LO (left) and the NLO corrections (right). Analogous distributions for higher-point correlators can be found in App.~\ref{app:singular_plots}. }
	\label{fig:AEEC_NPC_text}
\end{figure}

We begin by exploring the singular and non-singular distributions for the projected energy correlator observables in $\gamma+X$. This is important both for validating our factorization theorem and for understanding the size of the perturbative power corrections to it.
In \Fig{fig:AEEC_NPC_text}, we show the complete numerical calculation of the two-point energy correlator, compared with the singular prediction derived from our factorization theorem and the numerically computed hard function, at both LO and NLO. We also show the non-singular contribution, which is defined as the difference between the complete fixed-order calculation and the singular contribution. It is also referred to as the perturbative power correction. Analogous distributions up to the projected six-point correlator can be found in App.~\ref{app:singular_plots}.

First of all, we find good agreement between the complete numerical result and the singular result as predicted by our factorization theorem, in the limit $R_L \to 0$. This provides a highly non-trivial check on our factorization theorem. It also illustrates the numerical stability of the subtraction scheme.
Note that these plots show only the perturbative predictions. In reality, the distribution becomes non-perturbative at sufficiently small values of $R_L \sim \Lambda_{\rm QCD}/p_T$ and thus the result in \Fig{fig:AEEC_NPC_text} should not be interpreted as an actual prediction for the distribution below this value, but rather as a test of our factorization theorem.

We can fit the numerical fixed-order data with the following form,
\begin{equation}
\label{eq:pert_pc_fit}
\text{ENC}_\text{full}^{(n)}=\text{ENC}^{(n)}_\text{sing}\times f^{(n)}(R_L)\equiv\text{ENC}^{(n)}_\text{sing}\times\sum_{i=0}^\infty a_i^{(n)}R_L^{2i}\;,
\end{equation}
where $n$ represents the order $\mathcal{O}(\alpha_s^n)$, $\text{ENC}^{(n)}_\text{sing}$ is the singular expansion from the factorization theorem at order $n$ and $a_i^{(n)}$ are unknown coefficients. The non-singular contribution, namely the perturbative power corrections, is captured by the series expansion in $R_L^2$.

\begin{figure}
	\centering
	\includegraphics[width=0.45\textwidth]{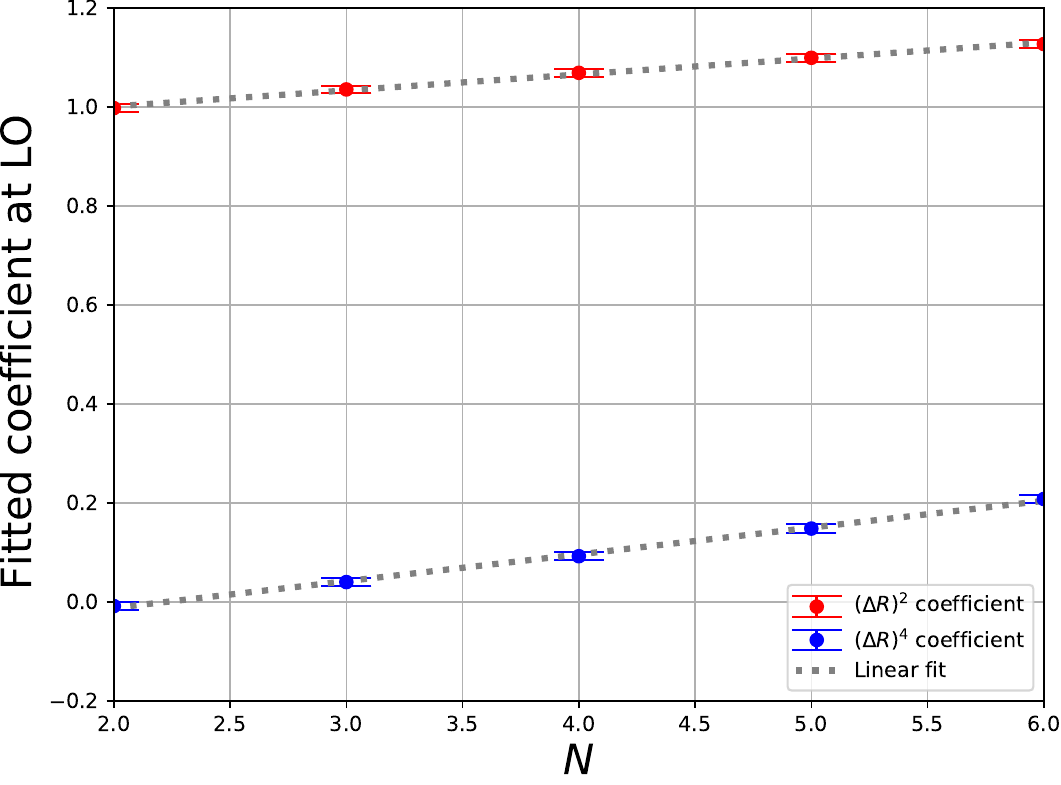}
	\includegraphics[width=0.45\textwidth]{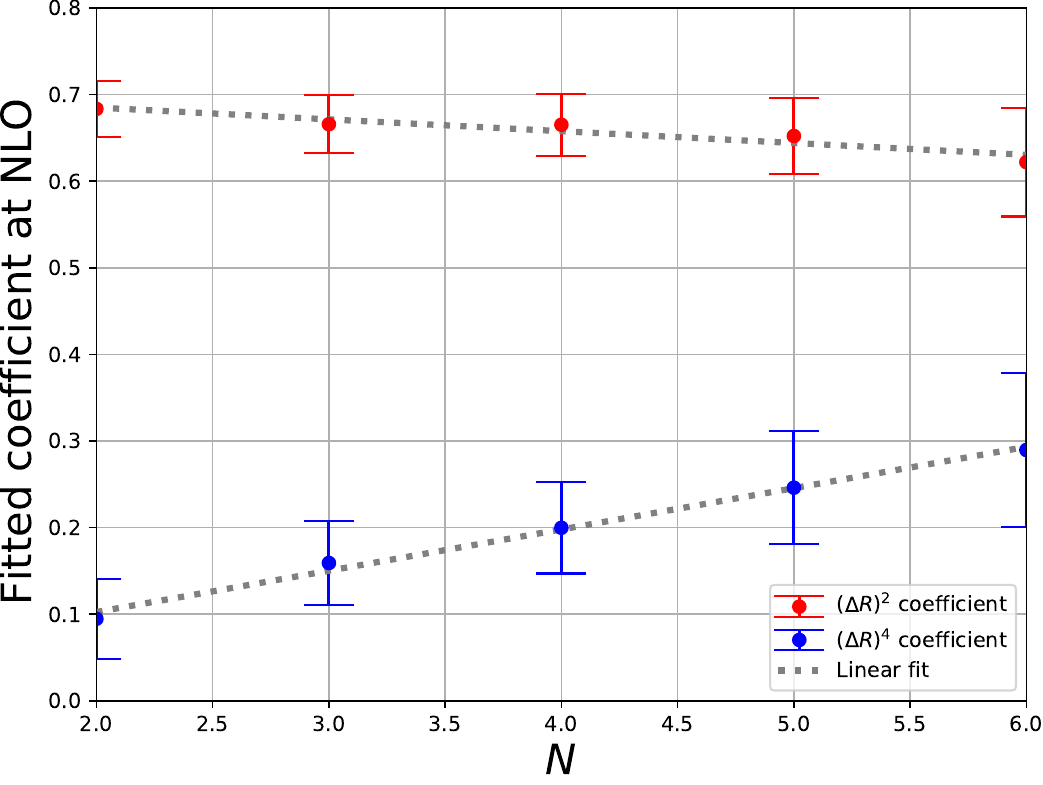}
	\caption{Coefficients $a_1^{(n)}$ (red) and $a_2^{(n)}$ (blue) at LO (left) and NLO (right) as a function of $N$. The values are obtained by fitting to numerical data from \textsc{Stripper}. The gray dashed lines show linear fits in $N$.}
	\label{fig:AEECPowerCorrectionsFit_text}
\end{figure}

The agreement in the $R_L\to 0$ limit shown in \Fig{fig:AEEC_NPC_text} implies that $a_0^{(n)}=1$ for any $n$. In Fig.~\ref{fig:AEECPowerCorrectionsFit_text}, we perform the fit for $a_1^{(n)}$ and $a_2^{(n)}$ at both LO and NLO, for $N=2$--$6$. In Ref.~\cite{Dixon:2019uzg}, it was found that setting $a_1^{(n)}=1/4$ and $a_2^{(n)}=1/24$ captures the non-singular contribution very well for the EEC in $e^+e^-$ collisions. Interestingly, for $pp\to \gamma+X$, we find $a_1^{(1)}\approx1$ and $a_2^{(1)}\approx 0$. This means that at LO, the coefficient of the next-to-leading power (NLP) correction is almost the same as that of the LP (singular) expansion, and the higher powers almost vanish. At NLO, we find $a_1^{(2)} \approx 0.7$ and $a_2^{(2)}\approx 0.1$, and thus the NLP coefficient is still very close to the LP one. As $N$ increases, the higher powers contribute more. Lastly, we find that at both orders the coefficients are, to a good approximation, linear in $N$: the gray dashed lines in \Fig{fig:AEECPowerCorrectionsFit_text} show linear fits, which describe the coefficients very well over the full range $N=2$--$6$. This linearity could potentially be used to recover the full $R_L$-dependence for any $N$ from the factorization theorem together with calculations at just two values of $N$.

\begin{figure}
	\centering	\includegraphics[width=0.47\textwidth]{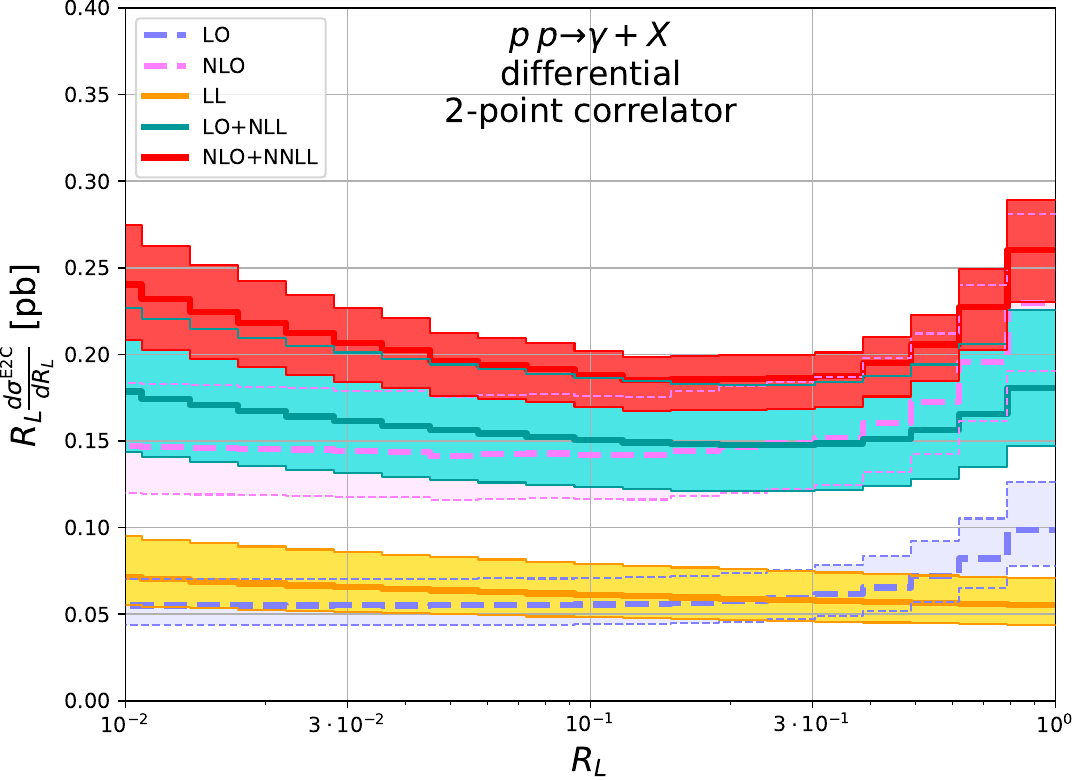}
\includegraphics[width=0.47\textwidth]{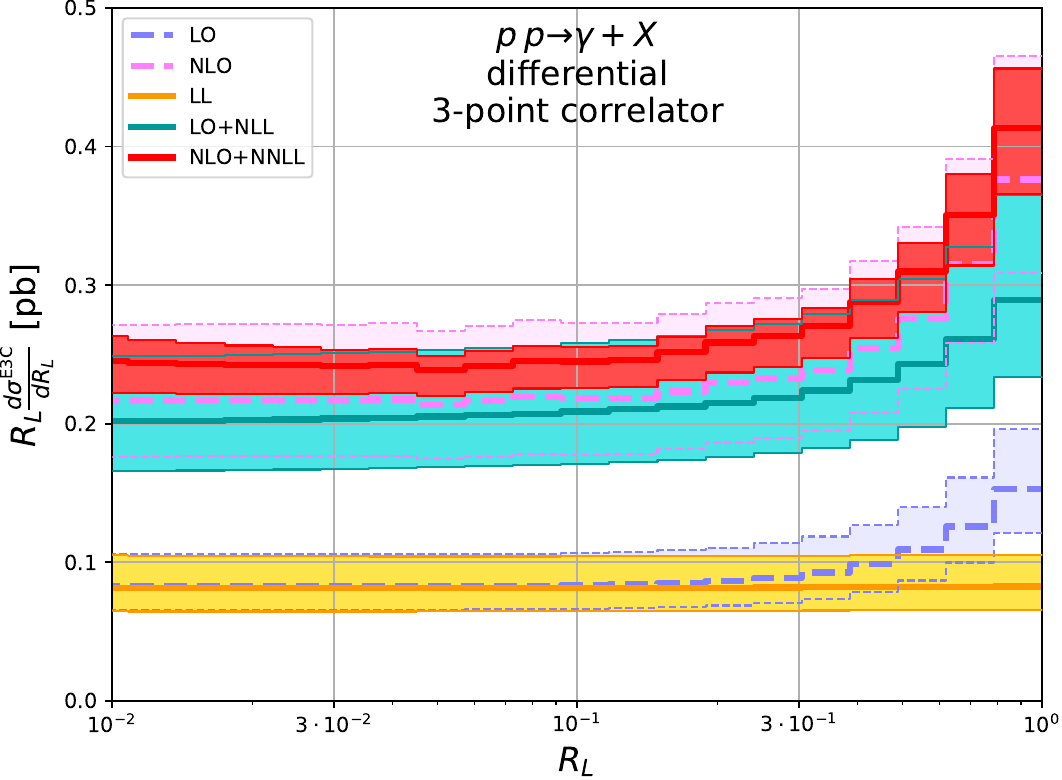}
	\caption{Differential matched distributions for the two-point (left) and three-point (right) correlators through NLO+NNLL. Analogous distributions for higher-point correlators can be found in App.~\ref{app:matched_plots}.}
	\label{fig:AEECN_full_text}
\end{figure}

\subsection{Matched resummation results}\label{sec:LHC}

In this subsection, we present results for matched projected energy correlators in $\gamma+X$ events at the LHC. We study in detail the effect of the matching, both on the full energy correlator distributions and on the ratio observables. This is of particular interest in light of recent strong coupling extractions from ratios of projected correlators~\cite{Chen:2023zlx,CMS:2024mlf}: the scaling region from which $\alpha_s$ is extracted can receive important contributions from non-singular terms, which must be accounted for in a precision extraction. Our results are the first complete matched calculations of projected energy correlators at this order at hadron colliders.

The matching formula for combining resummation and fixed-order is given in Eq.~\eqref{eq:matching}. In principle, we need to design a profile function (scale) to turn off the resummation properly when the singular contribution is no longer dominant~\cite{Abbate:2010xh,Hoang:2014wka,Benitez:2024nav,Benitez:2025vsp} (for an application of profile scales to the track-based EEC, see Ref.~\cite{Jaarsma:2025tck}).
Given that we are focusing primarily on the collinear region in this work, we do not turn off the resummation but instead add the non-singular corrections additively as in Eq.~\eqref{eq:matching}. Note that the non-singular contribution, $d\hat{\sigma}^{[N]}_{\rm ns}=d\hat{\sigma}^{[N]}_{\rm FO}-d\hat{\sigma}^{[N]}_{\rm sing}$, is precisely the perturbative power correction fit in Eq.~\eqref{eq:pert_pc_fit}.
We leave the design of profile scales to future work.

In \Fig{fig:AEECN_full_text} we show results for the matched energy correlators in $\gamma+X$ for two- and three-point energy correlators, and analogous results up to six-point projected correlators can be found in App.~\ref{app:matched_plots}. We present three resummation orders, LL, LO+NLL and NLO+NNLL. For comparison, we also include two fixed-order distributions. Aside from the lowest-order predictions, LO and LL, we observe good perturbative convergence: the scale uncertainty bands of the NLO, LO+NLL and NLO+NNLL predictions all overlap and the uncertainties are reduced by a factor of 2-3, down to about 10\%, when going from NLO or LO+NLL to NLO+NNLL. There are two exceptions to the good perturbative behavior. First, the NLO prediction starts to deviate from the resummed predictions in the small-angle limit. Of course, this is expected, since fixed-order calculations break down in that limit. The second exception is the large-angle limit, where we observe large perturbative corrections going from LO+NLL to NLO+NNLL. In this region, however, the distribution is dominated by the fixed-order contribution, for which large NLO corrections are not unusual. In the small angle regime, these distributions exhibit the characteristic enhancement associated with a scaling behavior in the collinear limit. This behavior is identical to their behavior in $e^+e^-$ collisions, or within jets.

\begin{figure}
	\centering
	\includegraphics[width=0.49\textwidth]{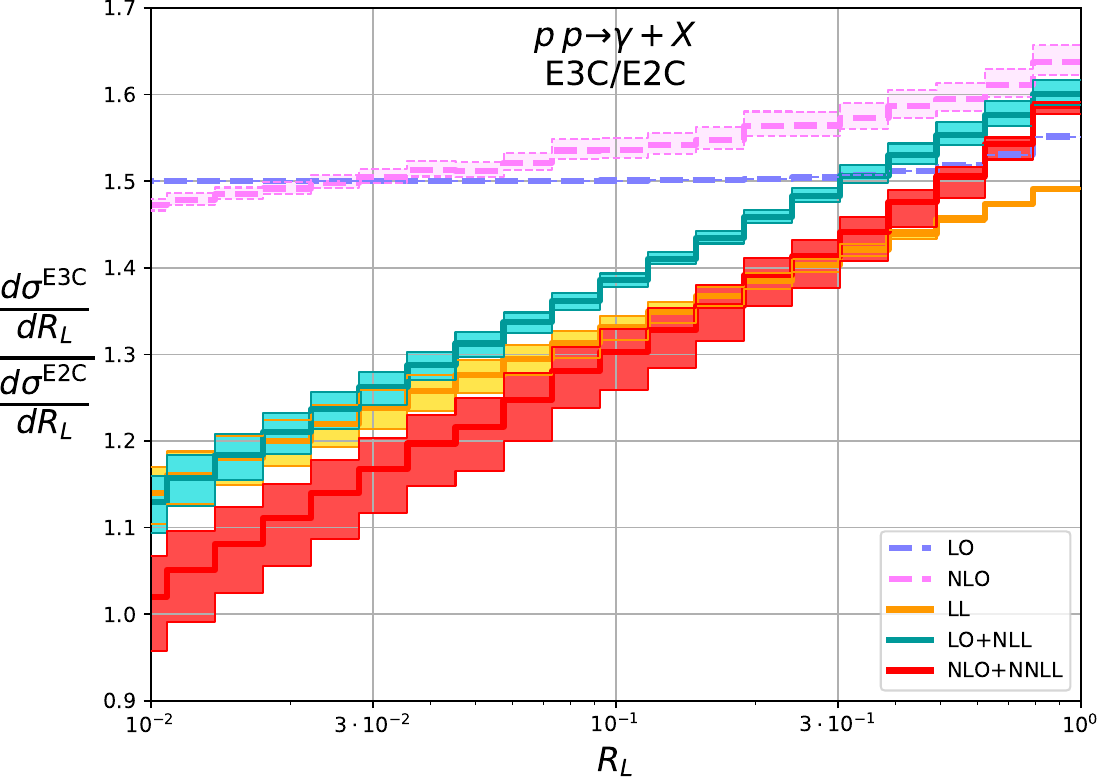}
	\includegraphics[width=0.49\textwidth]{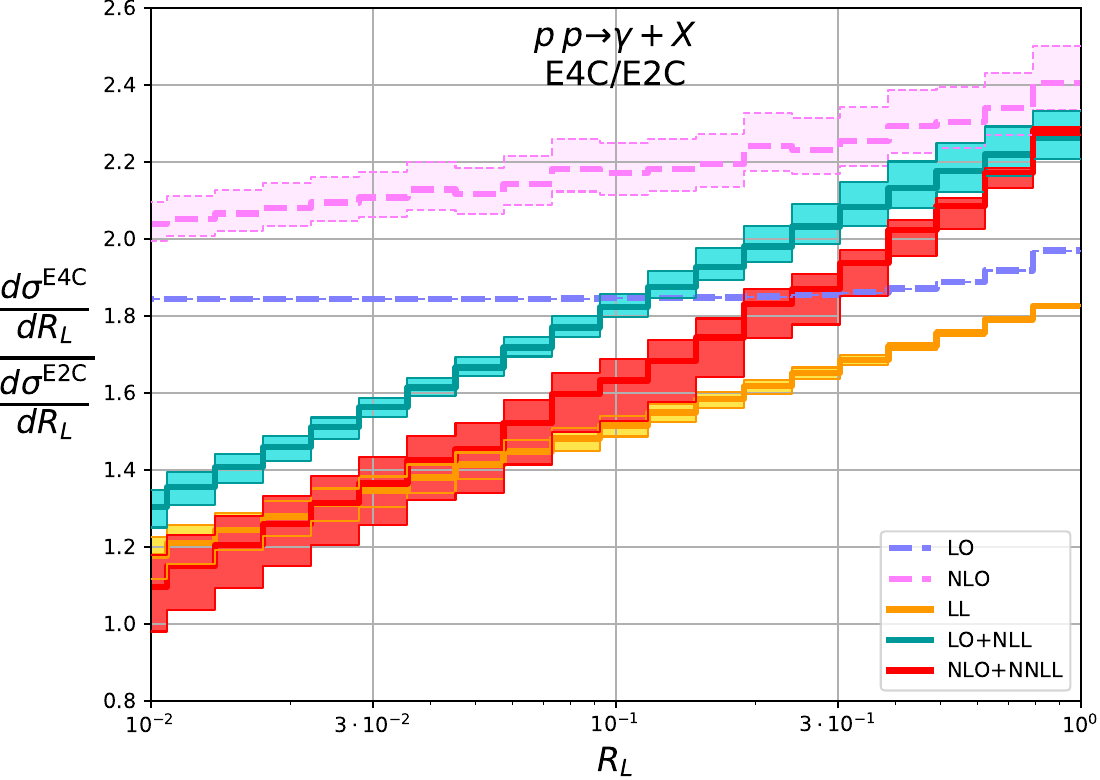}
	\includegraphics[width=0.49\textwidth]{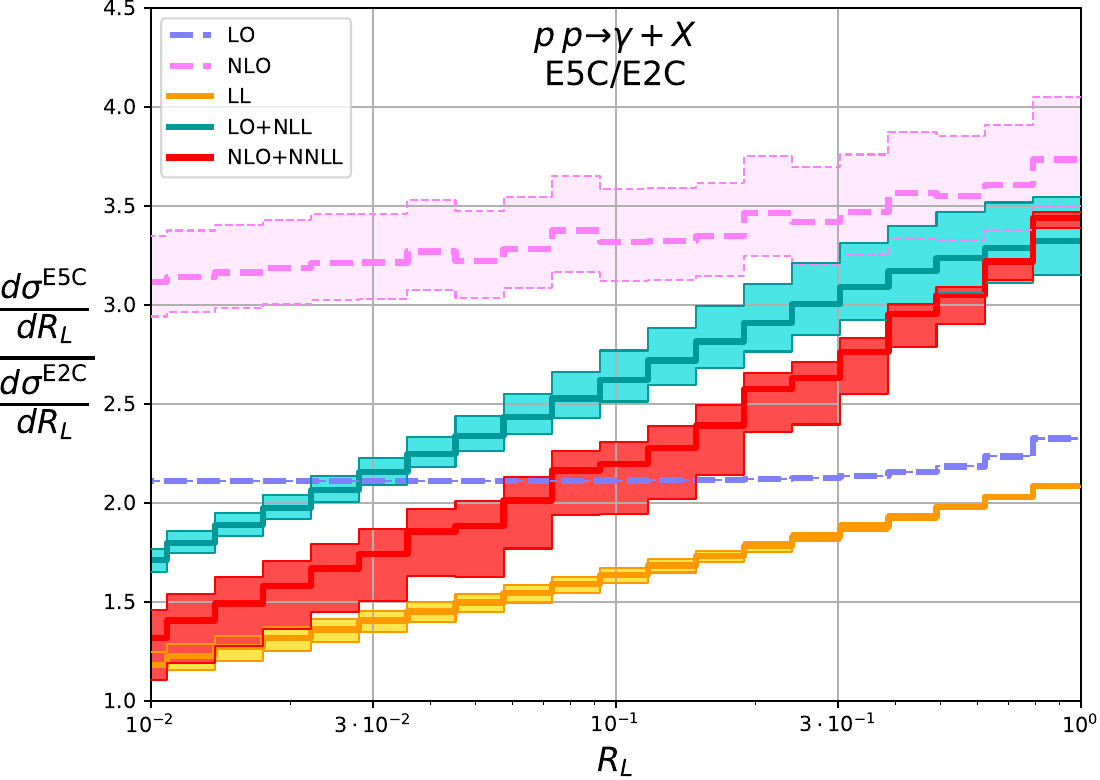}
	\includegraphics[width=0.49\textwidth]{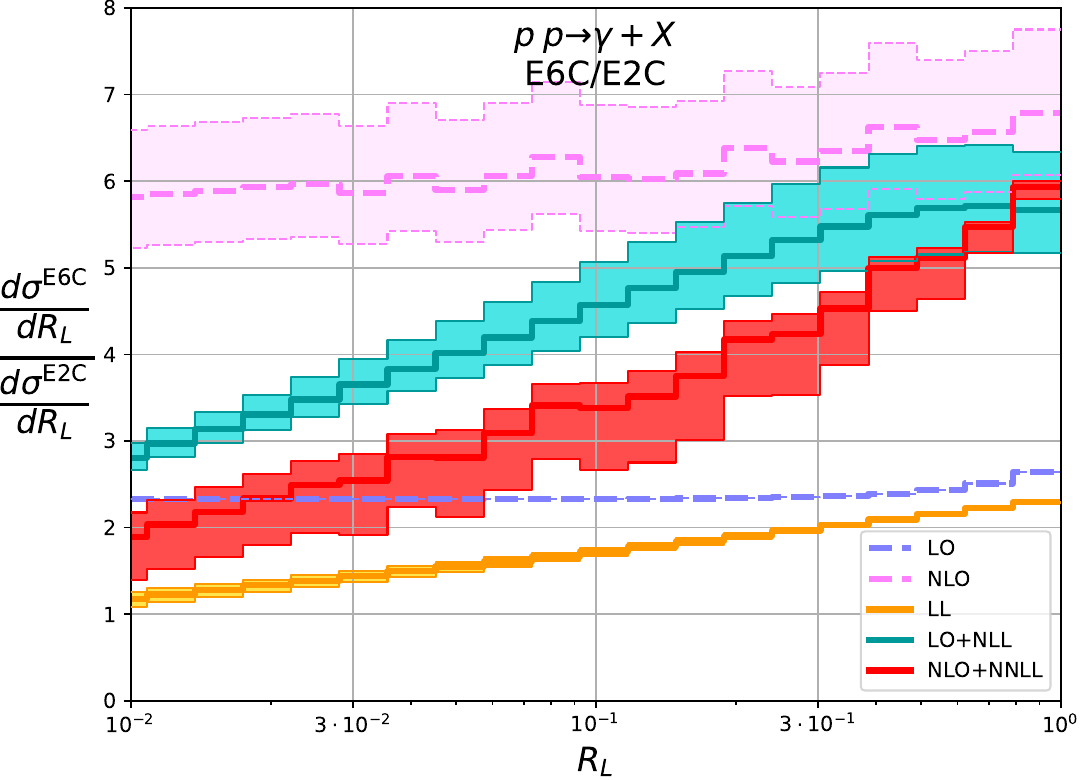}
	\caption{The ratios of the differential $N$-point correlators to the differential two-point correlator through NLO(+NNLL).}
	\label{fig:AENCfullRatio_text}
\end{figure}

In \Fig{fig:AENCfullRatio_text} we show ratios of the three- through six-point projected correlators with respect to the two-point correlator at different orders. These ratios were originally proposed in Ref.~\cite{Chen:2020vvp} to isolate the scaling behavior in the collinear limit of the energy correlators, analogous to three/two jet ratios.
Again, we include fixed-order curves together with the resummed distributions.
Note that the fixed-order results do not even qualitatively describe the shape of the distribution, indicating that resummation is essential to obtain the correct behavior of the energy correlators in the collinear limit.

\begin{figure}
	\centering
	\includegraphics[width=0.49\textwidth]{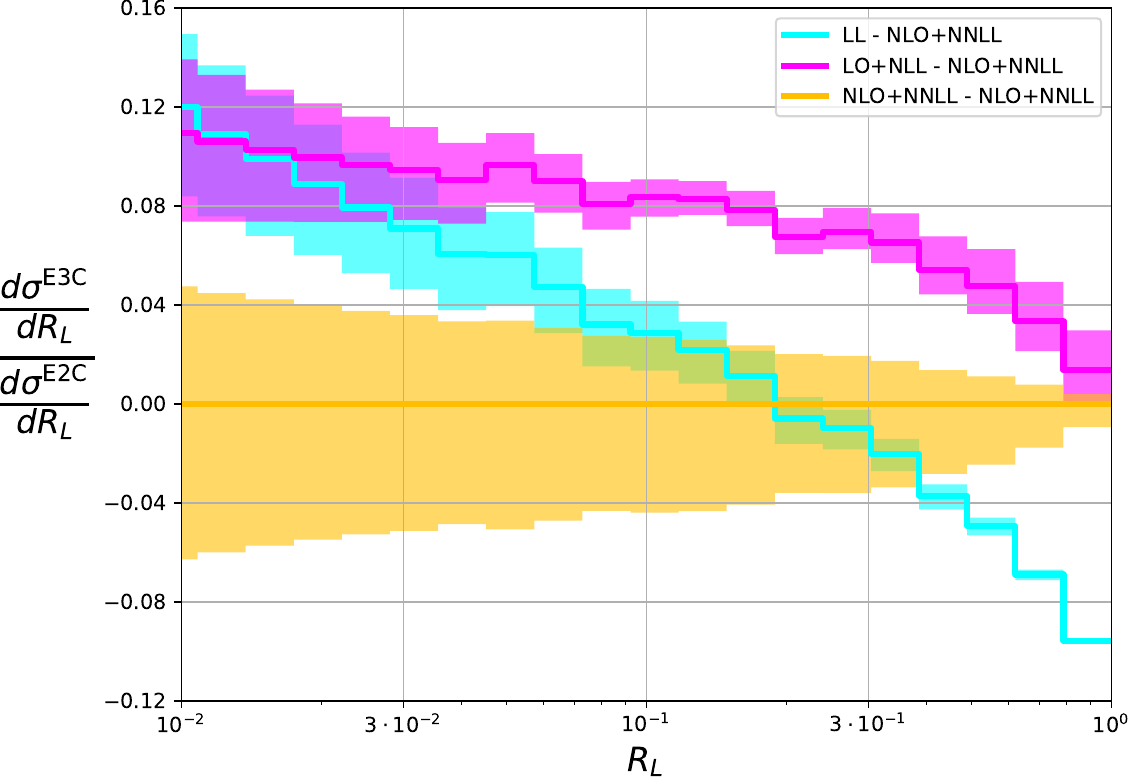}
	\includegraphics[width=0.49\textwidth]{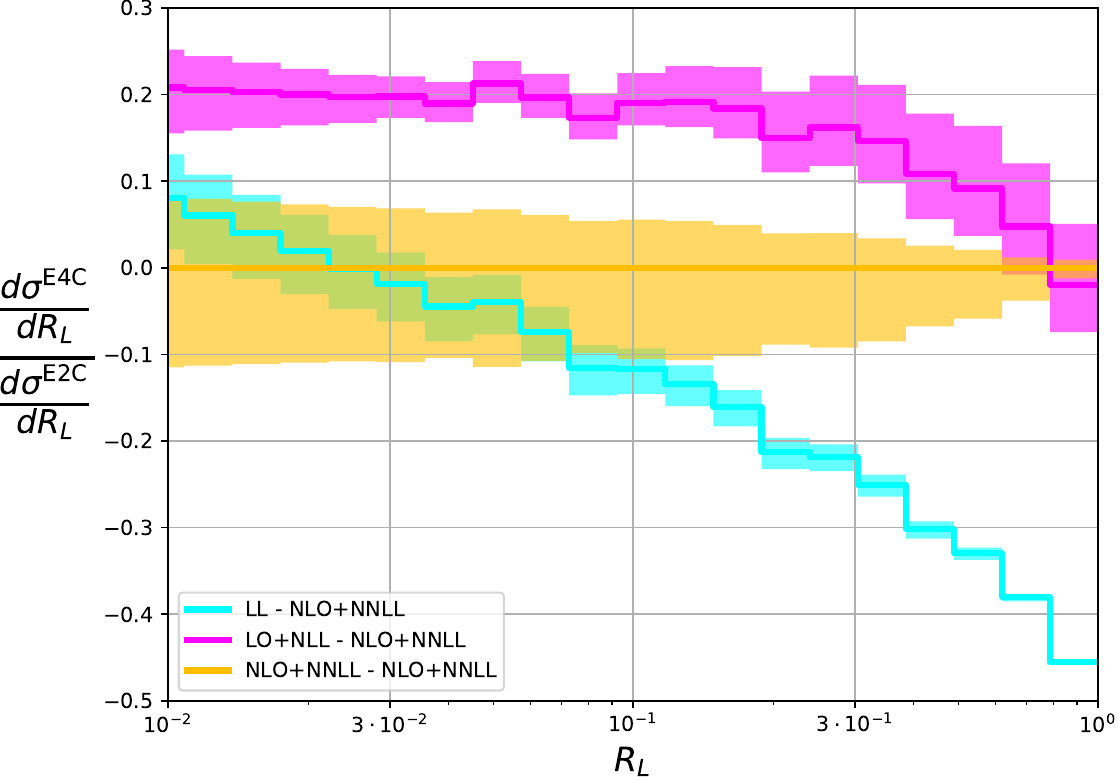}
	\caption{The ratios of the differential three- (left) and four-point (right) correlators to the differential 2-point correlator through NLO+NNLL, shown as differences relative to the NLO+NNLL curves. Analogous distributions for higher-point correlators can be found in App.~\ref{sec:ad_plots}.}
	\label{fig:AENC_diffs_text}
\end{figure}

To assess the stability of the slope, we compare these ratios $(d\sigma^{\text{ENC}}/dR_L)/(d\sigma^{\text{E2C}}/dR_L)$ at each order with the highest-order NLO+NNLL prediction. In \Fig{fig:AENC_diffs_text}, we show the cases of the three- and four-point correlators. For $10^{-2}<R_L<10^{-1}$, avoiding the non-perturbative and fixed-order regimes, the slope changes by about 10\% between LO+NLL and NLO+NNLL for the three-point case, indicating an uncertainty on the slope smaller than 10\% at NLO+NNLL, consistent with the scale uncertainties at that order, which correspond to an uncertainty on the slope of about 7\%. For the four-point case, the slope only changes negligibly between LO+NLL and NLO+NNLL, while the scale uncertainties indicate a roughly 5\% uncertainty on the slope at NLO+NNLL. For both the three-point and the four-point case, the change in slope between LL and LO+NLL is much larger, on the order of 30\%, than the change between LO+NLL and NLO+NNLL, suggesting good perturbative convergence.

\begin{figure}
	\centering
	\includegraphics[width=0.47\textwidth]{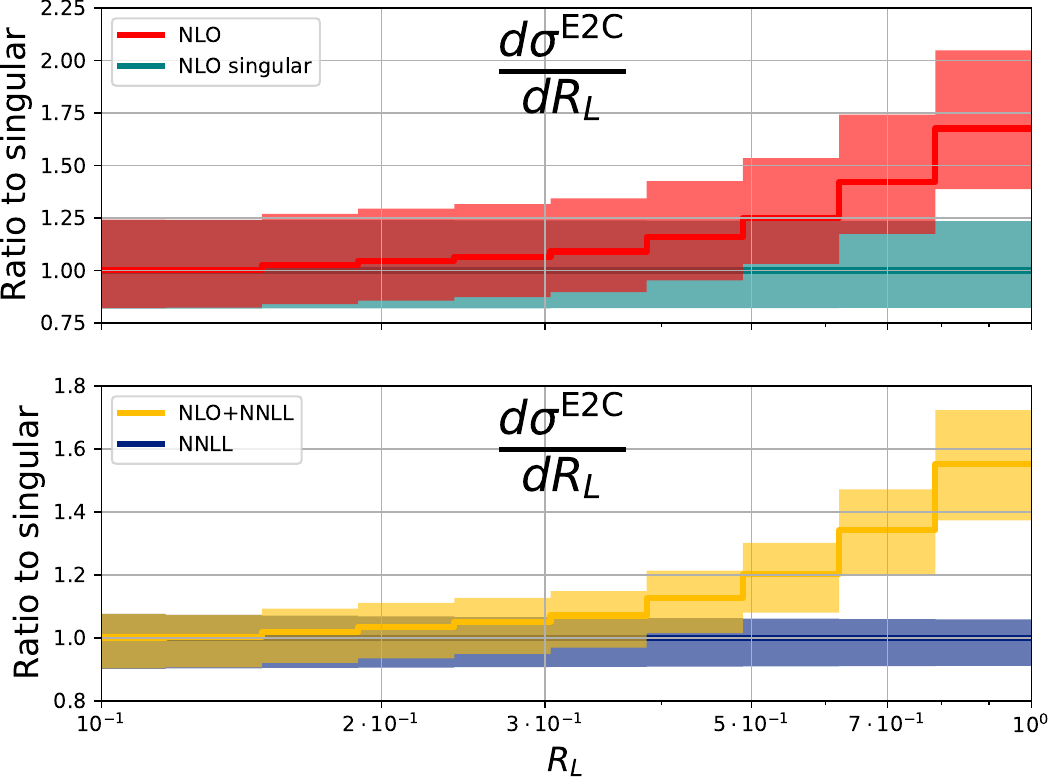}
	\includegraphics[width=0.47\textwidth]{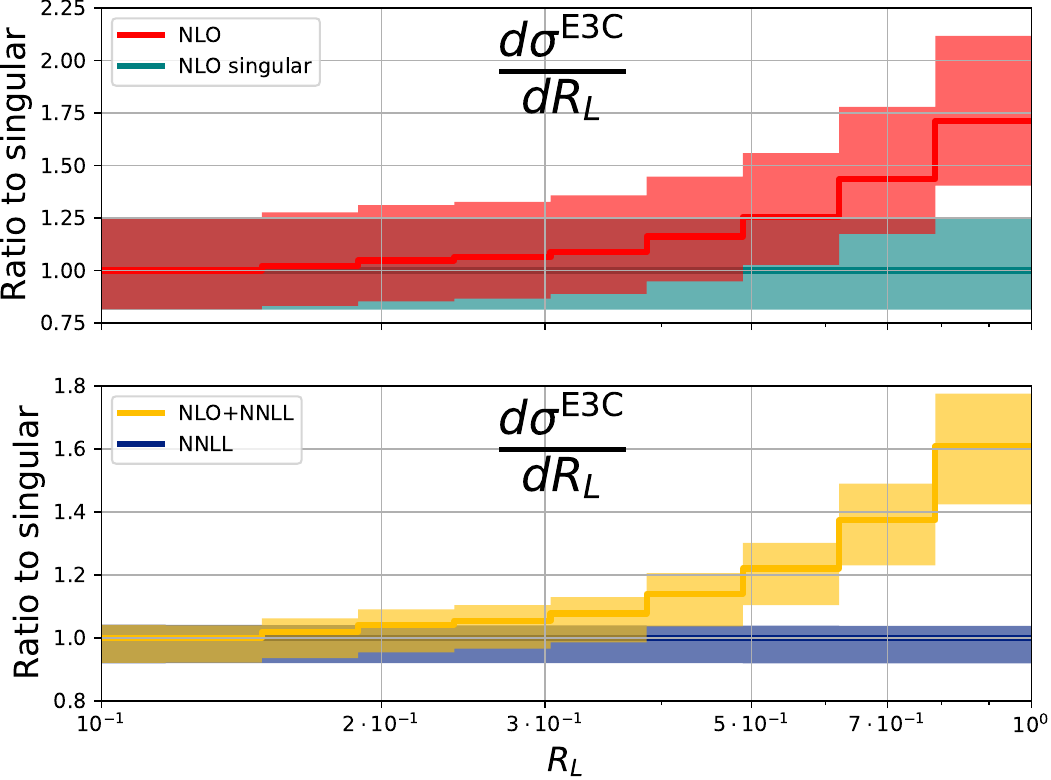}
	\caption{Impact of matching corrections in ENC resummation. The left panel shows EEC and the right panel is E3C. In the top panel, we compare the NLO fixed-order with its singular expansion, normalized to the latter. In the lower panel, we show the matched NLO+NNLL and pure NNLL resummations, both normalized to the latter. We find important non-singular corrections at larger $R_L$.}
	\label{fig:AdEECNPC_text}
\end{figure}

\begin{figure}
	\centering
	\includegraphics[width=0.47\textwidth]{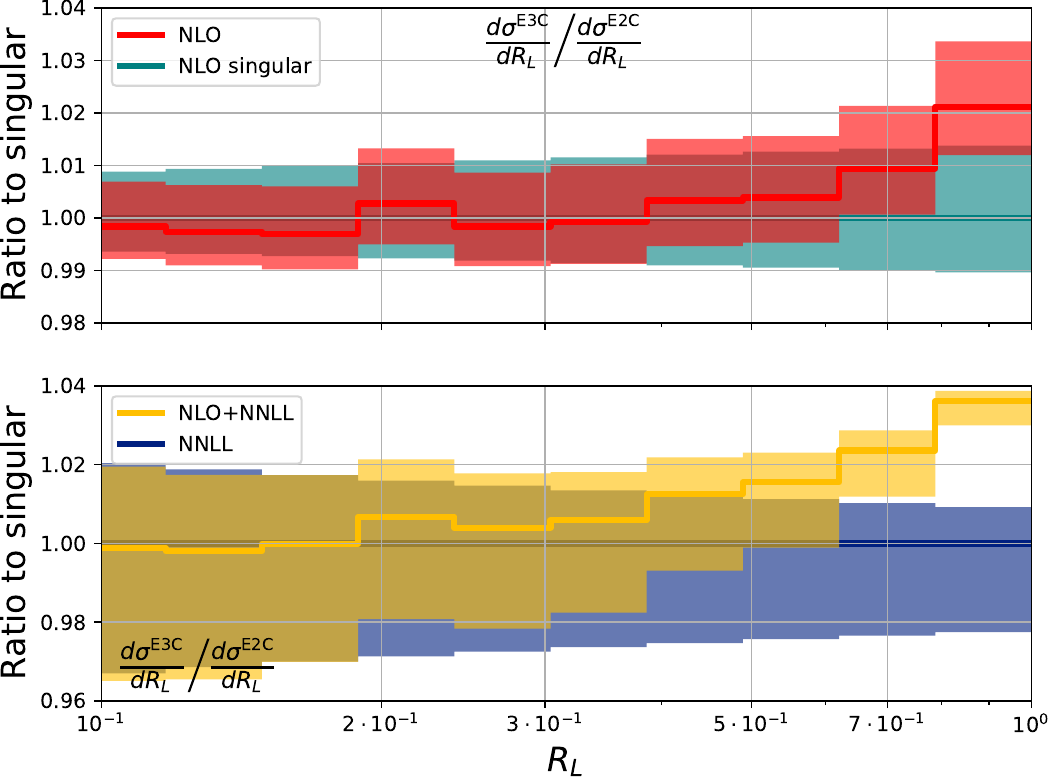}
	\includegraphics[width=0.47\textwidth]{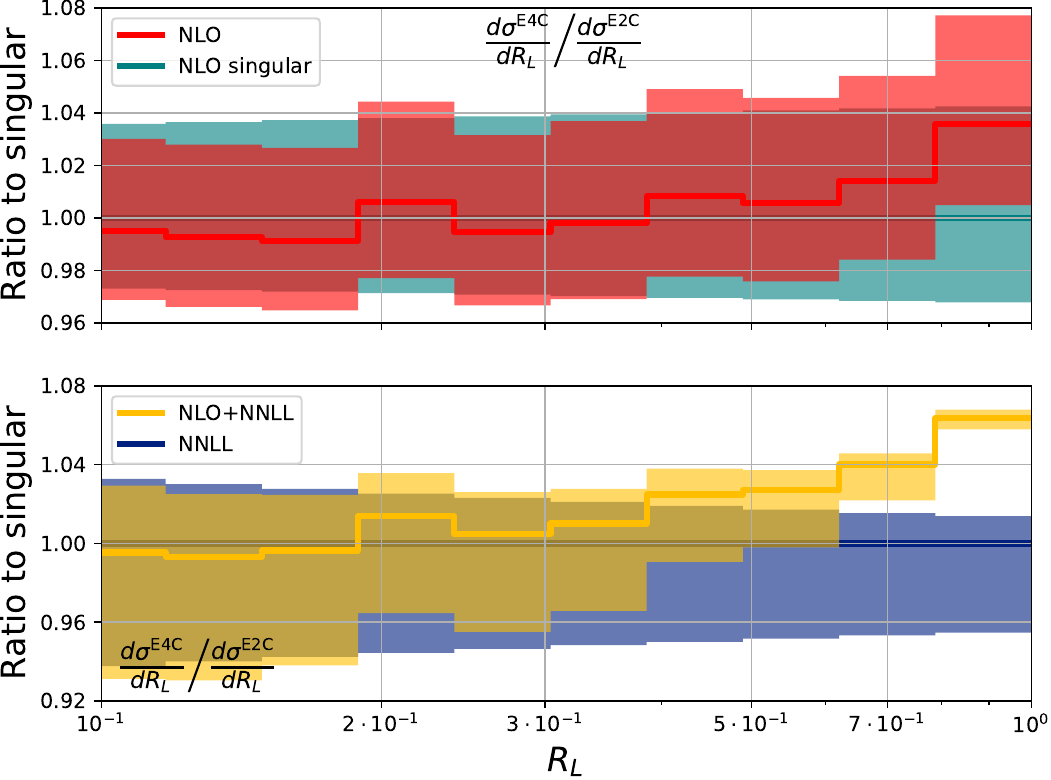}
	\caption{The impact of matching corrections to resummation similar to \Fig{fig:AdEECNPC_text}, but for the ratio correlators.}
	\label{fig:AdEECNRatioPC_text}
\end{figure}

In \Fig{fig:AdEECNPC_text}, we investigate the impact of matching corrections, namely the non-singular contributions. For both the EEC and E3C, we plot the pure NNLL resummation and the matched NLO+NNLL result, normalized to the former. For $R_L<0.3$, we find that the matched resummed distributions are close to the pure resummation, while at larger $R_L$, non-singular contributions from the fixed-order calculation become important. In particular, when $R_L$ is $\mathcal{O}(1)$, the matching correction reaches around $50\%$ for $N=2$ and $N=3$-point correlators. As $N$ increases, the non-singular correction also increases, approaching up to $80\%$ for the six-point correlator. The distributions beyond three-point are provided in App.~\ref{app:matching_impact}.  In \Fig{fig:AdEECNRatioPC_text}, we provide the identical figures for the ratio correlators. In both cases, we find that the non-singular corrections are significantly smaller: when $R_L$ is $\mathcal{O}(1)$, the E3C/EEC ratio receives $\lesssim 4\%$ and the E4C/EEC ratio $\lesssim 7\%$ corrections from matching.

\subsection{Complete Predictions with Non-Perturbative Corrections}\label{sec:NP_numerics}

\begin{figure}
	\centering
	\includegraphics[width=0.49\textwidth]{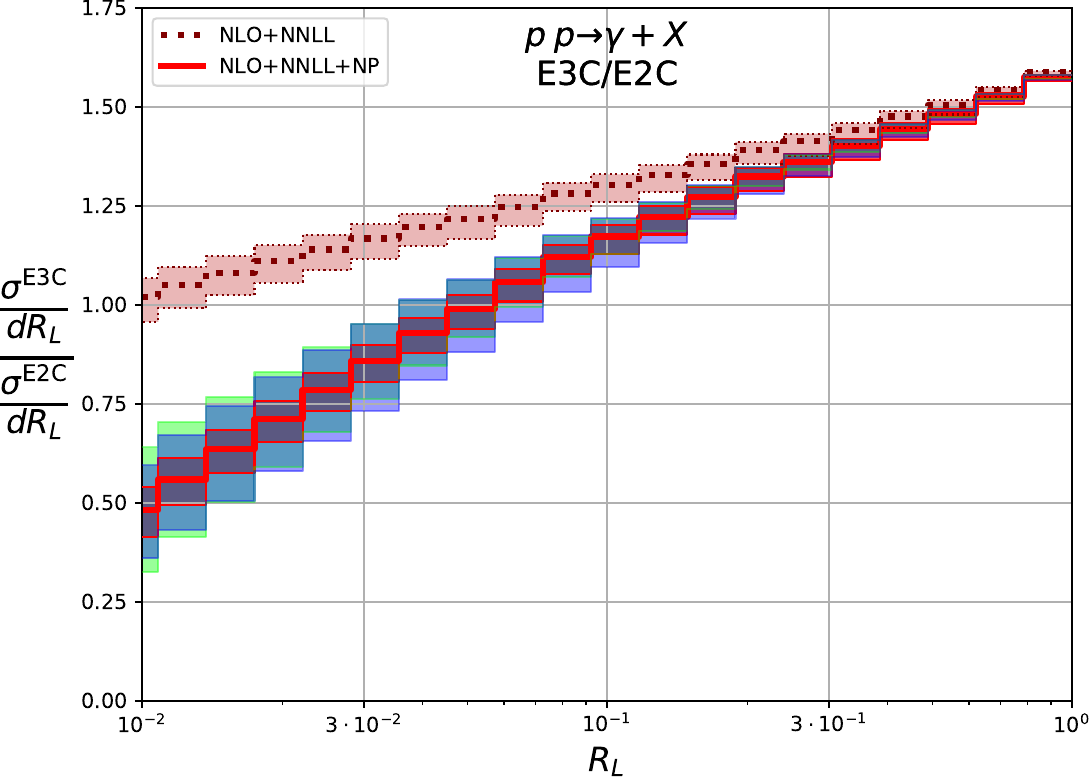}
	\includegraphics[width=0.49\textwidth]{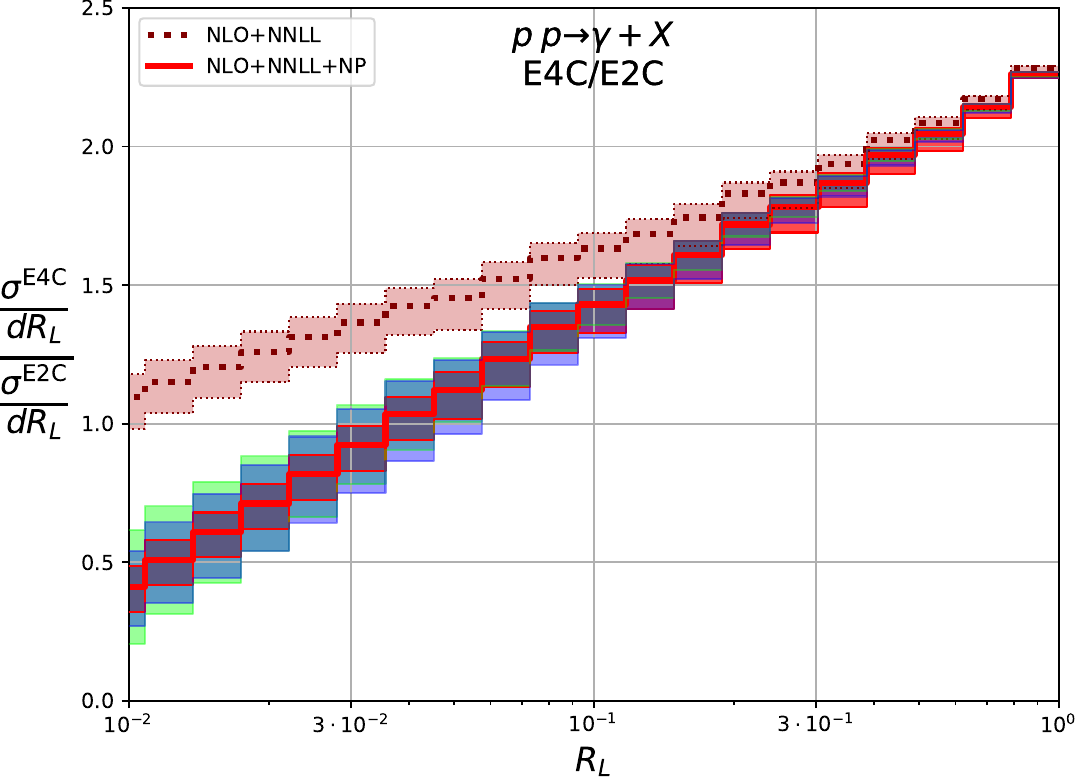}
	\includegraphics[width=0.49\textwidth]{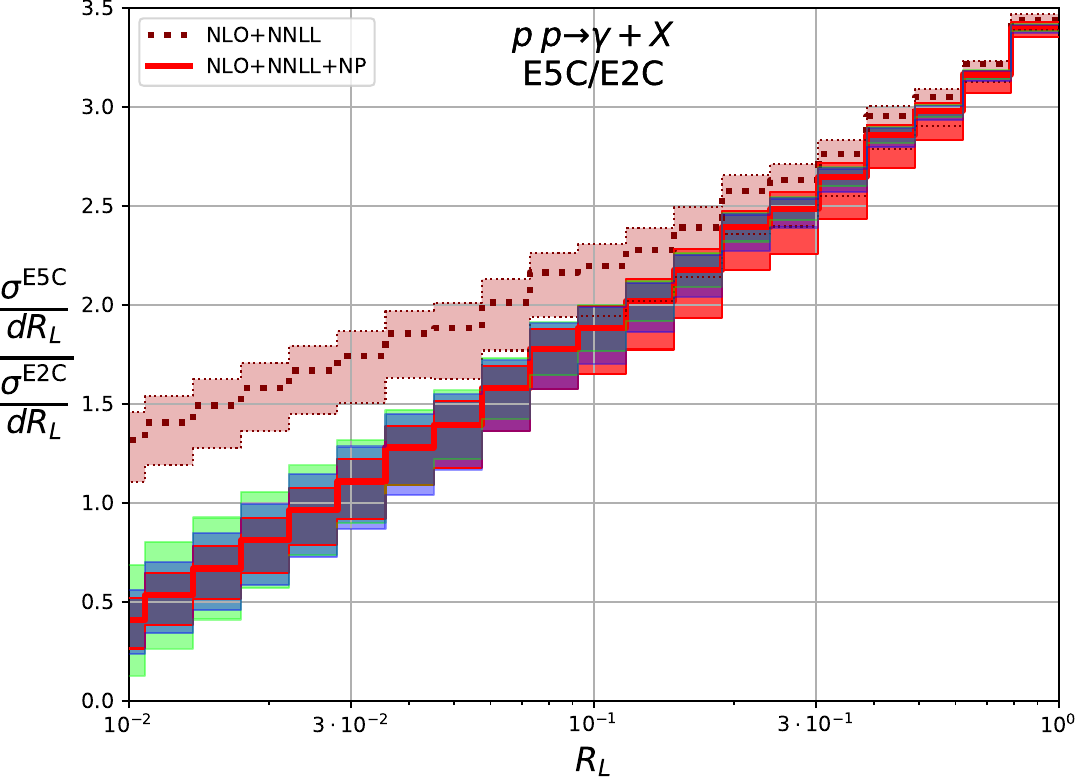}
	\includegraphics[width=0.49\textwidth]{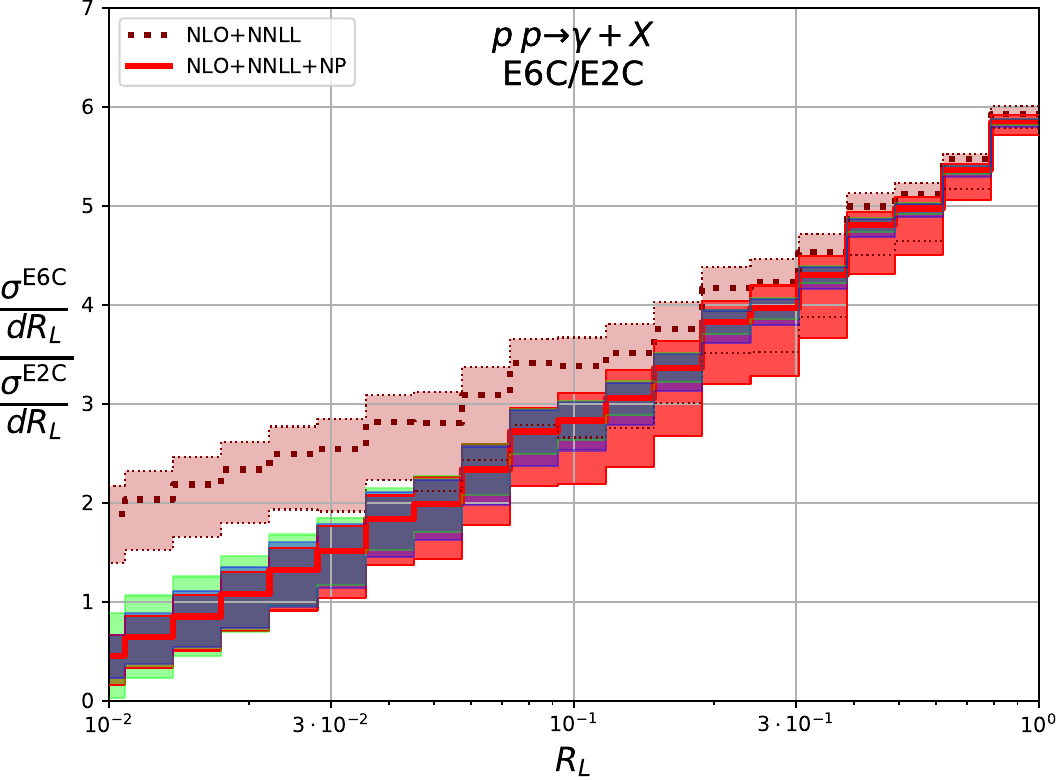}
	\caption{The ratios of the differential $N$-point correlators to the differential two-point correlator at NLO+NNLL with and without non-perturbative corrections. The red band indicates the scale uncertainty, while the green and blue bands indicate the uncertainties from the unknown higher-order constants of the non-perturbative jet functions and from the unknown value of $\overline{\Omega}_{1g}$, respectively, both estimated as described in \Sec{sec:NP_corrections}.}
	\label{fig:AENCfullRatioNPerrs_text}
\end{figure}

Finally, in \Fig{fig:AENCfullRatioNPerrs_text} we provide complete predictions for the ratios of the projected energy correlators in $\gamma + X$ events, including fixed-order matching, resummation, and non-perturbative corrections. These provide state-of-the-art predictions for energy correlators at hadron colliders. The non-perturbative corrections suppress the ratios at small $R_L$, where the leading power correction to the distribution, scaling as $\sim N\overline{\Omega}_{1\kappa}/(p_{\gamma,T}R_L^2)$, is largest, while leaving the large-angle region essentially unchanged; this effectively increases the slope of the ratio distributions. The effect grows with $N$ and is sizable, reaching several tens of percent at $R_L\sim 10^{-2}$ and remaining at the $\sim 10$--$20\%$ level in the scaling region around $R_L\sim 10^{-1}$. An accurate treatment of the non-perturbative corrections is thus essential for a reliable extraction of $\alpha_s$ from the slope of these ratios as neglecting them would lead to an erroneously larger extracted value of $\alpha_s$. Besides the scale uncertainty, our predictions include the two non-perturbative uncertainties discussed in \Sec{sec:NP_corrections}, from the modeled higher-order constants of the non-perturbative jet functions and from the variation of $\overline{\Omega}_{1g}$ around its Casimir-scaling value. We therefore conclude that for complete predictions of the projected energy correlators one needs matching, resummation, and non-perturbative corrections. The framework developed in this paper achieves all three, opening the door to precision phenomenology with energy correlators. 

\section{Conclusions and Outlook}\label{sec:conc}

In this paper we have proposed energy correlators measured on the radiation recoiling against a vector boson as a precision QCD observable at the LHC. Achieving precision predictions for this observable requires combining state-of-the-art ingredients from scattering amplitudes, numerical subtraction schemes, and factorization/resummation. To enable the calculation of the energy correlator for a broad set of final states at hadron colliders, we have developed a modular framework, combining analytic resummation with a numerical subtraction scheme based on \textsc{Stripper}~\cite{Czakon:2010td,Czakon:2014oma,Czakon:2019tmo}. This makes matched, resummed calculations of energy correlators possible in a wide class of processes for which the NLO/NNLO matrix elements are known.

We illustrated our framework by presenting results for the projected energy correlators at NLO+NNLL, incorporating the leading non-perturbative corrections. This is the first complete calculation of the energy correlator at a hadron collider that incorporates matching at this order. Since energy correlators are being used \cite{CMS:2024mlf} to measure the strong coupling constant at hadron colliders, it is essential to understand the importance of matching corrections. We performed detailed numerical studies of the impact of fixed-order matching and resummation on our results in an experimentally realistic phase space. We considered both the energy correlator distributions themselves and their ratios. For the distributions, the matching corrections grow to the $50$--$80\%$ level at wide angles, while for the ratios they are suppressed by roughly an order of magnitude; the scale uncertainties of the distributions at NLO+NNLL shrink to about $10\%$. The non-perturbative corrections are also significant, suppressing the ratios at small $R_L$ and thereby increase their slope, and neglecting them would bias an extraction of $\alpha_s$ toward larger values.

There are several clear ways in which our calculation can be extended. Most importantly, the primary goal of this paper, namely developing an interface between analytic resummation for energy correlators and numerical hard functions, was motivated by the desire to extend the calculations presented in this paper to include NNLO matching. The availability of NNLO $2\to 3$ scattering amplitudes (as discussed in detail in the introduction) makes this feasible, and we hope it will provide a beautiful illustration of their phenomenological importance. Numerically, this will be challenging, as it will require the evaluation of these amplitudes in singular regions of phase space. To assess this feasibility, we performed a detailed analysis of the structure of perturbative power corrections for the EEC observable. The size of perturbative power corrections, not captured by our leading power factorization theorem, dictates how far into the singular region the fixed-order calculation must be performed. Our study supports the feasibility of an NNLO calculation, since it will not need to be run too deep into singular regions.

Another clear way in which our analysis will need to be improved is to better understand the behavior of theoretical uncertainties for the ratios of projected energy correlators. In this paper, these were estimated using scale variations; however, this is clearly not ideal due to the strong correlations between the observables. A better approach is provided by Theory Nuisance Parameters (TNPs) \cite{Cridge:2025wwo,Tackmann:2024kci}, whose application we plan to investigate further in the future.

Although we have highlighted the clean nature of the $V+X$ process, our approach is also applicable to the study of energy correlators in identified jets. Since these have been directly used for extractions of the strong coupling constant, it will be important to study the effect of matching and perturbative power corrections for jets identified with the anti-$k_T$ algorithm \cite{Cacciari:2008gp}. This will be presented in future work.

We believe that the framework developed in this paper will enable the calculation of energy correlators for a wide variety of processes at the LHC and beyond.

\acknowledgments

We thank Jennifer Roloff, Matt Leblanc, Simon Rothman, Gavin Salam for useful discussions. This work was performed in part using the Cambridge Service for Data Driven Discovery (CSD3), part of which is operated by the University of Cambridge Research Computing on behalf of the STFC DiRAC HPC Facility (www.dirac.ac.uk). The DiRAC component of CSD3 was supported by STFC grants ST/P002307/1, ST/R002452/1 and ST/R00689X/1. T.G. is supported by STFC consolidated HEP theory grants ST/T000694/1 and ST/X000664/1. 
K.L is supported by the U.S. Department of Energy under contract DE-AC02-06CH11357. 
I.M. is supported by the DOE Early Career Award
DE-SC0025581, and the Sloan Foundation. 
R.P is funded by the European Union (ERC, STAPLE, 101222253). Views and opinions expressed are however those of the author(s) only and do not necessarily reflect those of the European Union or the European Research Council. Neither the European Union nor the granting authority can be held responsible for them.
X.Z. is supported by the MIT Pappalardo Fellowship.
This work was also supported in part by the Aspen Center for Physics under NSF grant PHY-2210452 and by the Kavli Institute for Theoretical Physics under NSF grant PHY-2309135. We also thank the Erwin Schrödinger International Institute for Mathematics and Physics for support.

\appendix

\section{Perturbative Ingredients}\label{sec:pert_ingredients}
In this appendix, we provide the expressions for ENC jet functions in Eq.~\eqref{eq:jet_ansatz}. 
First of all, the anomalous dimensions in collinear resummation are defined via the timelike splitting functions,
\begin{align}
	\gamma_{ij}^{(L)}(N)&\equiv -\int_0^1 \df x \,x^N \,P_{ij}^{(L)}(x)\,,\nn\\
	\dot \gamma_{ij}^{(L)}(N)&\equiv -\int_0^1 \df x\, \ln x \, x^N P_{ij}^{(L)}(x)\,,\nn\\
	\ddot \gamma_{ij}^{(L)}(N)&\equiv -\int_0^1 \df x\, \ln^2 x \, x^N P_{ij}^{(L)}(x)\,.
\end{align}
where we expand the splitting functions in $\alpha_s/(4\pi)$,
\begin{equation}
	P_{ij}(x)=\sum_{L=0}^{\infty}\left(\frac{\alpha_s}{4\pi}\right)^{L+1} P_{ij}^{(L)}(x) \,,
\end{equation}
The analytic expressions of $\gamma_{ij}^{(L)}$ up to three loops and up to $N=6$ (for 6-point correlators) can be found in Ref.~\cite{Lee:2026zyl}. 

As in Eq.~\eqref{eq:jet_ansatz}, the resummed jet functions can be expressed in terms of $\alpha_s(\mu)$ series, 
\begin{align}
\vec{J}^{[N]}\left(\ln\frac{R_LQ^2}{\mu^2},\mu \right)= \sum_{n=0}^\infty \sum_{m=0}^n \left(\frac{\alpha_s(\mu)}{4\pi}\right)^n \frac{\vec{j}_{n,m}^{[N]}}{m!} \ln^m \frac{R_LQ^2}{\mu^2}\,.
\end{align}
Here we present the coefficients $\vec{j}_{n,m}^{[N]}$ for the first three orders. For the quark channel, we find,
\begin{align}
    j^{[N]}_{q,0,0}&=j_{q,0}^{[N]}\,,\nn\\
    j^{[N]}_{q,1,1}&=\gamma _{gq}^{[0]} \, j_{g,0}^{[N]}+\gamma _{qq}^{[0]}\, j_{q,0}^{[N]}\,\nn\\
    j^{[N]}_{q,1,0}&=j_{q,1}^{[N]}\,,\nn\\
    j^{[N]}_{q,2,2}&=\gamma_{gg}^{[0]}\, \gamma_{gq}^{[0]} \,j_{g,0}^{[N]}-\beta_0\, \gamma_{gq}^{[0]} \,j_{g,0}^{[N]}+\gamma_{gq}^{[0]} \,\gamma_{qq}^{[0]} \,j_{g,0}^{[N]}+\gamma_{gq}^{[0]} \,\gamma_{qg}^{[0]} \,j_{q,0}^{[N]}-\beta _0\, \gamma_{qq}^{[0]}\, j_{q,0}^{[N]}+\left(\gamma_{qq}^{[0]}\right)^2\, j_{q,0}^{[N]}\,,\nn\\
    j^{[N]}_{q,2,1}&=\gamma _{gg}^{[0]}\, \dot{\gamma }_{gq}^{[0]}\, j_{g,0}^{[N]}+\gamma
   _{gq}^{[0]}\, \dot{\gamma }_{qq}^{[0]}\, j_{g,0}^{[N]}+\gamma
   _{qg}^{[0]}\, \dot{\gamma }_{gq}^{[0]}\, j_{q,0}^{[N]}+\gamma
   _{qq}^{[0]}\, \dot{\gamma }_{qq}^{[0]}\, j_{q,0}^{[N]}+\gamma
   _{gq}^{[0]}\, j_{g,1}^{[N]}+\gamma _{qq}^{[0]}\,
   j_{q,1}^{[N]}\nn\\
   &+\gamma _{gq}^{[1]} \,j_{g,0}^{[N]}+\gamma
   _{qq}^{[1]}\, j_{q,0}^{[N]}-\beta _0\, j_{q,1}^{[N]}\,,\nn\\
   j^{[N]}_{q,2,0}&=j_{q,2}^{[N]}\,,\nn\\
   j^{[N]}_{q,3,3}&=-3 \beta _0 \,\gamma _{gg}^{[0]} \,\gamma_{gq}^{[0]}\, j_{g,0}^{[N]}+\left(\gamma
   _{gg}^{[0]}\right)^2\, \gamma _{gq}^{[0]} \,j_{g,0}^{[N]}+\gamma
   _{gg}^{[0]}\, \gamma _{gq}^{[0]} \,\gamma _{qq}^{[0]}\,
   j_{g,0}^{[N]}+2 \beta _0^2\, \gamma _{gq}^{[0]}\, j_{g,0}^{[N]}\nn\\
   &+\left(\gamma
   _{gq}^{[0]}\right)^2\, \gamma _{qg}^{[0]} \,j_{g,0}^{[N]}-3 \beta _0\, \gamma
   _{gq}^{[0]}\, \gamma _{qq}^{[0]} \,j_{g,0}^{[N]}+\gamma
   _{gq}^{[0]}\, \left(\gamma _{qq}^{[0]}\right)^2 \,j_{g,0}^{[N]}+\gamma
   _{gg}^{[0]}\, \gamma _{gq}^{[0]} \,\gamma _{qg}^{[0]}\,
   j_{q,0}^{[N]}-3 \beta _0\, \gamma _{gq}^{[0]}\, \gamma _{qg}^{[0]}\,
   j_{q,0}^{[N]}\nn\\
   &+2 \gamma _{gq}^{[0]}\, \gamma _{qg}^{[0]}\, \gamma
   _{qq}^{[0]}\, j_{q,0}^{[N]}+2 \beta _0^2 \,\gamma _{qq}^{[0]}\,
   j_{q,0}^{[N]}-3 \beta _0\, \left(\gamma _{qq}^{[0]}\right)^2\, j_{q,0}^{[N]}+\left(\gamma
   _{qq}^{[0]}\right)^3\, j_{q,0}^{[N]}\,,\nn\\
   j^{[N]}_{q,3,2}&=-3 \beta _0\, \gamma _{gg}^{[0]}\, \dot{\gamma }_{gq}^{[0]}\,
   j_{g,0}^{[N]}+\gamma _{gg}^{[0]} \,\gamma _{gq}^{[0]}\,
   \dot{\gamma }_{gg}^{[0]} \,j_{g,0}^{[N]}+\left(\gamma _{gg}^{[0]}\right)^2\,
   \dot{\gamma }_{gq}^{[0]} \,j_{g,0}^{[N]}+\gamma _{gg}^{[0]}\,
   \gamma _{qq}^{[0]}\, \dot{\gamma }_{gq}^{[0]}\,
   j_{g,0}^{[N]}\nn\\
   &+\gamma _{gg}^{[0]} \,\gamma _{gq}^{[0]}\,
   \dot{\gamma }_{qq}^{[0]} \,j_{g,0}^{[N]}+\gamma _{gq}^{[0]}\,
   \gamma _{qg}^{[0]} \,\dot{\gamma }_{gq}^{[0]}\,
   j_{g,0}^{[N]}+\left(\gamma _{gq}^{[0]}\right)^2 \,\dot{\gamma }_{qg}^{[0]}\,
   j_{g,0}^{[N]}-3 \beta _0\, \gamma _{gq}^{[0]} \,\dot{\gamma
   }_{qq}^{[0]}\, j_{g,0}^{[N]}\nn\\
   &+2 \gamma _{gq}^{[0]}\, \gamma
   _{qq}^{[0]} \,\dot{\gamma }_{qq}^{[0]}\, j_{g,0}^{[N]}+\gamma
   _{gg}^{[0]} \,\gamma _{qg}^{[0]} \,\dot{\gamma
   }_{gq}^{[0]}\, j_{q,0}^{[N]}+\gamma _{gq}^{[0]} \,\gamma
   _{qg}^{[0]} \,\dot{\gamma }_{gg}^{[0]}\, j_{q,0}^{[N]}-3 \beta _0\,
   \gamma _{qg}^{[0]}\, \dot{\gamma }_{gq}^{[0]}\, j_{q,0}^{[N]}\nn\\
   &+2
   \gamma _{qg}^{[0]} \,\gamma _{qq}^{[0]}\, \dot{\gamma
   }_{gq}^{[0]}\, j_{q,0}^{[N]}+\gamma _{gq}^{[0]}\, \gamma
   _{qq}^{[0]} \,\dot{\gamma }_{qg}^{[0]}\, j_{q,0}^{[N]}+\gamma
   _{gq}^{[0]} \,\gamma _{qg}^{[0]} \,\dot{\gamma
   }_{qq}^{[0]}\, j_{q,0}^{[N]}-3 \beta _0 \,\gamma _{qq}^{[0]}\,
   \dot{\gamma }_{qq}^{[0]} \,j_{q,0}^{[N]}+2 \left(\gamma _{qq}^{[0]}\right)^2\,
   \dot{\gamma }_{qq}^{[0]}\, j_{q,0}^{[N]}\nn\\
   &+\gamma _{gg}^{[0]}\,
   \gamma _{gq}^{[1]} \,j_{g,0}^{[N]}+\gamma _{gq}^{[0]} \,\gamma
   _{gg}^{[1]}\, j_{g,0}^{[N]}+\gamma _{qq}^{[0]} \,\gamma
   _{gq}^{[1]}\, j_{g,0}^{[N]}+\gamma _{gq}^{[0]}\, \gamma
   _{qq}^{[1]}\, j_{g,0}^{[N]}+\gamma _{qg}^{[0]}\, \gamma
   _{gq}^{[1]}\, j_{q,0}^{[N]}+\gamma _{gq}^{[0]}\, \gamma
   _{qg}^{[1]}\, j_{q,0}^{[N]}\nn\\
   &+2 \gamma _{qq}^{[0]} \,\gamma _{qq}^{[1]}\,
   j_{q,0}^{[N]}-\beta _1\, \gamma _{gq}^{[0]}\, j_{g,0}^{[N]}-\beta _1 \,\gamma
   _{qq}^{[0]} \,j_{q,0}^{[N]}+\gamma _{gg}^{[0]}\, \gamma
   _{gq}^{[0]} \,j_{g,1}^{[N]}-3 \beta _0 \,\gamma _{gq}^{[0]}\,
   j_{g,1}^{[N]}+\gamma _{gq}^{[0]} \,\gamma _{qq}^{[0]}\,
   j_{g,1}^{[N]}\nn\\
   &+\gamma _{gq}^{[0]}\, \gamma _{qg}^{[0]}\,
   j_{q,1}^{[N]}-3 \beta _0 \,\gamma _{qq}^{[0]} \,j_{q,1}^{[N]}+\left(\gamma
   _{qq}^{[0]}\right)^2 \, j_{q,1}^{[N]}-2 \beta _0 \, \gamma _{gq}^{[1]}\,
   j_{g,0}^{[N]}-2 \beta _0\, \gamma _{qq}^{[1]}\, j_{q,0}^{[N]}+2 \beta _0^2\,
   j_{q,1}^{[N]}\,,\nn\\
   j^{[N]}_{q,3,1}&=+\dot{\gamma }_{gg}^{[0]}\, \dot{\gamma }_{gq}^{[0]}\,
   j_{g,0}^{[N]} \,\gamma _{gg}^{[0]}+\dot{\gamma }_{gq}^{[0]}\,
   \dot{\gamma }_{qq}^{[0]} \,j_{g,0}^{[N]}\, \gamma
   _{gg}^{[0]}-\frac{1}{2} \beta _0 \,\ddot{\gamma }_{gq}^{[0]}\,
   j_{g,0}^{[N]} \,\gamma _{gg}^{[0]}+\frac{1}{2} \gamma _{gq}^{[0]}\,
   \ddot{\gamma }_{qq}^{[0]} \,j_{g,0}^{[N]} \,\gamma
   _{gg}^{[0]}\nn\\
   &+\dot{\gamma }_{gq}^{[1]} \, j_{g,0}^{[N]} \, \gamma
   _{gg}^{[0]}+\dot{\gamma }_{gq}^{[0]}\, j_{g,1}^{[N]}\, \gamma
   _{gg}^{[0]}+\frac{1}{2} \gamma _{qg}^{[0]}\, \ddot{\gamma
   }_{gq}^{[0]}\, j_{q,0}^{[N]} \, \gamma _{gg}^{[0]}+\frac{1}{2}
   \ddot{\gamma }_{gq}^{[0]} \,j_{g,0}^{[N]}\, \left(\gamma _{gg}^{
   [0]}\right)^2+\dot{\gamma }_{gq}^{[0]}\, j_{g,0}^{[N]}\, \gamma
   _{gg}^{[1]}\nn\\
   &+\gamma_{gq}^{[0]}\, \dot{\gamma
   }_{gq}^{[0]}\, \dot{\gamma }_{qg}^{[0]}\, j_{g,0}^{[N]}+\gamma
   _{gq}^{[1]}\, \dot{\gamma }_{qq}^{[0]}\,
   j_{g,0}^{[N]}+\frac{1}{2} \gamma _{gq}^{[0]}\, \gamma _{qg}^{[0]}\,
   \ddot{\gamma }_{gq}^{[0]} \,j_{g,0}^{[N]}-\frac{1}{2} \beta _0 \,\gamma
   _{gq}^{[0]}\, \ddot{\gamma }_{qq}^{[0]}\,
   j_{g,0}^{[N]}+\frac{1}{2} \gamma _{gq}^{[0]} \,\gamma _{qq}^{[0]}\,
   \ddot{\gamma }_{qq}^{[0]} \,j_{g,0}^{[N]}\nn\\
   &+\gamma
   _{gq}^{[0]}\, \left(\dot{\gamma }_{qq}^{[0]}\right)^2\, j_{g,0}^{[N]}+\gamma
   _{gq}^{[0]}\, \dot{\gamma }_{qq}^{[1]}\, j_{g,0}^{[N]}+\gamma
   _{gq}^{[2]}\, j_{g,0}^{[N]}-\beta _0 \,\dot{\gamma }_{gq}^{[0]}\,
   j_{g,1}^{[N]}+\gamma _{gq}^{[0]} \,\dot{\gamma }_{qq}^{[0]}\,
   j_{g,1}^{[N]}+\gamma _{gq}^{[1]} \,j_{g,1}^{[N]}\nn\\
   &+\gamma
   _{gq}^{[0]} \, j_{g,2}^{[N]}+\gamma _{qg}^{[0]}\, \dot{\gamma
   }_{gg}^{[0]}\, \dot{\gamma }_{gq}^{[0]}\, j_{q,0}^{[N]}+\gamma
   _{qg}^{[1]}\, \dot{\gamma }_{gq}^{[0]}\, j_{q,0}^{[N]}+\gamma
   _{qq}^{[0]} \,\dot{\gamma }_{gq}^{[0]}\, \dot{\gamma
   }_{qg}^{[0]}\, j_{q,0}^{[N]}+\gamma _{qg}^{[0]}\, \dot{\gamma
   }_{gq}^{[0]} \,\dot{\gamma }_{qq}^{[0]}\, j_{q,0}^{[N]}\nn\\
   &+\gamma
   _{qq}^{[1]} \,\dot{\gamma }_{qq}^{[0]}\,
   j_{q,0}^{[N]}-\frac{1}{2} \beta _0 \,\gamma _{qg}^{[0]}\, \ddot{\gamma
   }_{gq}^{[0]}\, j_{q,0}^{[N]}+\frac{1}{2} \gamma _{qg}^{[0]} \,\gamma
   _{qq}^{[0]} \,\ddot{\gamma }_{gq}^{[0]}\,
   j_{q,0}^{[N]}+\frac{1}{2} \gamma _{gq}^{[0]} \,\gamma _{qg}^{[0]}\,
   \ddot{\gamma }_{qq}^{[0]} \,j_{q,0}^{[N]}\nn\\
   &-\frac{1}{2} \beta _0 \,\gamma
   _{qq}^{[0]} \,\ddot{\gamma }_{qq}^{[0]}\,
   j_{q,0}^{[N]}+\frac{1}{2} \left(\gamma _{qq}^{[0]}\right)^2\, \ddot{\gamma
   }_{qq}^{[0]}\, j_{q,0}^{[N]}+\gamma _{qq}^{[0]} \,\left(\dot{\gamma
   }_{qq}^{[0]}\right)^2\, j_{q,0}^{[N]}+\gamma _{qg}^{[0]} \,\dot{\gamma
   }_{gq}^{[1]} \,j_{q,0}^{[N]}+\gamma _{qq}^{[0]} \,\dot{\gamma
   }_{qq}^{[1]}\, j_{q,0}^{[N]}\nn\\
   &+\gamma _{qq}^{[2]}\,
   j_{q,0}^{[N]}+\gamma _{qg}^{[0]} \,\dot{\gamma }_{gq}^{[0]}\,
   j_{q,1}^{[N]}+\gamma _{qq}^{[0]} \,\dot{\gamma }_{qq}^{[0]}\,
   j_{q,1}^{[N]}-\beta _0\, \dot{\gamma }_{qq}^{[0]}\, j_{q,1}^{[N]}+\gamma
   _{qq}^{[1]}\, j_{q,1}^{[N]}-\beta _1 \, j_{q,1}^{[N]}+\gamma
   _{qq}^{[0]} \,j_{q,2}^{[N]}-2 \beta _0 \,j_{q,2}^{[N]}\,.
\end{align}
where we have omitted the $N$ dependence in the anomalous dimensions.
For the gluon channel, we have the same expressions with the following replacements:
\begin{align}
    j_{q,i}^{[N]} &\leftrightarrow  j_{g,i}^{[N]},\quad i=0,1,2\nn\\
    \gamma^{[i]}_{qq} \leftrightarrow \gamma^{[i]}_{gg}, \quad \gamma^{[i]}_{qg} \leftrightarrow \gamma^{[i]}_{gq},\quad \dot\gamma^{[i]}_{qq} &\leftrightarrow \dot\gamma^{[i]}_{gg}, \quad \dot\gamma^{[i]}_{qg} \leftrightarrow \dot\gamma^{[i]}_{gq},\quad \ddot\gamma^{[i]}_{qq} \leftrightarrow \ddot\gamma^{[i]}_{gg}, \quad \ddot\gamma^{[i]}_{qg} \leftrightarrow \ddot\gamma^{[i]}_{gq}\,.
\end{align}
$j_{q,i}^{[N]}$ and $j_{g,i}^{[N]}$ with $i=0,1,2$ are the perturbative jet constants. At tree-level, we have $j_{q,0}^{[N]}=1$ and $j_{g,0}^{[N]}=1$. At one-loop, the result with full $N$ dependence is~\cite{Chen:2020vvp}
\begin{align}
    j_{q,1}^{[N]}&=C_F \left(\frac{(N-1) (N (13 N+37)+12)}{N (N+1)^2}-4 (\psi ^{(0)}(N)+\gamma )^2-\frac{12 (\psi
   ^{(0)}(N)+\gamma )}{N+1}+12 \psi ^{(1)}(N)-2 \pi ^2\right)\nn\\
   j_{g,1}^{[N]}&=C_A \Bigg\{\frac{1}{9 N (N+1)^2 (N+2)^2 (N+3)^2}\Bigg[2 N (67 N^6+804 N^5+3634 N^4+7380 N^3\nn\\
   & +4723 N^2-5520 N-9 \pi ^2 (N+1)^2 (N+2)^2
   (N+3)^2-8712)-4752\Bigg]-4 (\psi ^{(0)}(N)+\gamma )^2\nn\\
   &-\frac{8 (N (2
   N+9)+11) (\psi ^{(0)}(N)+\gamma )}{(N+1) (N+2) (N+3)}+12 \psi ^{(1)}(N)\Bigg\}\nn\\
   &+\frac{n_f}{9 (N+1)^2 (N+2)^2 (N+3)^2} \left(-23
   N^6-276 N^5-1190 N^4-2376 N^3-1703 N^2\right.\nn\\
   &\left.+1644 N+\frac{864}{N}+36 (N+1) (N+2) (N+3) (N (N+3)+4) (\psi
   ^{(0)}(N)+\gamma )+3060\right)\,.
\end{align}
At two-loop, we don't have the closed form in $N$; instead, we compute them semi-analytically up to $N=6$~\cite{Lee:2026zyl}
\begin{align}
    j_{q,2}^{[3]}&=C_F T_F n_f (98.416094\pm 0.001584) + C_F^2 (171.153969 \pm 0.018553)\notag\\
	&+ C_F C_A (-210.211343 \pm 0.008717)\,,\notag\\
    j_{q,2}^{[4]}&=C_F T_F n_f (124.967567\pm 0.002166) + C_F^2 (254.438849 \pm 0.018860)\notag\\
	&+ C_F C_A (-269.979451 \pm 0.009042)\,,\notag\\
    j_{q,2}^{[5]}&=C_F T_F n_f (147.891331 \pm 0.002602) + C_F^2 (339.713439 \pm 0.021279)\notag\\
	&+ C_F C_A (-323.170146 \pm 0.009702)\,,\notag\\
    j_{q,2}^{[6]}&=C_F T_F n_f (168.459145 \pm 0.003037) + C_F^2 (426.020756 \pm 0.024441)\notag\\
	&+ C_F C_A (-371.824995 \pm 0.011163)\,,\\
    j_{g,2}^{[3]}&=C_F T_F n_f ( -16.427368 \pm 0.001614 )+ C_A T_F n_f ( 140.389919 \pm 0.003094 )\notag\\
    &+C_A^2 ( -47.854829 \pm 0.021790 )+n_f^2 T_F^2\left(\frac{4688}{375}-\frac{8 \pi ^2}{15}\right)  \,,\notag\\
    j_{g,2}^{[4]}&=C_F T_F n_f ( -21.592489 \pm 0.001800 )+ C_A T_F n_f ( 182.634493 \pm 0.003985 )\notag\\
	&+C_A^2 ( -27.054523 \pm 0.022288 )+ n_f^2 T_F^2 \left( \frac{52689656}{3472875}-\frac{608 \pi ^2}{945} \right)\,,\notag\\
    j_{g,2}^{[5]}&=C_F T_F n_f ( -25.583483 \pm 0.001906 )+ C_A T_F n_f ( 219.793744 \pm 0.007128 )\notag\\
	&+C_A^2 ( 2.207608 \pm 0.036991 )+ n_f^2 T_F^2 \left( \frac{11872972}{694575}-\frac{136 \pi ^2}{189} \right)\,,\notag\\
    j_{g,2}^{[6]}&=C_F T_F n_f ( -28.753959 \pm 0.002008 )+ C_A T_F n_f ( 253.496395 \pm 0.005500 )\notag\\
	&+C_A^2 ( 37.086330 \pm 0.026721 )+ n_f^2 T_F^2 \left( \frac{348270772}{18753525}-\frac{440 \pi ^2}{567} \right)\,.
\end{align}

\section{Additional Plots}\label{sec:ad_plots}

In this appendix, we provide additional plots supplementing those in the main text: the comparisons shown there are extended to the complete set of projected correlators up to six points and, where noted, to additional perturbative orders and to cumulative distributions. The appendix is organized into subsections mirroring the structure of \Sec{sec:pheno}.

\subsection[Singular and Non-Singular Distributions]{Singular and Non-Singular Distributions}\label{app:singular_plots}

\begin{figure}[h]
	\centering
	\includegraphics[width=0.31\textwidth]{AEEC/PowerCorrectionCheckLO2.pdf}
	\includegraphics[width=0.31\textwidth]{AEEC/PowerCorrectionCheckNLO2.pdf}\\
	\includegraphics[width=0.31\textwidth]{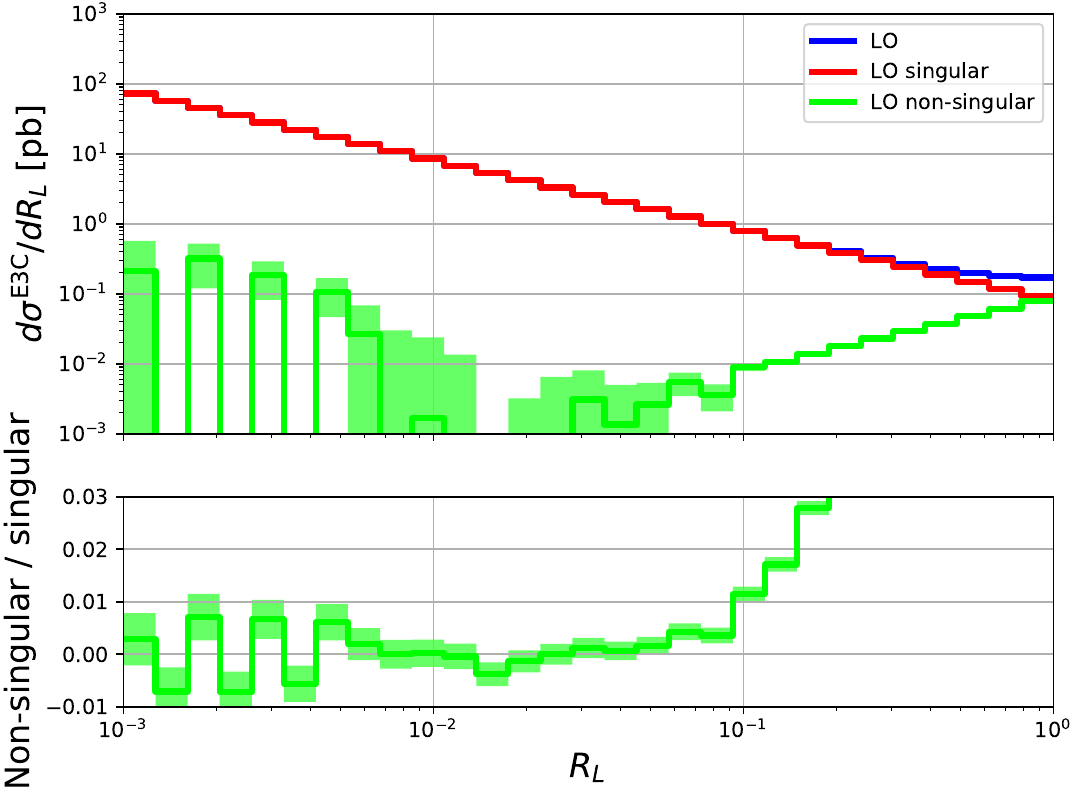}
	\includegraphics[width=0.31\textwidth]{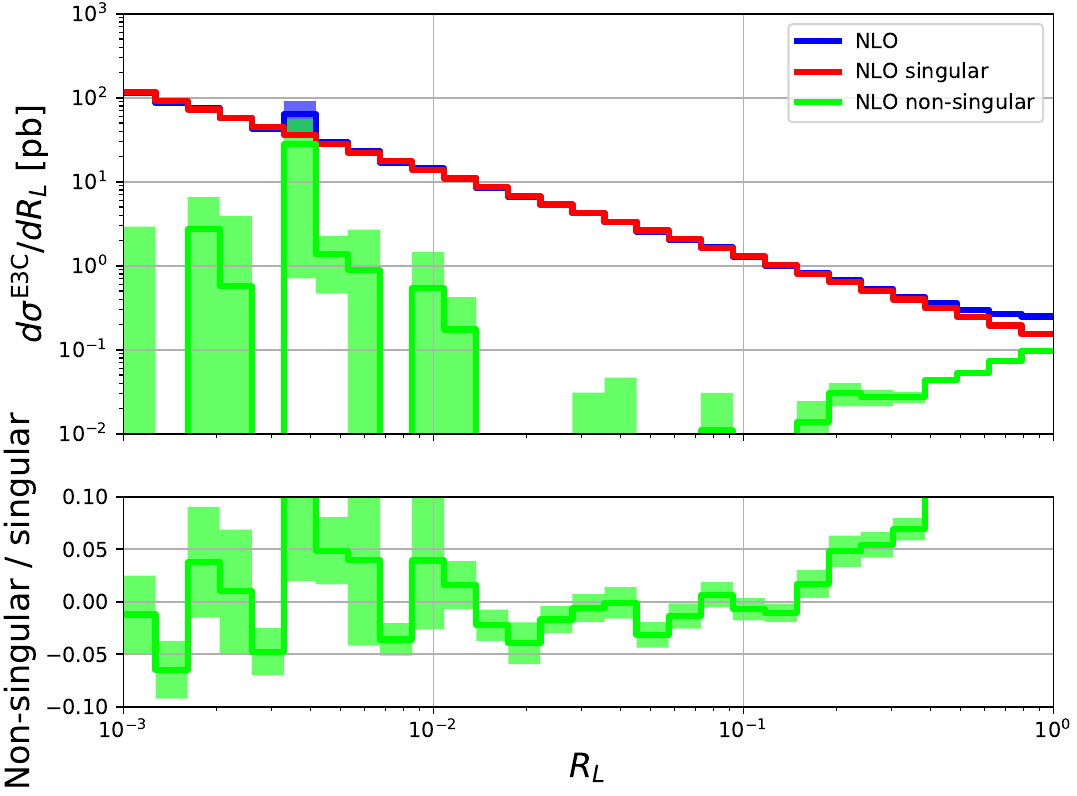}\\
	\includegraphics[width=0.31\textwidth]{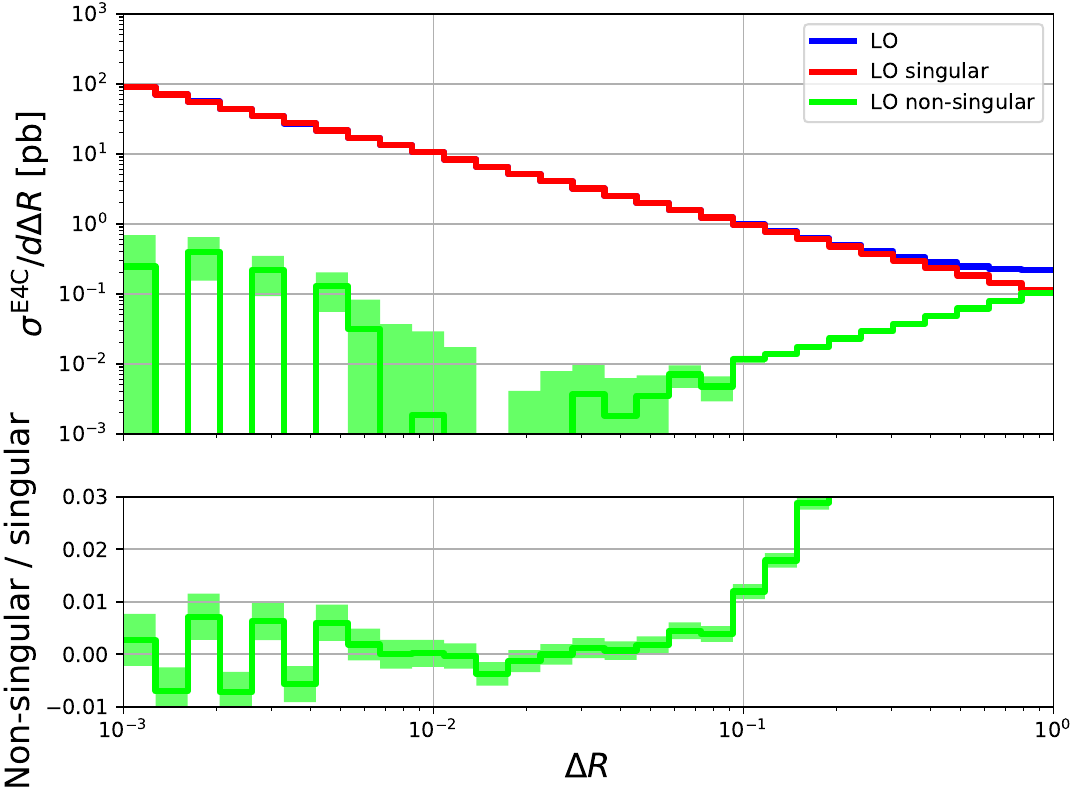}
	\includegraphics[width=0.31\textwidth]{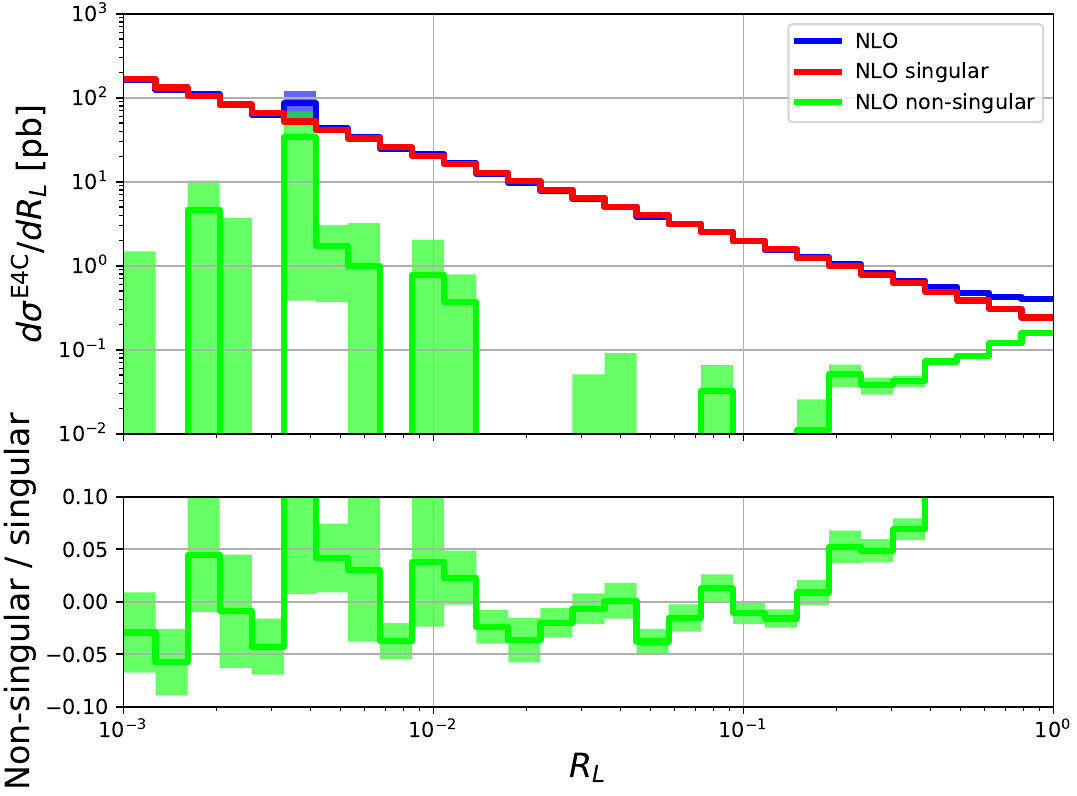}\\
	\includegraphics[width=0.31\textwidth]{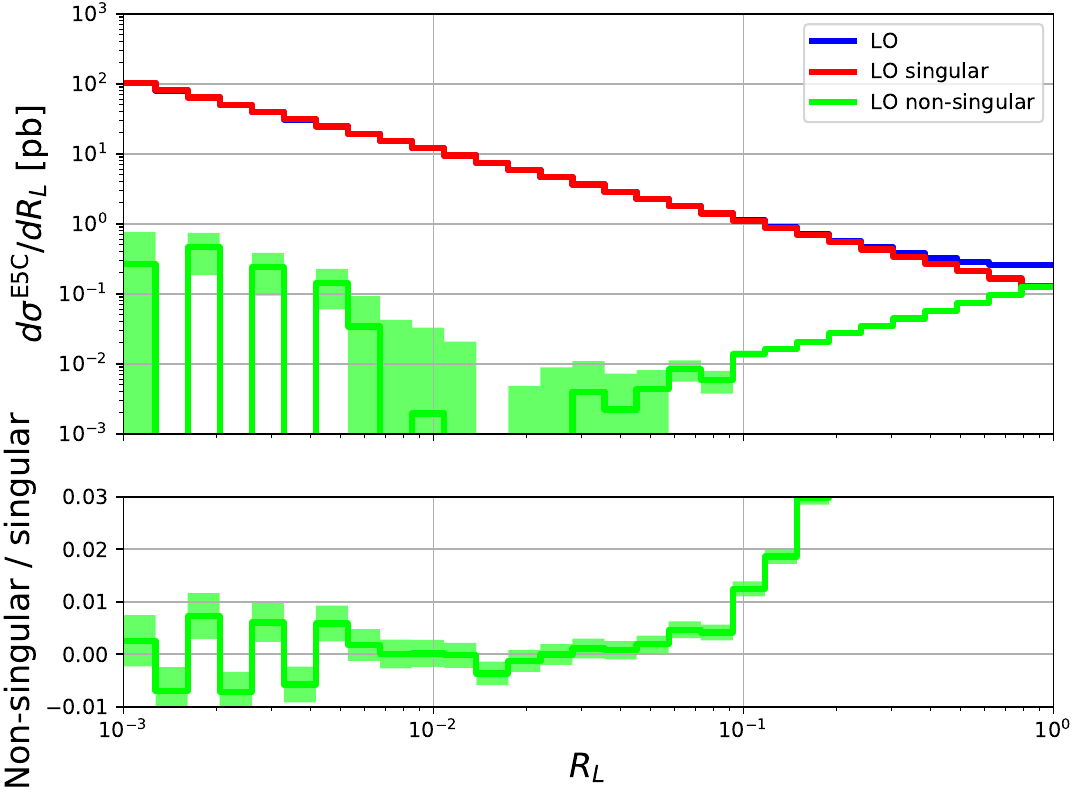}
	\includegraphics[width=0.31\textwidth]{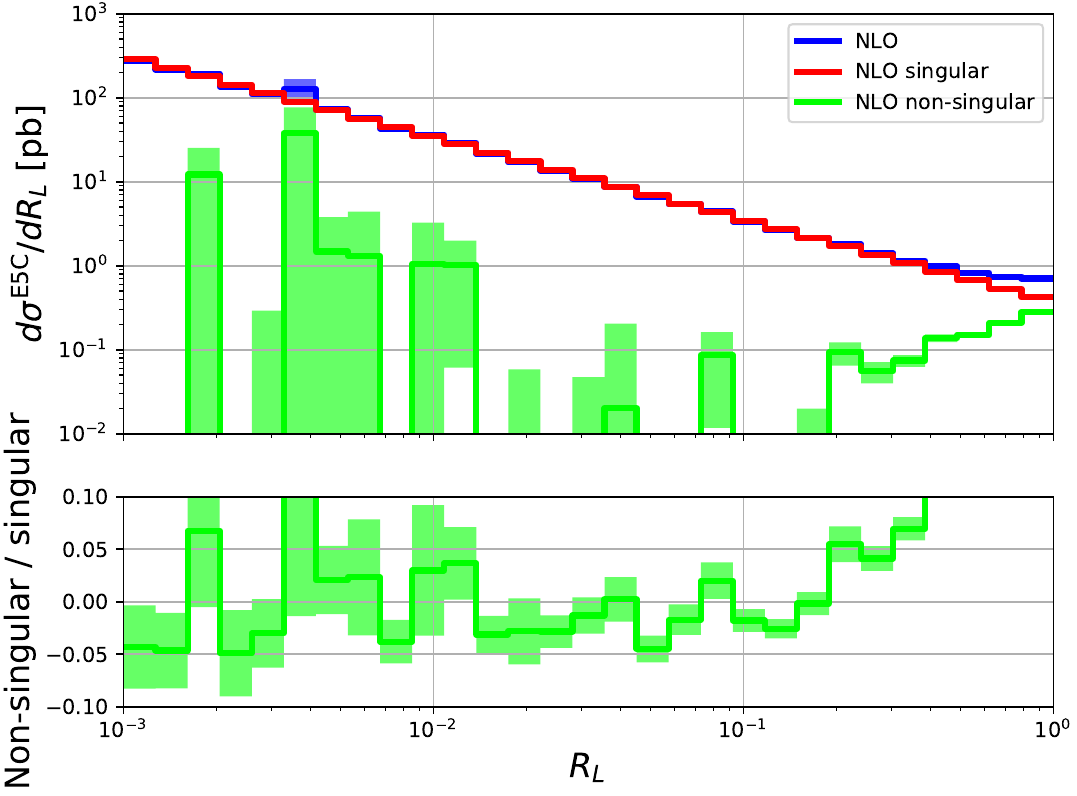}\\
	\includegraphics[width=0.31\textwidth]{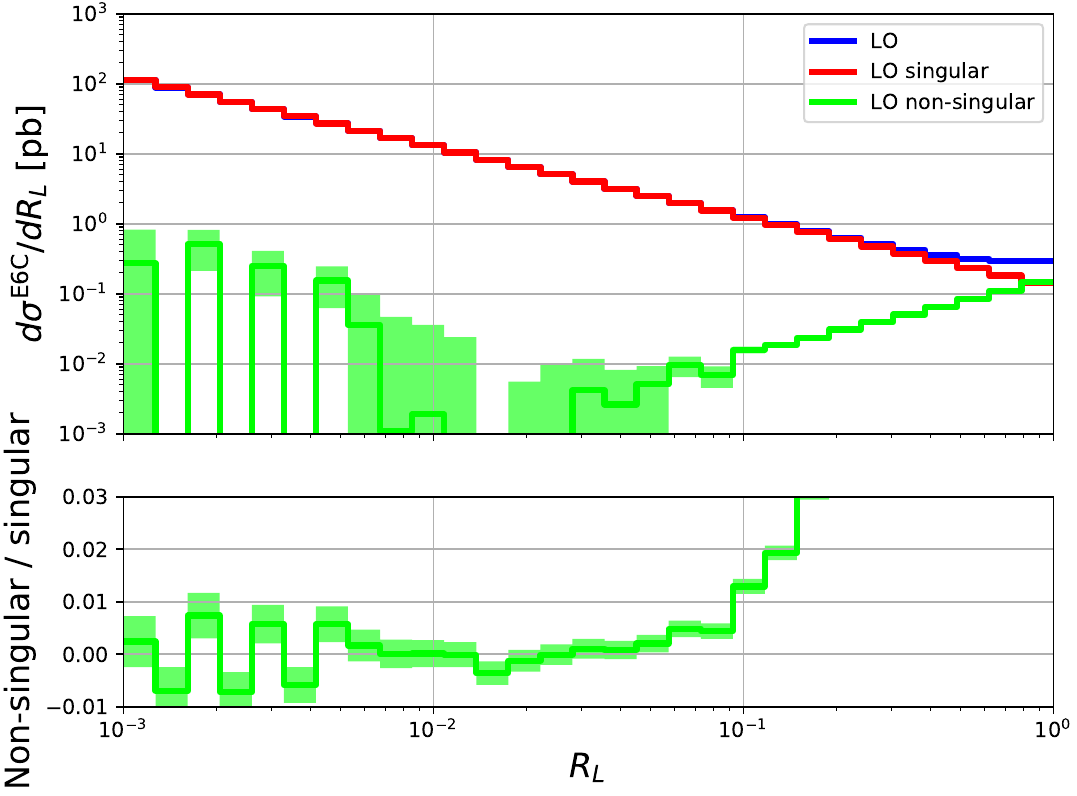}
	\includegraphics[width=0.31\textwidth]{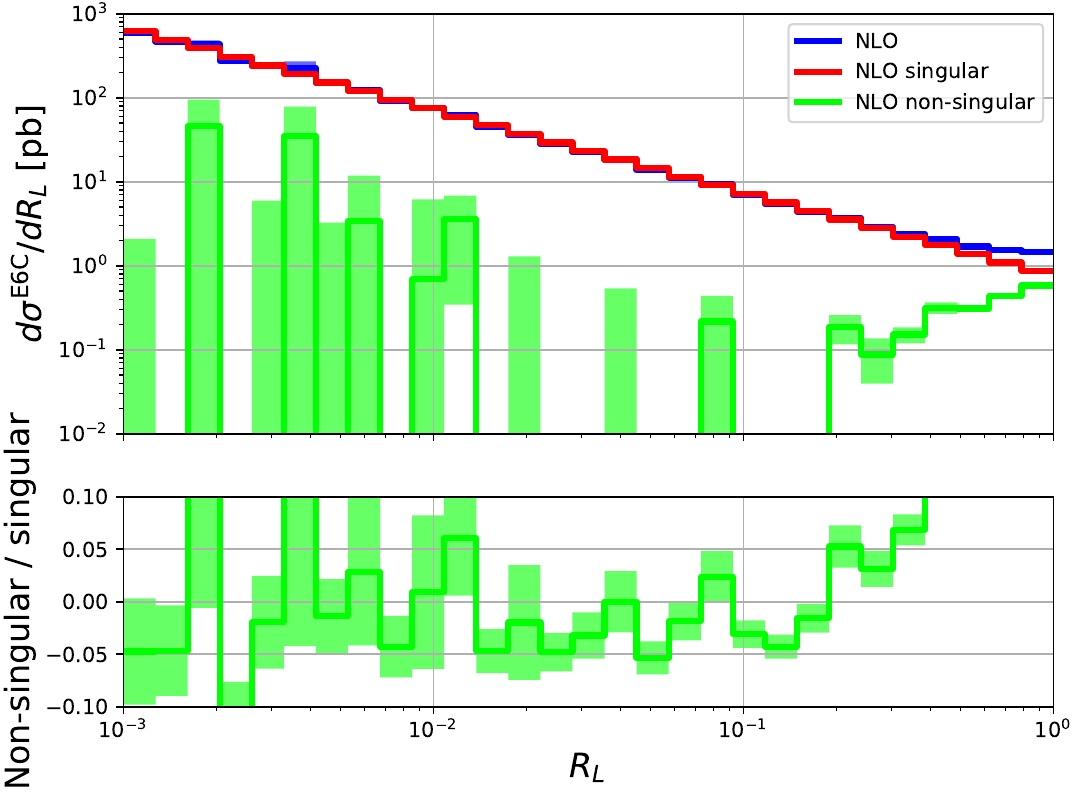}
	\caption{The comparisons of the full, singular and non-singular projected $N$-point correlators, for $N=2$--$6$ (top to bottom), at LO (left) and NLO (right). }
	\label{fig:PCcheckN_app}
\end{figure}

In \Fig{fig:PCcheckN_app}, we extend the comparison of the full, singular, and non-singular distributions in \Fig{fig:AEEC_NPC_text} to all projected correlators up to six points.

\clearpage

\subsection[Matched Distributions]{Matched Distributions}\label{app:matched_plots}

\begin{figure}[h]
	\centering
	\includegraphics[width=0.31\textwidth]{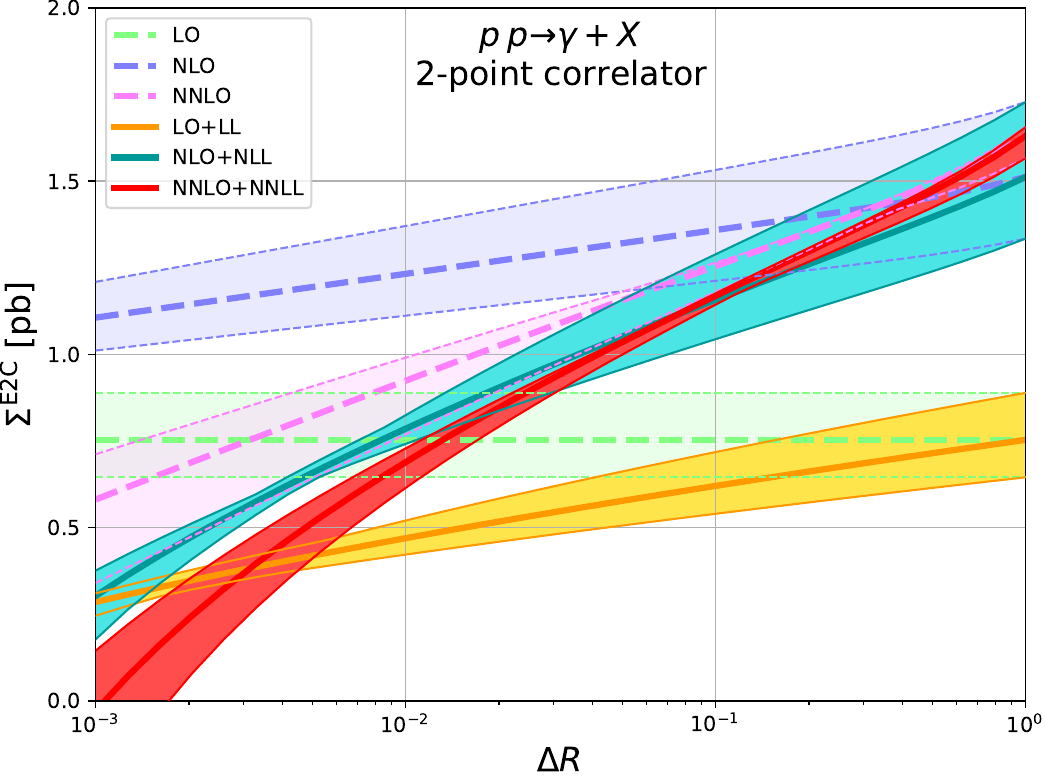}
	\includegraphics[width=0.31\textwidth]{AEEC/AdEEC2.pdf}\\
	\includegraphics[width=0.31\textwidth]{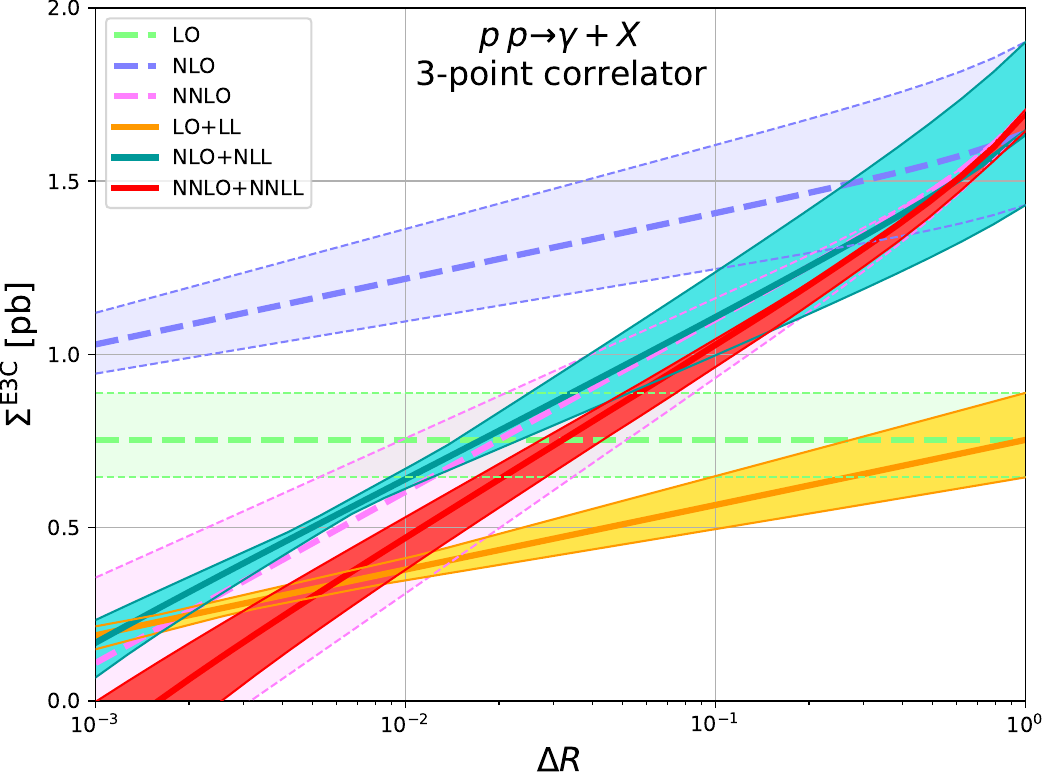}
	\includegraphics[width=0.31\textwidth]{AEEC/AdEEC3.pdf}\\
	\includegraphics[width=0.31\textwidth]{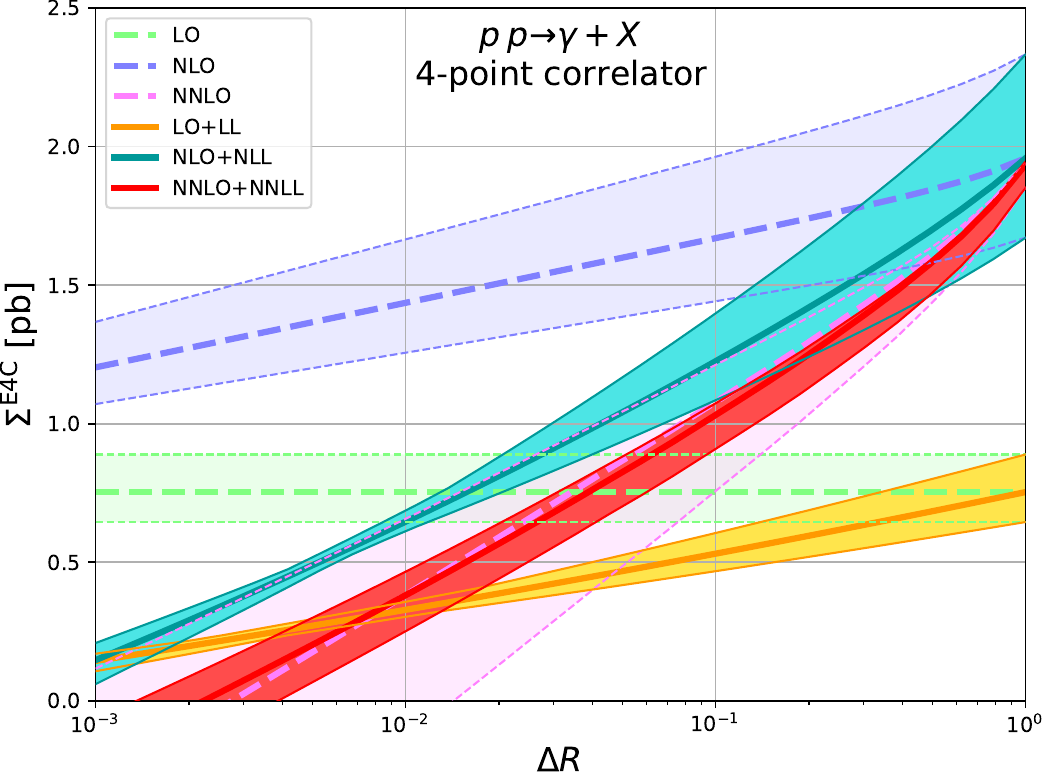}
	\includegraphics[width=0.31\textwidth]{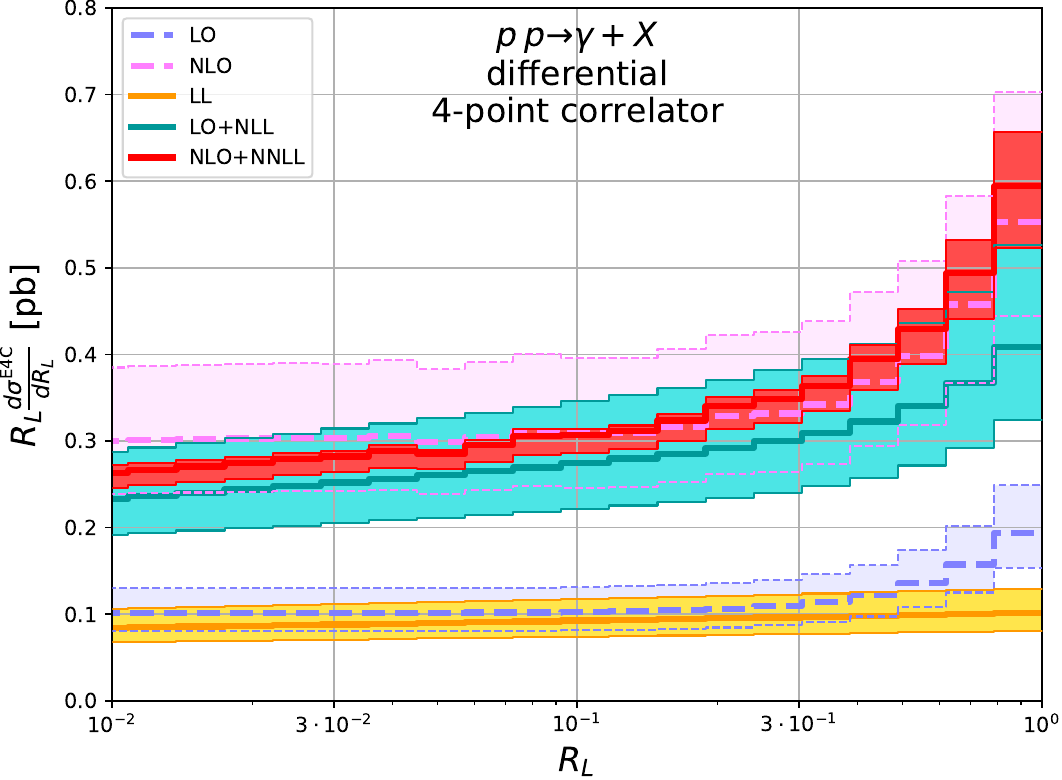}\\
	\includegraphics[width=0.31\textwidth]{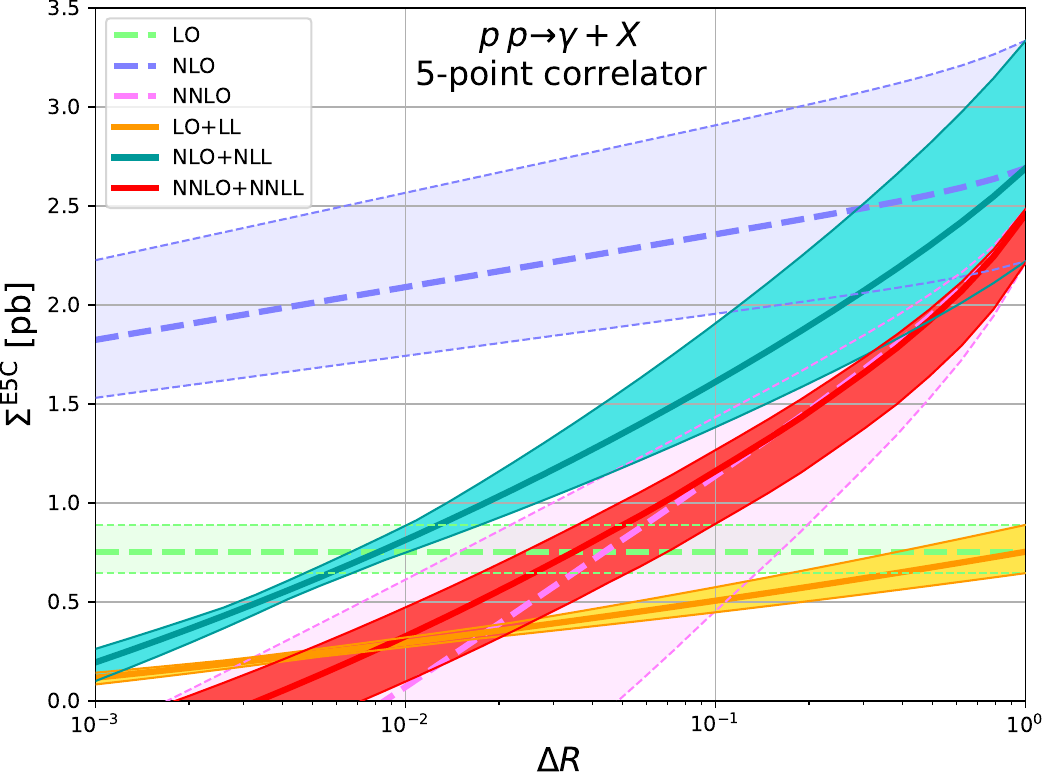}
	\includegraphics[width=0.31\textwidth]{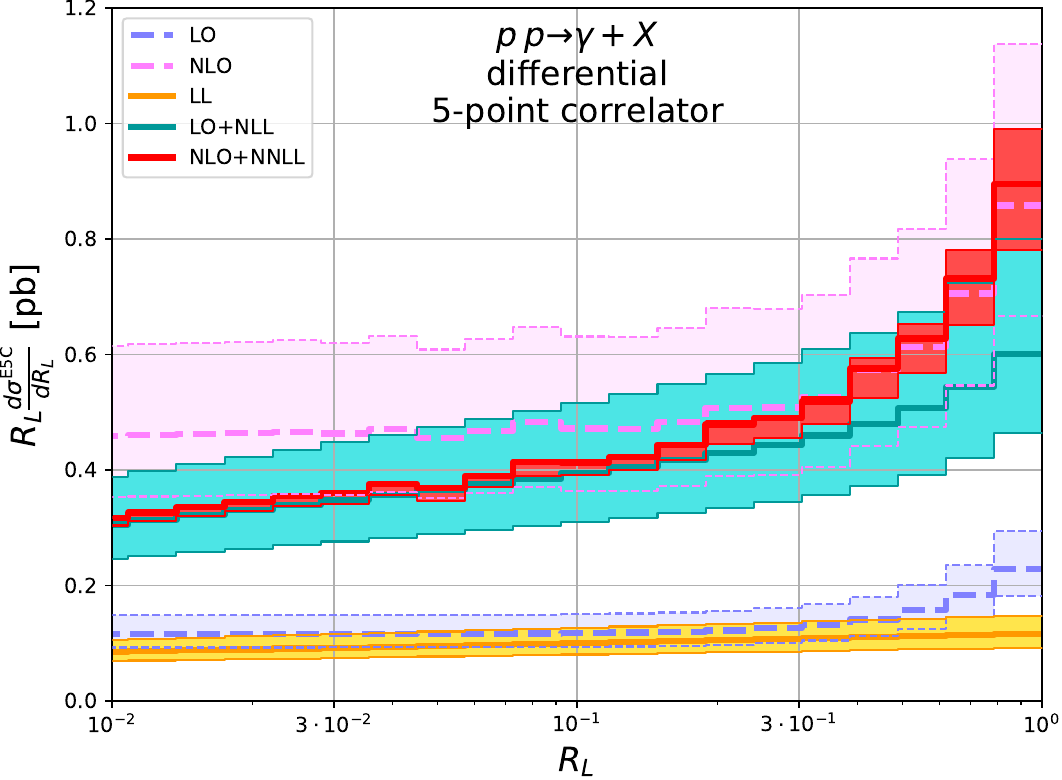}\\
	\includegraphics[width=0.31\textwidth]{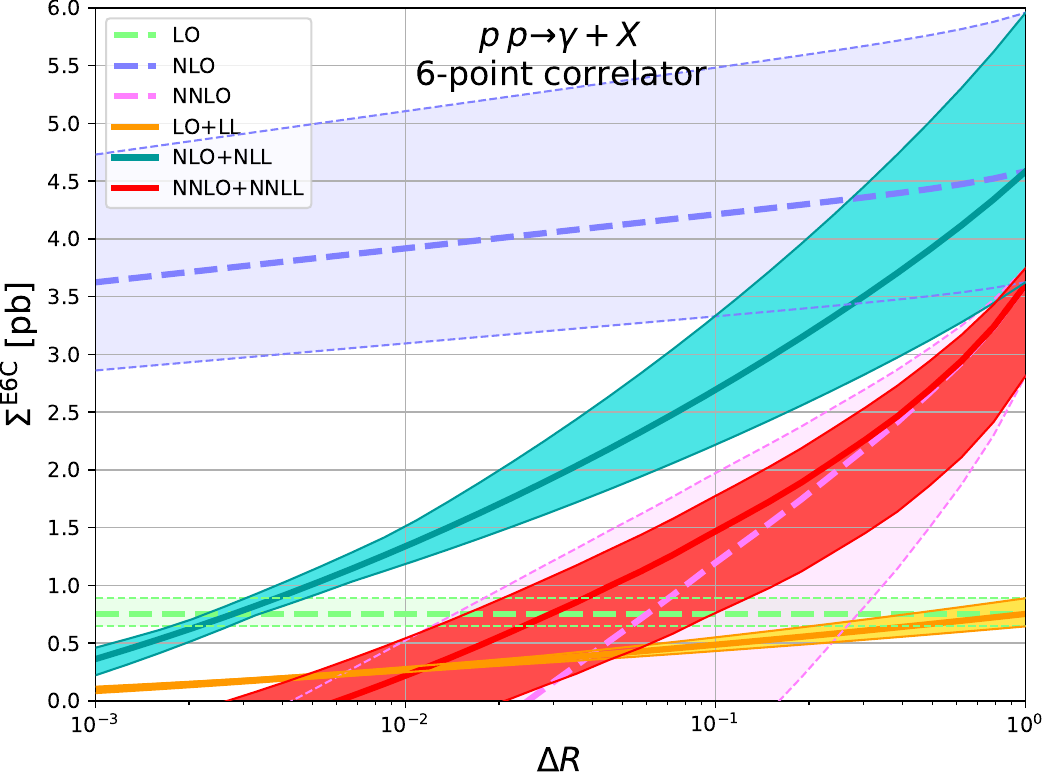}
	\includegraphics[width=0.31\textwidth]{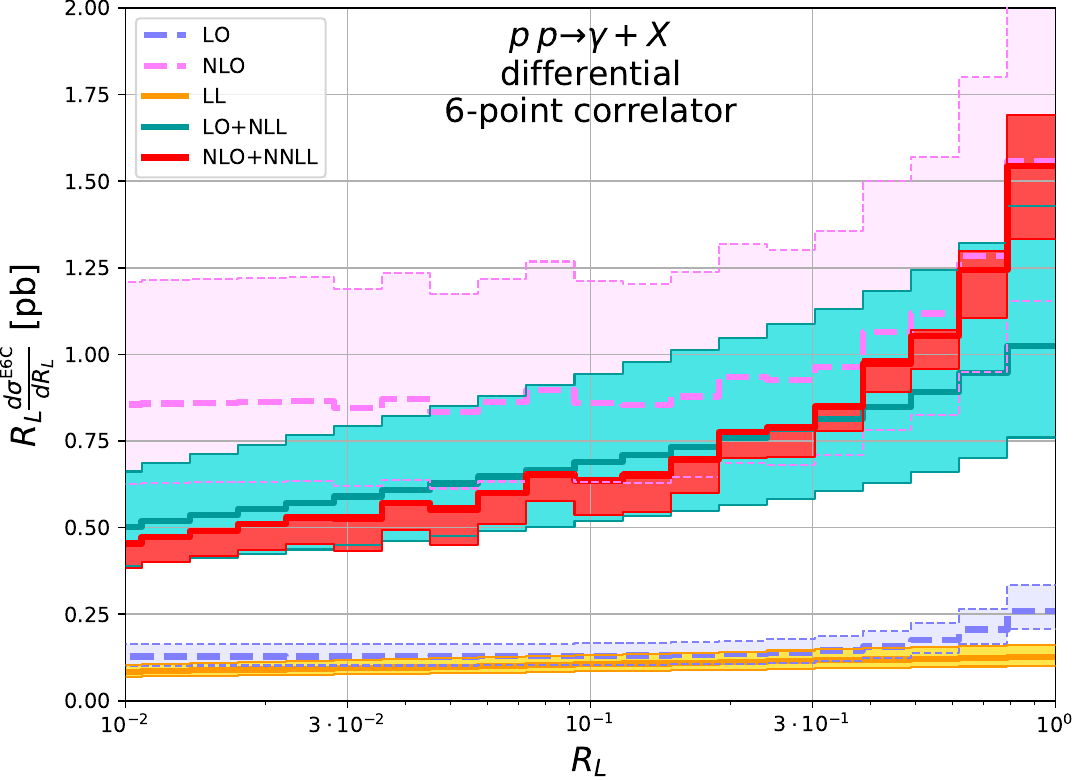}
	\caption{The cumulative (left) and differential (right) $N$-point correlators through (N)NLO(+NNLL), for $N=2$--$6$ (top to bottom).}
	\label{fig:AENCN_app}
\end{figure}

In \Fig{fig:AENCN_app}, we show the cumulative and differential matched distributions for all projected correlators up to six points, extending \Fig{fig:AEECN_full_text}.

\clearpage

\subsection[Slope Stability of the Ratio Distributions]{Slope Stability of the Ratio Distributions}\label{app:ratio_stability}

\begin{figure}[h]
	\centering
	\includegraphics[width=0.49\textwidth]{AEEC/AdEEC3RatioSubtracted.pdf}
	\includegraphics[width=0.49\textwidth]{AEEC/AdEEC4RatioSubtracted.pdf}
	\includegraphics[width=0.49\textwidth]{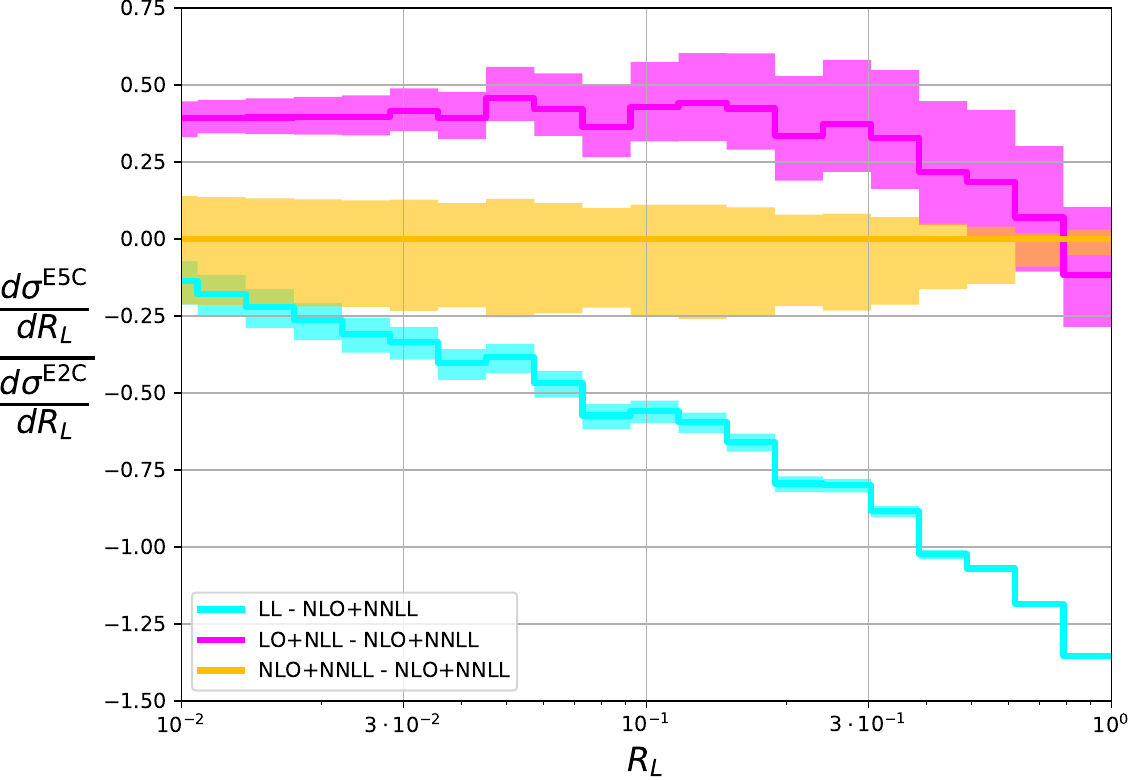}
	\includegraphics[width=0.49\textwidth]{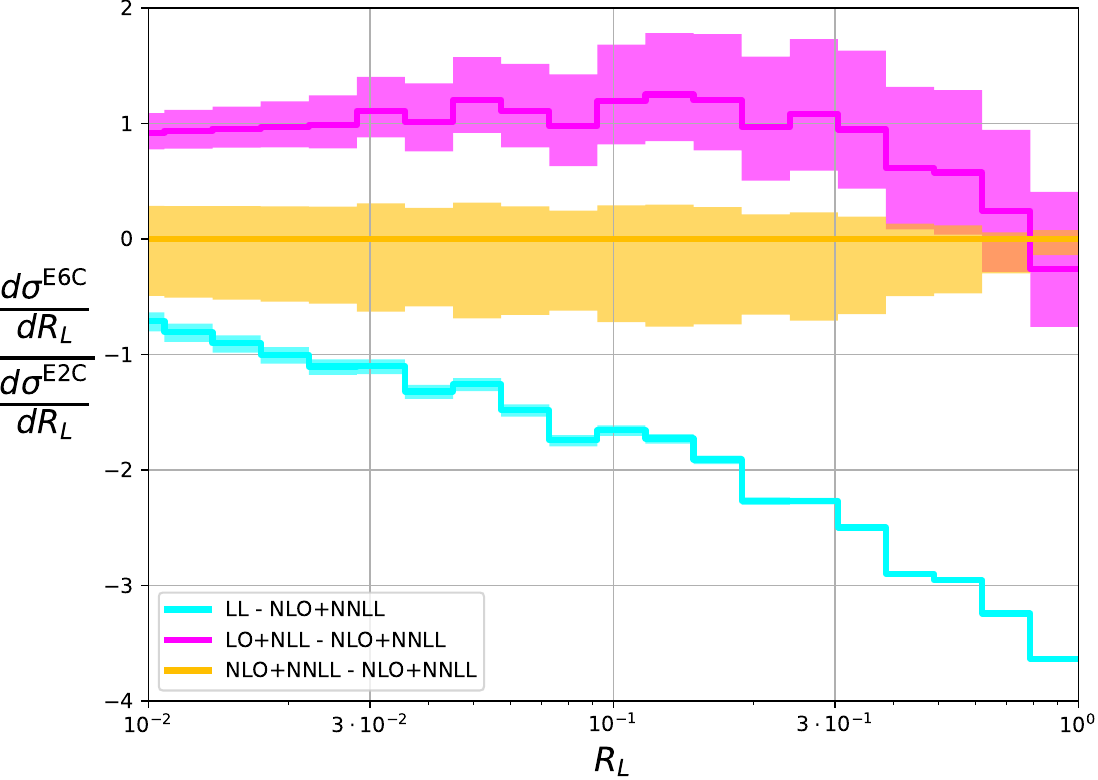}
	\caption{The ratios of the differential $N$-point correlators to the differential two-point correlator through NLO+NNLL, shown as differences relative to the NLO+NNLL curves.}
	\label{fig:AENCRatioSubtracted_app}
\end{figure}

In \Fig{fig:AENCRatioSubtracted_app}, the ratio distributions at each order are compared with the highest-order NLO+NNLL predictions for all ratio correlators ($N=3$--$6$), extending \Fig{fig:AENC_diffs_text}.

\clearpage

\subsection[Impact of Matching Corrections]{Impact of Matching Corrections}\label{app:matching_impact}

\begin{figure}[h]
	\centering
	\includegraphics[width=0.31\textwidth]{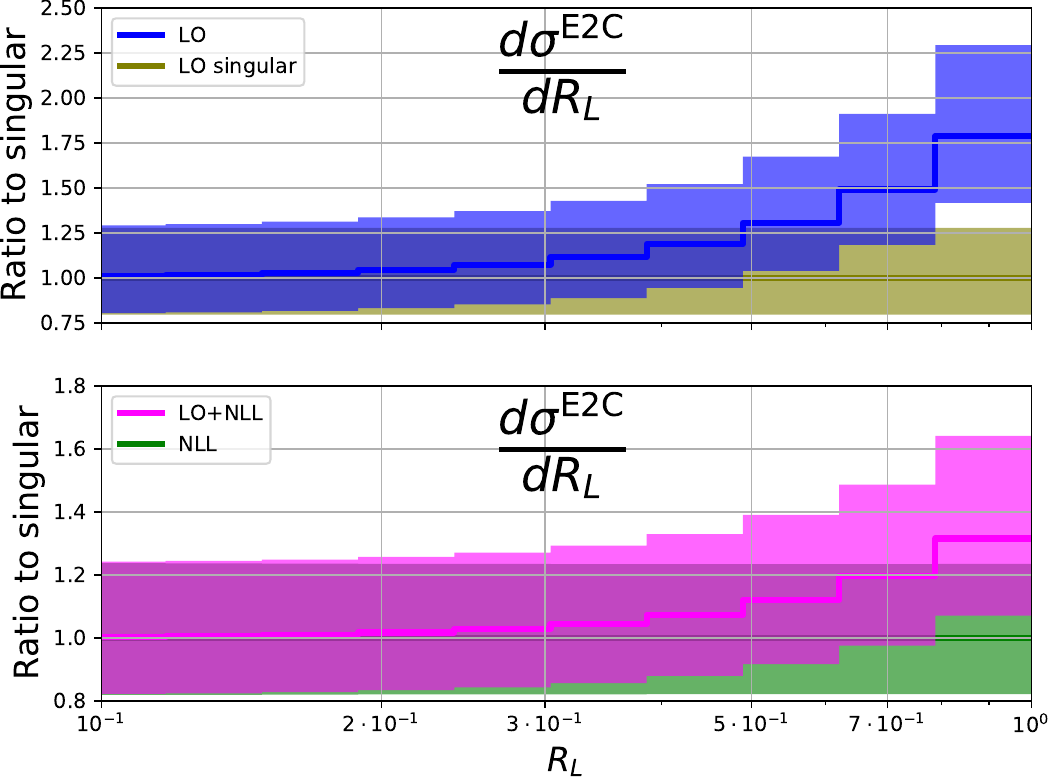}
	\includegraphics[width=0.31\textwidth]{AEEC/AdEEC2NLOPC.pdf}\\
	\includegraphics[width=0.31\textwidth]{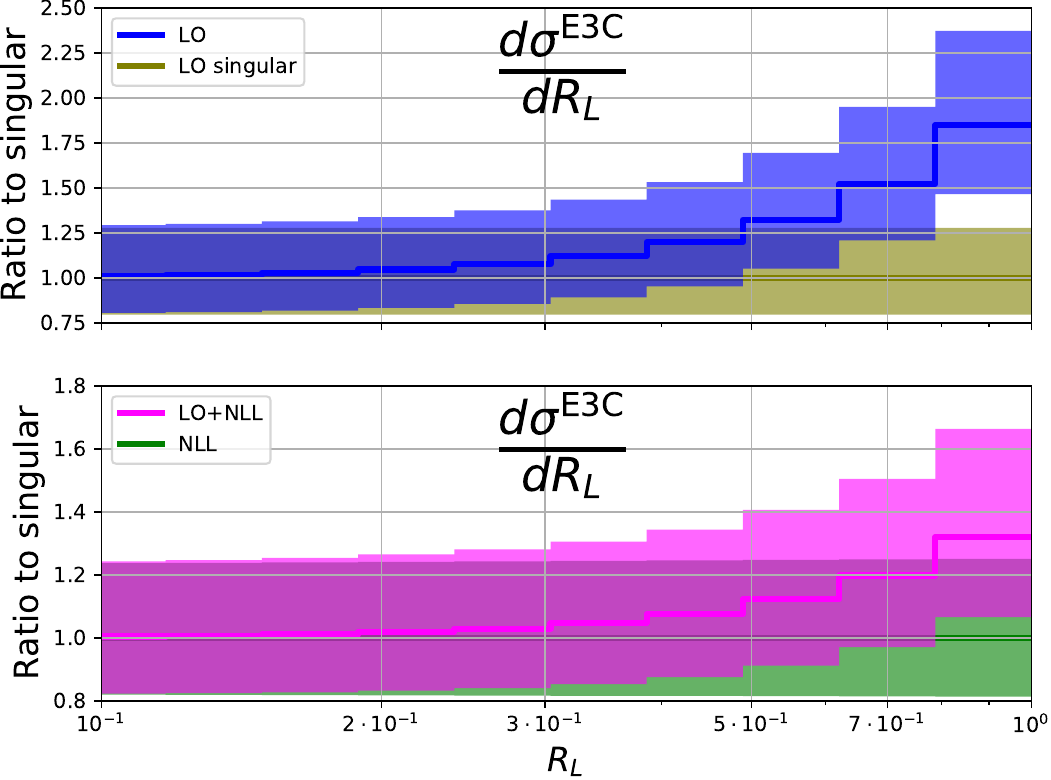}
	\includegraphics[width=0.31\textwidth]{AEEC/AdEEC3NLOPC.pdf}\\
	\includegraphics[width=0.31\textwidth]{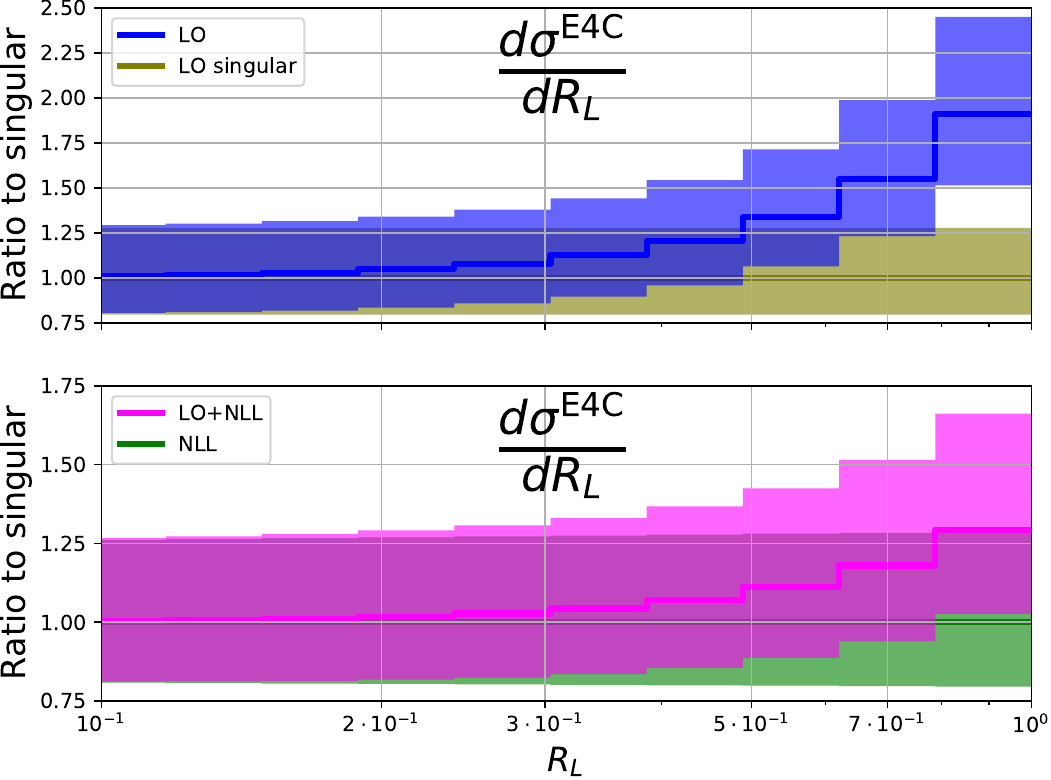}
	\includegraphics[width=0.31\textwidth]{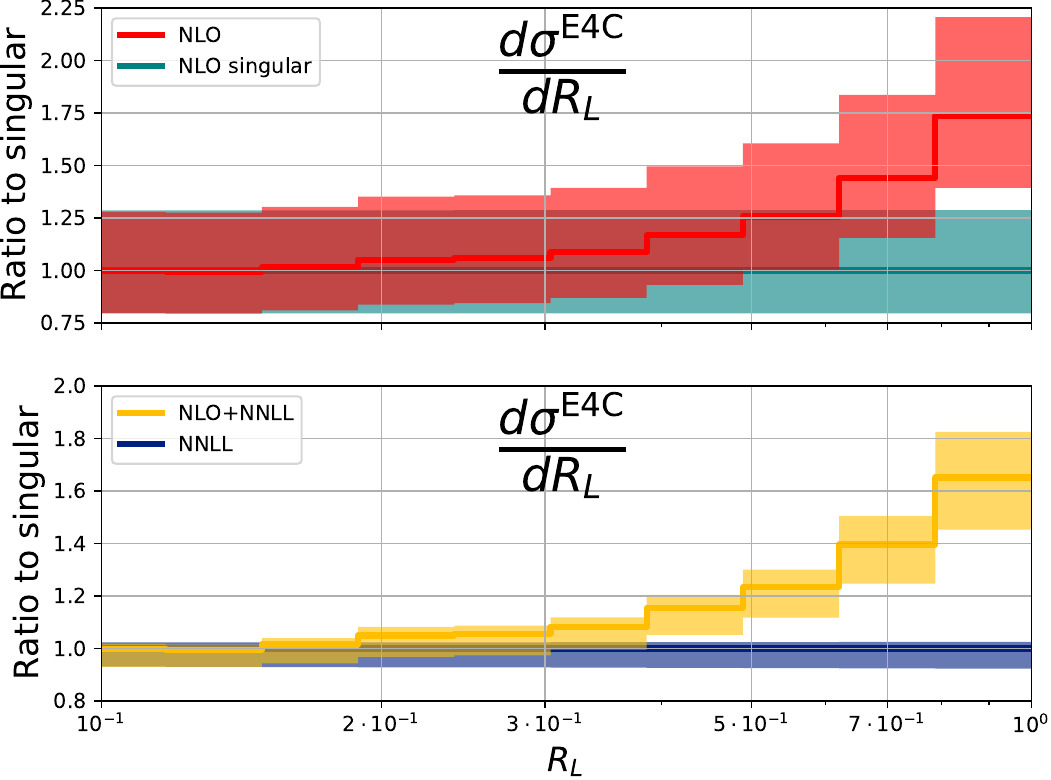}\\
	\includegraphics[width=0.31\textwidth]{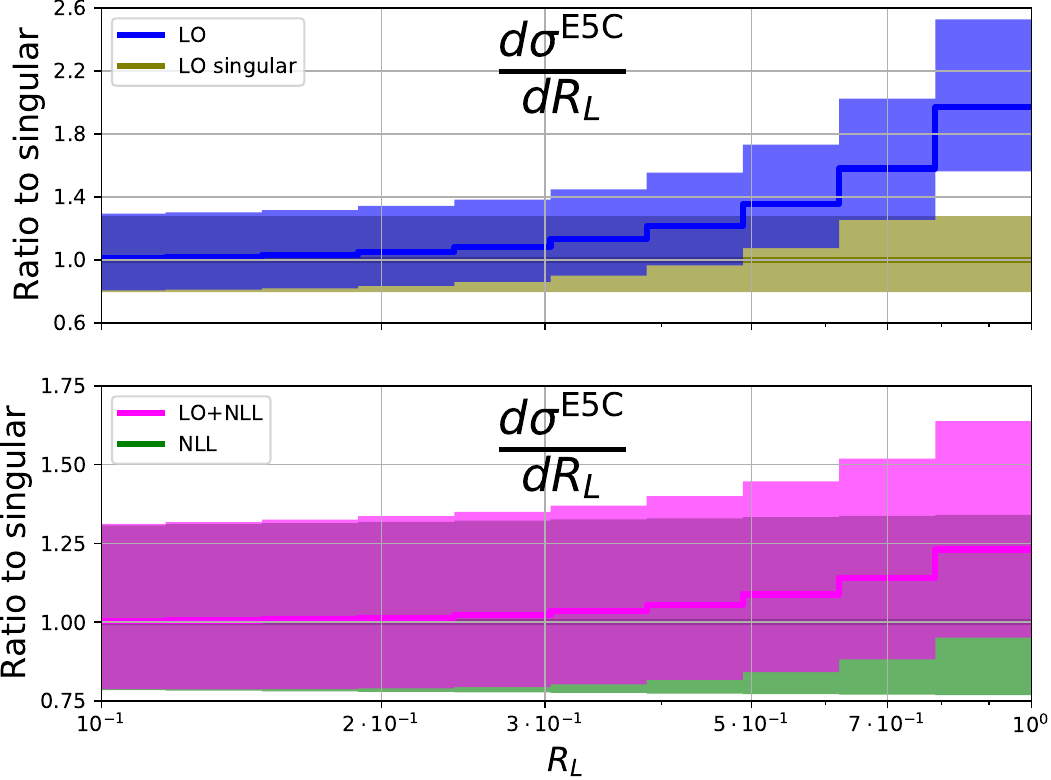}
	\includegraphics[width=0.31\textwidth]{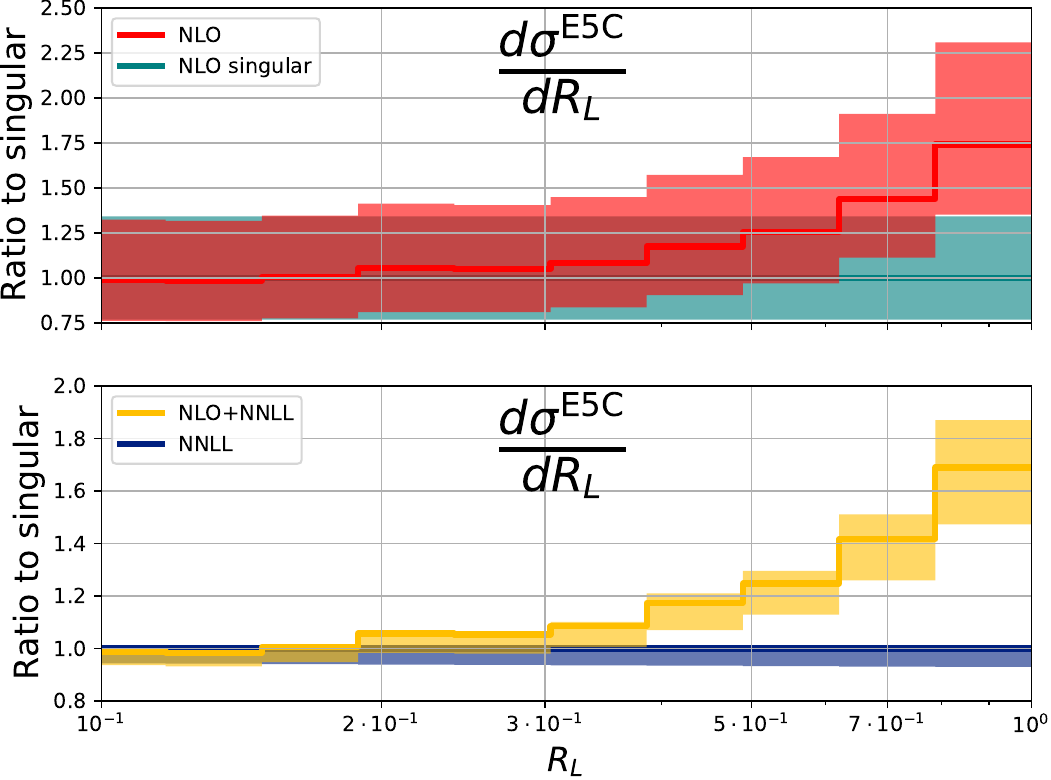}\\
	\includegraphics[width=0.31\textwidth]{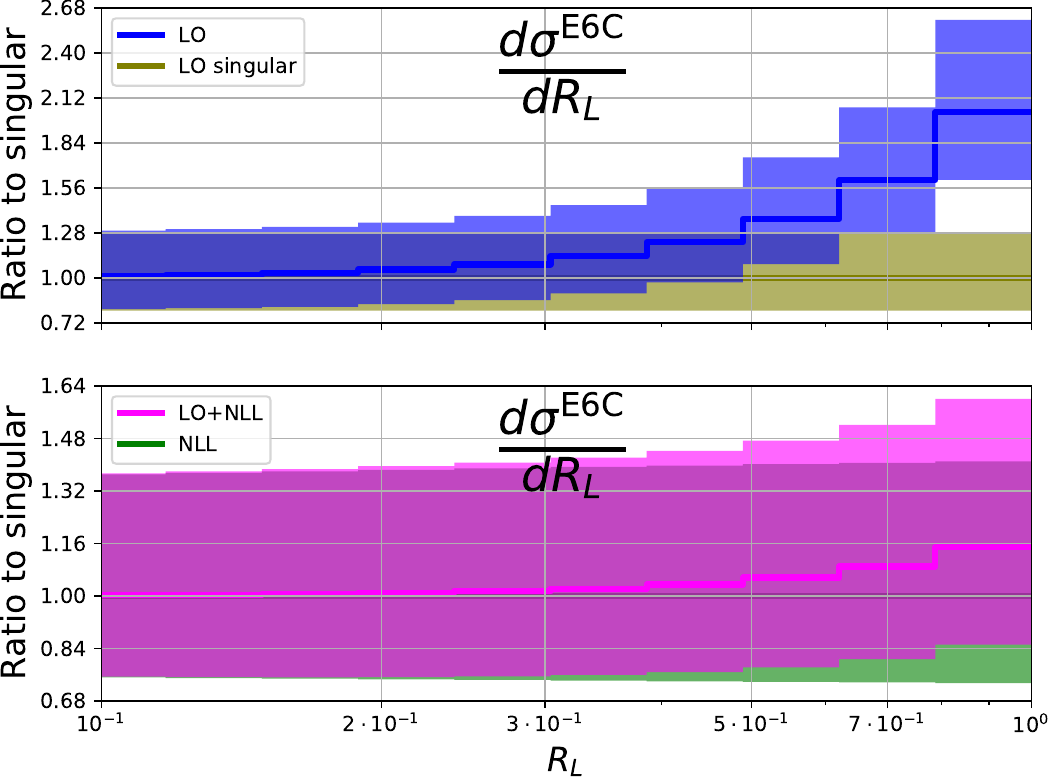}
	\includegraphics[width=0.31\textwidth]{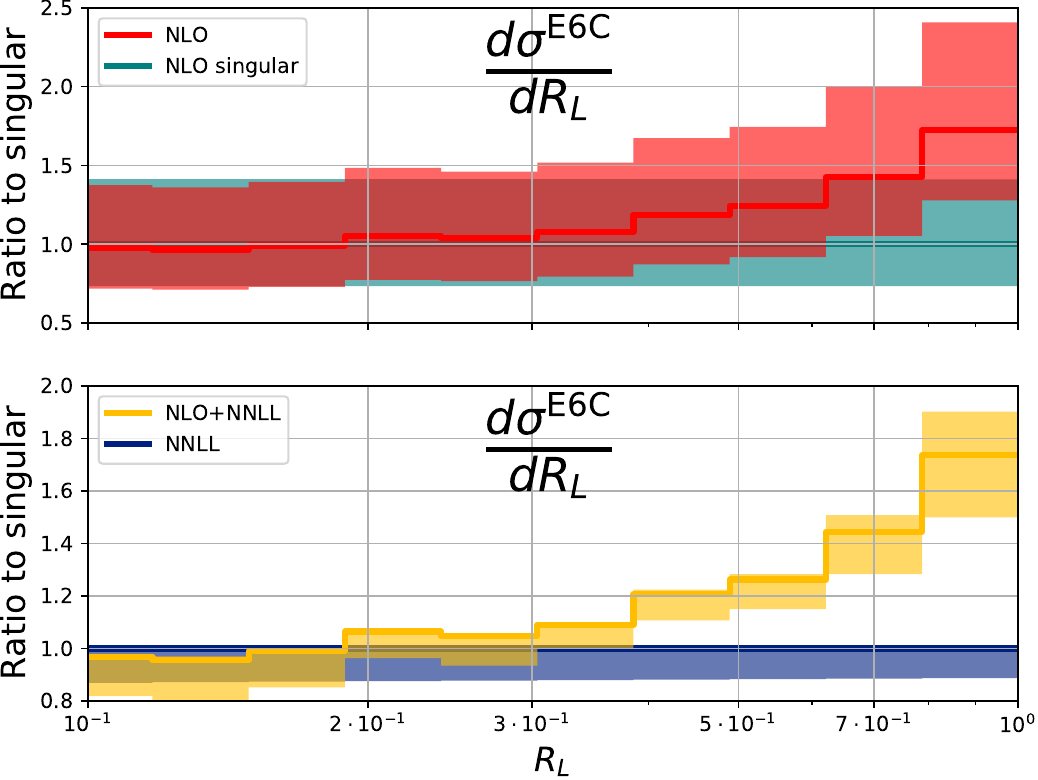}
	\caption{The effect of the matching (non-singular) corrections on the differential ENCs at LO(+NLL) (left) and NLO(+NNLL) (right), for $N=2$--$6$ (top to bottom).}
	\label{fig:AdEECNPC}
\end{figure}

\begin{figure}
	\centering
	\includegraphics[width=0.34\textwidth]{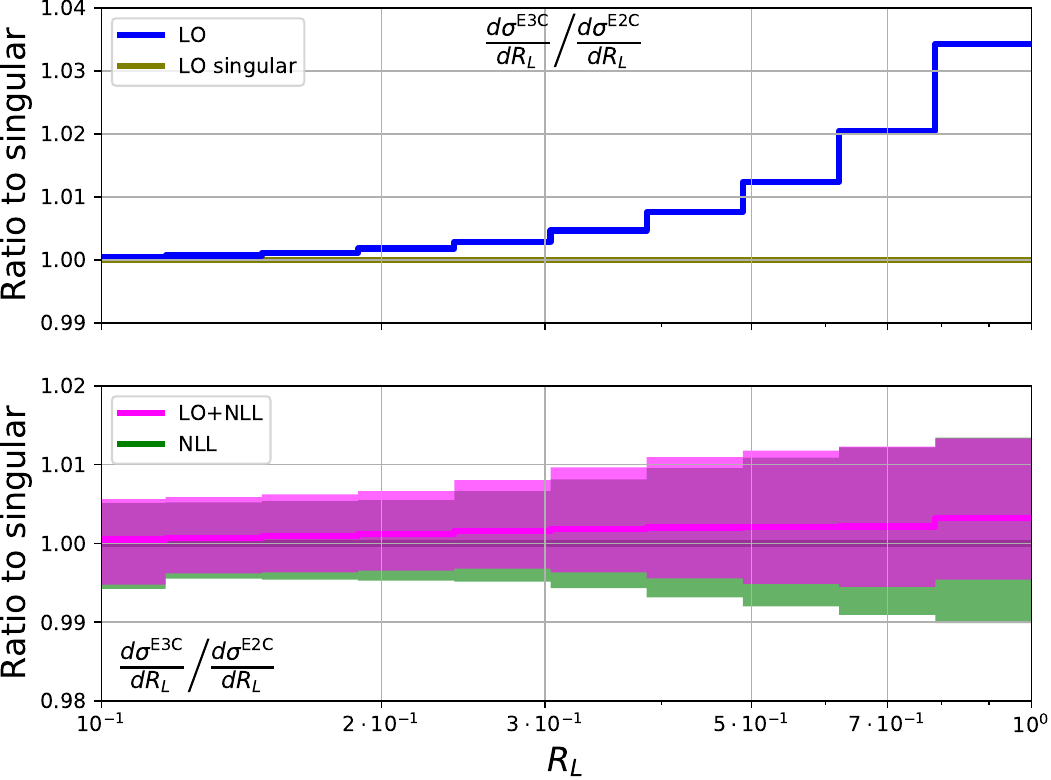}
	\includegraphics[width=0.34\textwidth]{AEEC/AdEEC3NLORatioPC.pdf}
	\includegraphics[width=0.34\textwidth]{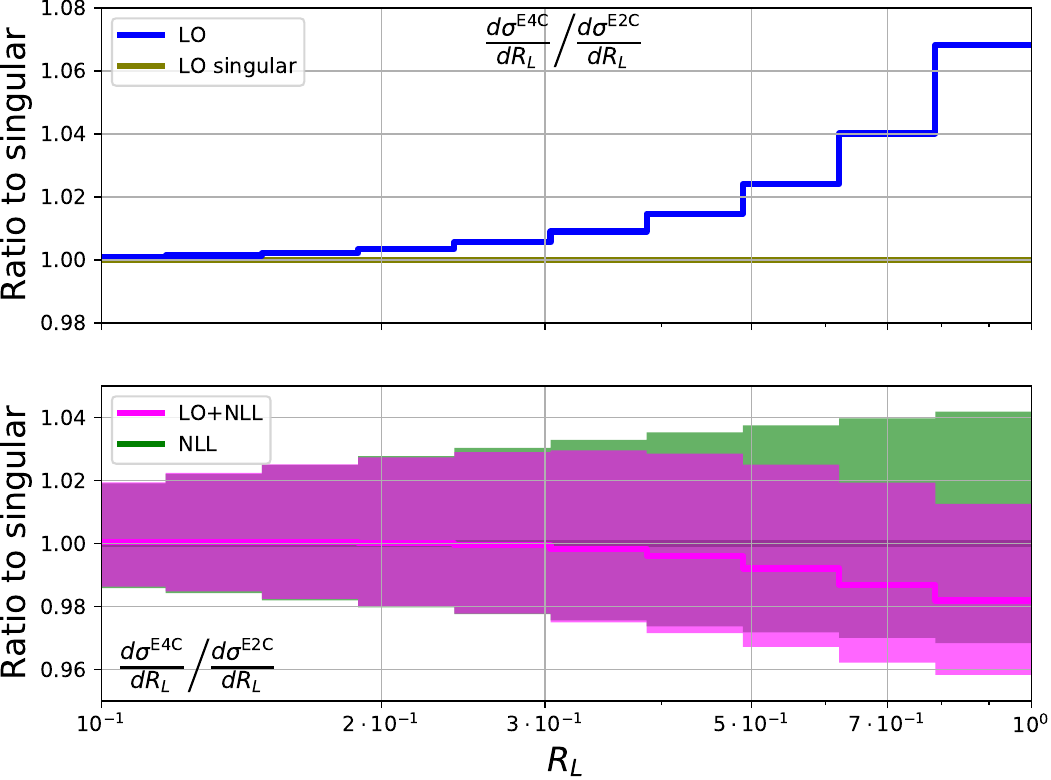}
	\includegraphics[width=0.34\textwidth]{AEEC/AdEEC4NLORatioPC.pdf}
	\includegraphics[width=0.34\textwidth]{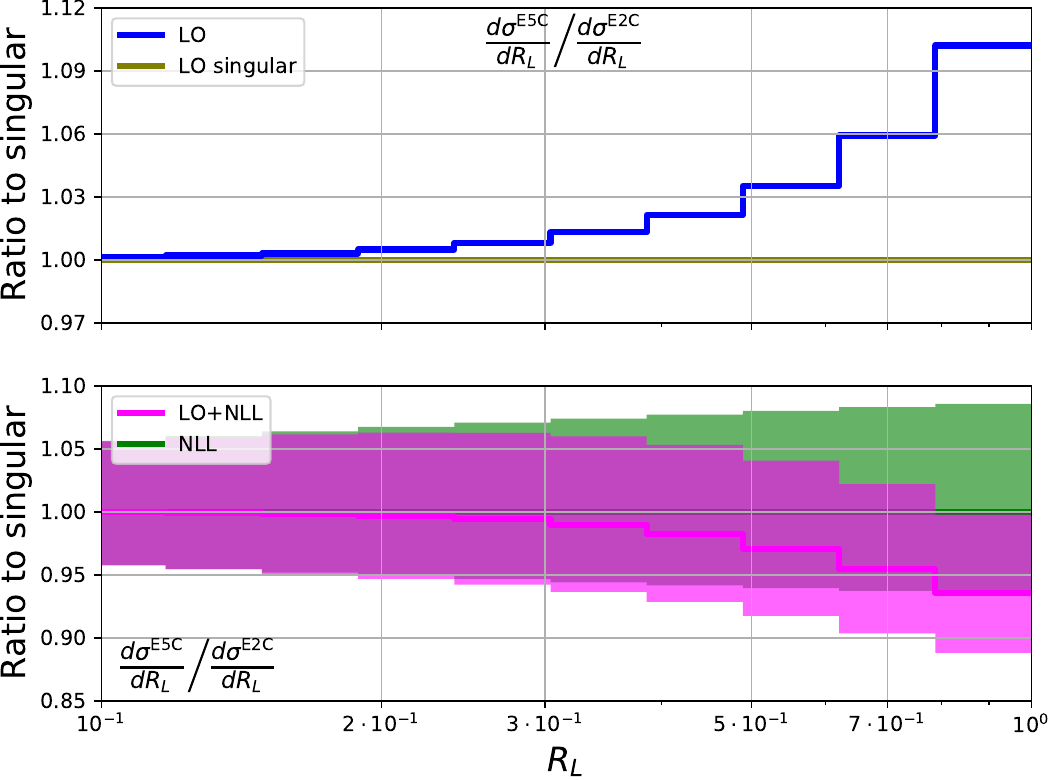}
	\includegraphics[width=0.34\textwidth]{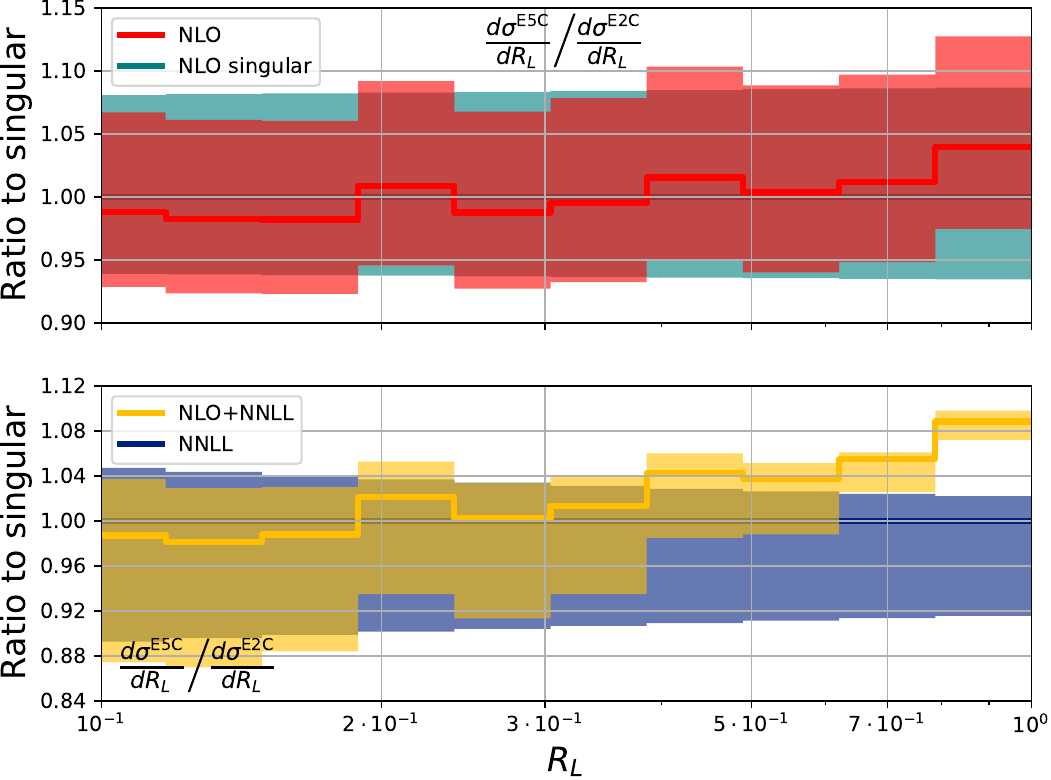}
	\includegraphics[width=0.34\textwidth]{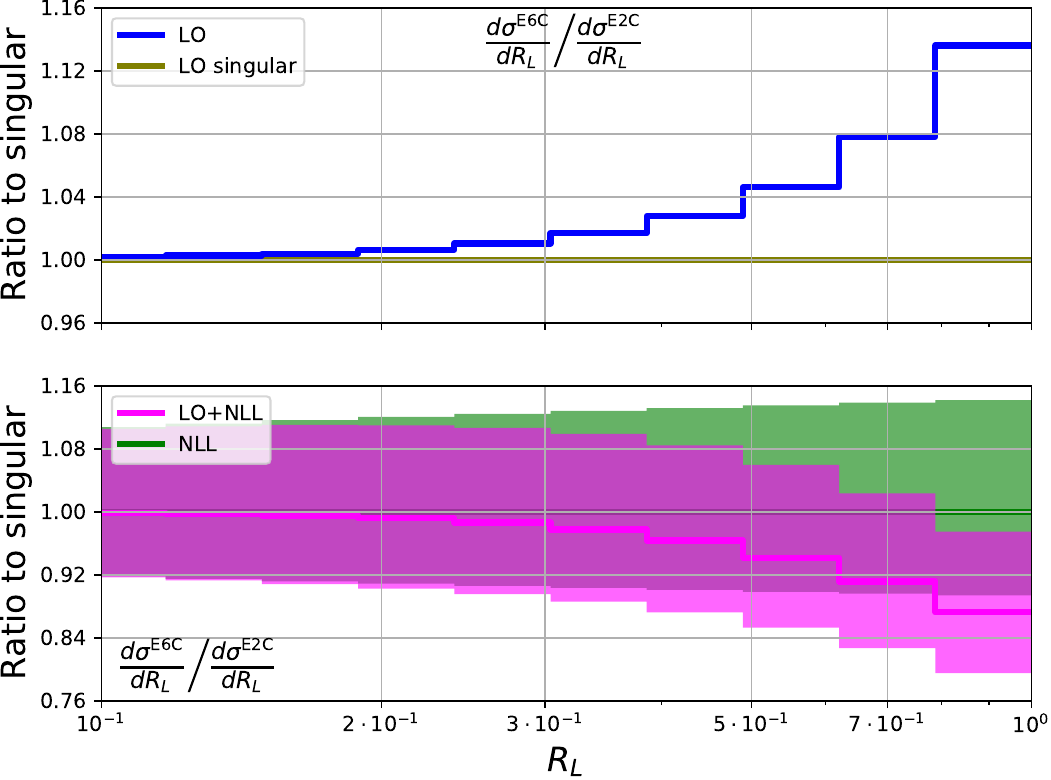}
	\includegraphics[width=0.34\textwidth]{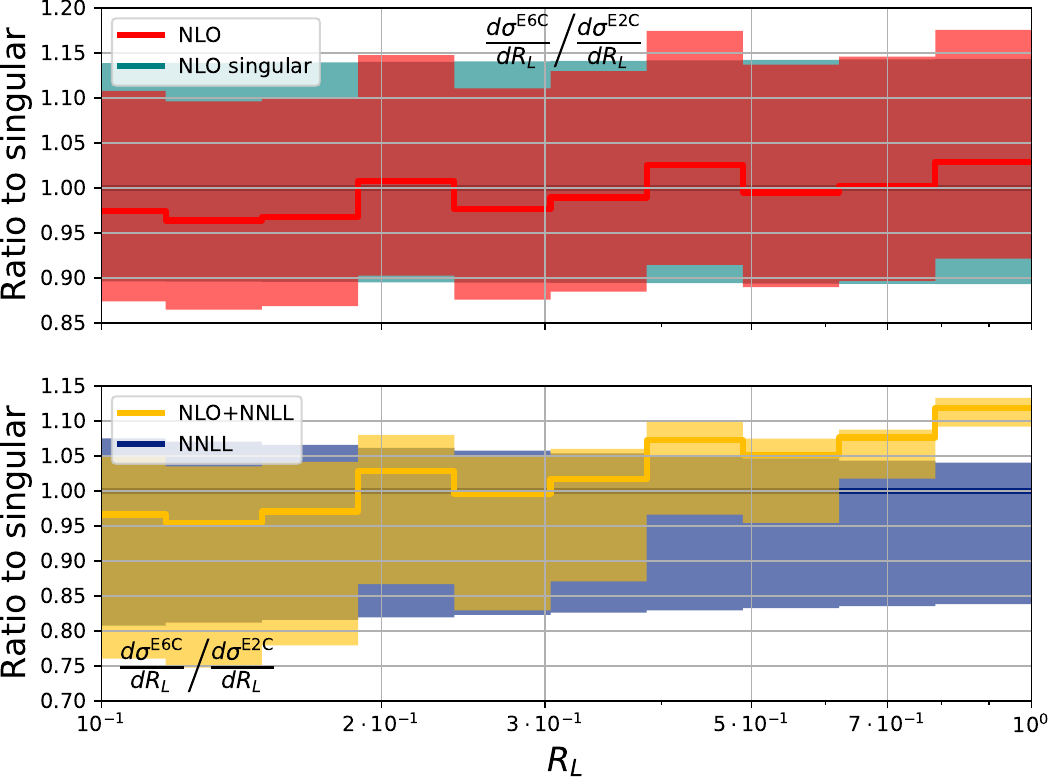}
	\caption{The effect of the matching (non-singular) corrections on the differential ENC/EEC ratios at LO(+NLL) (left) and NLO(+NNLL) (right), for $N=3$--$6$ (top to bottom).}
	\label{fig:AdEECNRatioPC}
\end{figure}

In \Fig{fig:AdEECNPC} and \Fig{fig:AdEECNRatioPC}, we show the impact of the matching corrections on the differential distributions ($N=2$--$6$) and on the ratio correlators ($N=3$--$6$), extending \Fig{fig:AdEECNPC_text} and \Fig{fig:AdEECNRatioPC_text}.

\clearpage

\clearpage

\bibliographystyle{JHEP}
\bibliography{EEC_ref.bib}

\end{document}